\documentclass[letterpaper, aps, prd, twocolumn, superscriptaddress, showpacs, nofootinbib,floatfix]{revtex4}
\pdfoutput=1

\usepackage{graphicx}
\usepackage{bm}
\usepackage{amssymb}
\usepackage{amsmath}
\usepackage{amsfonts}
\usepackage{dcolumn}
\usepackage{natbib}
\usepackage{color}
\usepackage{calc}

\newcommand{\raiseentry}[1]{\smash{\raise 0.7 em \hbox{#1}}}

\newcommand{\scri}{\ensuremath{\mathcal{J}}}

\newcommand{\Ibardotdotdot}{\dddot{I}} 
\newcommand{\Ibardotdot}{\ddot{I}} 
\newcommand{\Ibardotdottilde}{\tilde{\ddot{I}}} 
\newcommand{\Ytwo}{{{}^{(-2)}Y}}

\newcommand{\lowentry}[1]{\smash{\lower 1.5 ex \hbox{#1}}}

\newcommand{\scis}[2]{#1 \!\! \times \!\! 10^{#2}}

\def\apj{Astrophys. J.}
\def\apjl{Astrophys. J. Lett.}

\def\aap{Astron. Astrophys. }

\def\physrep{Phys. Rep. }
\def\mnras{Mon. Not. Roy. Astron. Soc. }

\def\prl{Phys. Rev. Lett.}
\def\prd{Phys. Rev. D.}

\def\cqg{Class. Quantum Grav.}

\newenvironment{equationarray}
{\arraycolsep 0.14 em
\begin{eqnarray}}
{\end{eqnarray}}

\newenvironment{equationarray*}
{\arraycolsep 0.14 em
\begin{eqnarray*}}
{\end{eqnarray*}}

\newcommand{\code}[1]{{\tt #1}}

\begin{document}

\title{Correlated Gravitational Wave and Neutrino Signals\\ from
   General-Relativistic Rapidly Rotating Iron Core Collapse}

\author{C.~D.~Ott}
\thanks{Alfred P. Sloan Research Fellow}
\email{cott@tapir.caltech.edu}
\affiliation{TAPIR, MC 350-17, California Institute of Technology, 
1200 E California Blvd., Pasadena, CA 91125, USA}
\affiliation{Kavli Institute for the Physics and Mathematics of the Universe, 
The University of Tokyo, Kashiwa, Japan 277-8583}
\affiliation{Center for Computation \& Technology, Louisiana State
  University, 216 Johnston Hall, Baton Rouge, LA 70803, USA}

\author{E.~Abdikamalov}
\affiliation{TAPIR, MC 350-17, California Institute of Technology, 
1200 E California Blvd., Pasadena, CA 91125, USA}

\author{E.~O'Connor}
\affiliation{TAPIR, MC 350-17, California Institute of Technology, 
1200 E California Blvd., Pasadena, CA 91125, USA}

\author{C.~Reisswig}
\affiliation{TAPIR, MC 350-17, California Institute of Technology, 
1200 E California Blvd.,
Pasadena, CA 91125, USA}

\author{R.~Haas}
\affiliation{TAPIR, MC 350-17, California Institute of Technology, 
1200 E California Blvd., Pasadena, CA 91125, USA}

\author{P.~Kalmus} \affiliation{LIGO Laboratory, MC 100-36, California
  Institute of Technology, 1200 E California Blvd., Pasadena, CA
  91125, USA} \affiliation{TAPIR, MC 350-17, California Institute of
  Technology, 1200 E California Blvd., Pasadena, CA 91125, USA}

\author{S.~Drasco}
\affiliation{California Polytechnic State University,
1 Grand Ave., San Luis Obispo, CA 93407, USA}
\affiliation{TAPIR, MC 350-17, California Institute of Technology, 
1200 E California Blvd., Pasadena, CA 91125, USA}

\author{A. Burrows} \affiliation{Department of Astrophysical Sciences,
  4 Ivy Lane, Princeton, NJ, 08544, USA}

\author{E. Schnetter}
\affiliation{Perimeter Institute for Theoretical Physics, 31 Caroline
  St.~N., Waterloo, ON N2L 2Y5, Canada}
\affiliation{Department of Physics, University of Guelph, 50 Stone
  Road East, Guelph, ON N1G 2W1, Canada}
\affiliation{Center for Computation \& Technology, Louisiana State
  University, 216 Johnston Hall, Baton Rouge, LA 70803, USA}

\date{April 2, 2012}


\begin{abstract}
  We present results from a new set of 3D general-relativistic
  hydrodynamic simulations of rotating iron core collapse. We assume
  octant symmetry and focus on axisymmetric collapse, bounce, the
  early postbounce evolution, and the associated gravitational wave
  (GW) and neutrino signals.  We employ a finite-temperature nuclear
  equation of state, parameterized electron capture in the collapse
  phase, and a multi-species neutrino leakage scheme after bounce.
  The latter captures the important effects of deleptonization,
  neutrino cooling and heating and enables approximate predictions for
  the neutrino luminosities in the early evolution after core
  bounce. We consider $12$-$M_\odot$ and $40$-$M_\odot$ presupernova
  models and systematically study the effects of (\emph{i}) rotation,
  (\emph{ii}) progenitor structure, and (\emph{iii}) postbounce
  neutrino leakage on dynamics, GW, and, neutrino signals. We
  demonstrate, that the GW signal of rapidly rotating core collapse is
  practically independent of progenitor mass and precollapse
  structure.  Moreover, we show that the effects of neutrino leakage
  on the GW signal are strong only in nonrotating or slowly rotating
  models in which GW emission is not dominated by inner core dynamics.
  In rapidly rotating cores, core bounce of the centrifugally-deformed
  inner core excites the fundamental quadrupole pulsation mode of the
  nascent protoneutron star. The ensuing global oscillations ($f \sim
  700-800\,\mathrm{Hz}$) lead to pronounced oscillations in the GW
  signal and correlated strong variations in the rising luminosities
  of antineutrino and heavy-lepton neutrinos. We find these features
  in cores that collapse to protoneutron stars with spin periods
  $\lesssim$$2.5\,\mathrm{ms}$ and rotational energies sufficient to
  drive hyper-energetic core-collapse supernova explosions.
  Hence, GW or neutrino observations of a core collapse
    event could deliver strong evidence for or against rapid core
    rotation.  Joint GW + neutrino observations would allow to make
    statements with high confidence. Our estimates suggest that the
  GW signal should be detectable throughout the Milky Way by advanced
  laser-interferometer GW observatories, but a water-Cherenkov
  neutrino detector would have to be of near-megaton size to observe
  the variations in the early neutrino luminosities from a core
  collapse event at 1\,kpc.
\end{abstract}


\maketitle


\section{Introduction}
\label{section:introduction}

The collapse of massive stars ($8\,M_\odot \lesssim M \lesssim
130\,M_\odot$ at zero-age main-sequence [ZAMS]) proceeds in a generic
way: the electron-degenerate core separates into a
homologously collapsing \emph{inner core} and a supersonically
collapsing outer core. At densities near nuclear, the nuclear equation
of state (EOS) stiffens, leading to core bounce and the sudden
reversal of the inner core's infall velocity, imparting kinetic energy
onto the supernova shock. The latter initially rapidly propagates
outward into the still infalling and sonically disconnected outer
core, but soon, weakened by the work afforded to the dissociation of
heavy nuclei and by neutrino losses, succumbs to the outer core's ram
pressure, stalls, and turns into an accretion shock. The shock must be
revived to drive an explosion and produce a core-collapse supernova,
leaving behind a neutron star; otherwise the protoneutron star (PNS)
quickly, within $\sim$$0.5-3~\mathrm{s}$, accumulates sufficient mass
to collapse to a black hole~\cite{janka:07,oconnor:11}.

The currently favored mechanism for shock revival (see, e.g.,
\cite{janka:07,mueller:12a,bruenn:09,nordhaus:10,takiwaki:11b}) relies
on a combination of neutrino energy deposition in the region behind
the shock with multi-dimensional convective and advective-acoustic
instabilities. However, alternative mechanisms exist (e.g.,
\cite{sagert:09,burrows:06}). In particular, massive stars with rapid
rotation in their core may experience magnetorotational explosions in
which magnetic fields, amplified by rapid rotation (e.g., via the
magnetorotational instability \cite{balbus:91,obergaulinger:09}), drive
powerful bipolar explosions
\cite{bisno:70,leblanc:70,burrows:07b,takiwaki:11}.  Most massive
stars are expected to have slowly spinning cores
\cite{ott:06spin,heger:05}, but of order $1\%$ \cite{woosley:06} may
have sufficient core angular momentum for a magnetorotational
explosion. This special group of rapidly spinning massive stars has
also been proposed as progenitors of long gamma-ray bursts (GRBs)
\cite{woosley:06,yoon:06}, since all currently discussed long-GRB
models require rapid rotation in combination with a compact,
hydrogen-poor stellar envelope to produce the characteristic Type Ib/c
supernova features observed in GRB afterglows (e.g.,
\cite{wb:06}). Rapid rotation could also play a key role in very
energetic non-GRB core-collapse supernovae (e.g., \cite{soderberg:10})
whose total energy output may be enhanced by magnetar spindown via
dipole radiation \cite{kasen:10,woosley:10} or by a magnetohydrodynamic
(MHD) propeller ejecting fallback material~\cite{piro:11}.

Probing the role of rotation at the heart of core-collapse supernovae
and long GRBs with electromagnetic waves is difficult, since they only
provide an image of optically-thin regions far away from the
core. Much more direct information is carried by neutrinos and
gravitational waves.  Both are emitted inside the core where the
deciding dynamics occur and travel to observers on Earth essentially
unimpeded by intervening material. Neutrinos carry primarily
thermodynamic and structural information about their source (e.g.,
\cite{marek:09b,ott:08,brandt:11,lund:10}) while gravitational waves
(GWs), owing to their quadrupole nature, provide complementing
information on the multi-dimensional dynamics of the core
(e.g.,\cite{ott:09,marek:09b,yakunin:10,murphy:09,kotake:11b}). The
combined observation of neutrinos and GWs from the next nearby core
collapse event may place stringent constraints on rotation in the core
in particular and on core-collapse supernova physics in general (see,
e.g., \cite{roever:09,marek:09b,ott:09b,logue:12} and references
therein). This, however, will require robust theoretical neutrino and
GW signal predictions to be contrasted with observations.

Reliable modeling of the GW signal of rotating core collapse and
bounce requires, in the ideal case, 3D general relativity,
microphysical EOS, progenitor data from stellar evolutionary
calculations, energy-dependent multi-species neutrino transport, and
MHD. Such complete models of stellar collapse still pose a formidable
technical and computational challenge, but much progress in
understanding the GW signature of core collapse and the subsequent
core-collapse supernova evolution has been made in the past decade.
We refer the reader to recent overview articles by
Kotake~\cite{kotake:11b} and Ott~\cite{ott:09} for a comprehensive
discussion of the range of potential GW emission processes in
core-collapse supernovae and focus in the following only on GW
emission from rotating core collapse, bounce, and the early postbounce
evolution.

Pioneering axisymmetric (2D) work on the GW signal of rotating core
collapse and bounce relied on Newtonian gravity with
\cite{mueller:82,moenchmeyer:91,kotake:03,ott:04} or without
\cite{zwerger:97,yamadasato:95} the inclusion of microphysical aspects
such as a finite-temperature nuclear EOS and/or neutrino
treatment. These studies identified three types of GW signal
morphologies: \emph{Type I} signals exhibit a prominent
  large negative spike associated with the tremendous deceleration
  and re-expansion of the inner core at bounce and a subsequent short
  ring-down feature emitted as the core settles into its new
  equilibrium.  \emph{Type II} show multiple pronounced peaks
correlated with repeated large-scale postbounce contractions and
re-expansions (``multiple bounces'') of extremely rapidly rotating
inner cores that experience bounce due to the centrifugal
force. \emph{Type III} signals are associated with very small inner
core mass, leading to a low-amplitude bounce spike with typically a
positive peak value.

Early 2D general-relativistic (GR) models also used simplified
microphysics \cite{dimmelmeier:02,shibata:04}, but more recent ones
\cite{dimmelmeier:07,dimmelmeier:08,cerda:08} employed microphysical
EOS and deleptonization during collapse (but not after bounce). The
latter set of studies showed that, due to a combination of
deleptonization during collapse (which reduces the inner core mass)
and general relativistic gravity, the GW signal of rotating core
collapse has a generic morphology corresponding to the previously
identified Type I signal shape. The multiple-bounce Type II dynamics
and GW signals do not occur, while Type III signals may result from
the accretion-induced collapse of rapidly rotating massive O-Ne white
dwarfs \cite{abdikamalov:10}. These studies also demonstrated that the
peak signal amplitude and characteristic GW frequency are determined
primarily by the mass and angular momentum of the inner core at
bounce. Waveform details are expected to depend on the nuclear EOS,
progenitor core structure and thermodynamics, angular momentum
distribution in the core, and
deleptonization history during collapse
\cite{dimmelmeier:08,abdikamalov:10,scheidegger:10b}.

MHD effects on the dynamics and GW signature of rotating collapse and
bounce have been considered by
\cite{kotake:04,obergaulinger:06a,obergaulinger:06b,shibata:06,cerda:08,scheidegger:08,scheidegger:10b,takiwaki:11}. While
MHD can be crucial in driving an explosion in the postbounce phase,
these studies showed that rotating collapse, bounce and the very early
postbounce evolution (within $\lesssim 20-25\,\mathrm{ms}$ of bounce)
and the associated GW signal are essentially unaffected by MHD
effects unless the precollapse magnetization is set to extreme values
$\gtrsim 10^{12}\,\mathrm{G}$. Even in extreme cases, current stellar
evolutionary studies suggest $B \lesssim 10^9\,\mathrm{G}$
\cite{heger:05,woosley:06}.

3D core collapse simulations were first attempted by Bonazzola \&
Marck~\cite{bonazzola:93}, who studied only the collapse phase. The
first set of studies investigating the GW signal of bounce and
postbounce evolution in 3D started from 2D simulations of the collapse
phase, relied on simplified microphysics and were carried out in
Newtonian \cite{rmr:98} or GR gravity \cite{shibata:05}.  They focused
on extremely rapidly rotating configurations and found nonaxisymmetric
dynamics developing in models with ratios of rotational kinetic energy
to gravitational energy $T/|W|\gtrsim 27\%$ (see, e.g.,
\cite{andersson:03} for a review of rotational instabilities). More
recent 3D simulations that started at the onset of collapse and
included microphysics and deleptonization before bounce
\cite{ott:07prl,ott:07cqg,scheidegger:08,scheidegger:10b} showed that
the collapse phase proceeds axisymmetrically even if the progenitor
core is very rapidly spinning. They also found the development of
nonaxisymmetric dynamics after bounce at low values of $T/|W|$ of
$\sim$5-13\% via a dynamical rotational shear instability
\cite{watts:05,fu:11}.

Several 2D and 3D simulations of rapidly rotating core collapse
included neutrino cooling and deleptonization in the postbounce phase
via approximate leakage schemes
\cite{moenchmeyer:91,kotake:03,kotake:04,scheidegger:10b,takiwaki:11}
or multi-group flux-limited neutrino diffusion
\cite{walder:05,ott:06spin,burrows:07b}. However, the effect of
cooling/deleptonization on the nonaxisymmetric postbounce dynamics 
and associated GW emission was
investigated only by Scheidegger~et~al.~\cite{scheidegger:10b}.  They
found up to ten times stronger GW emission if deleptonization/cooling
was included. The influence of deleptonization and cooling on
the earlier essentially axisymmetric dynamics and GW emission has not
been addressed to date. Furthermore, the neutrino signature of rapidly
rotating core collapse has been studied to date only on the basis of
1.5D (spherical symmetry plus approximate rotation)
\cite{thompson:05} that cannot account for the centrifugal deformation
of the PNS and GW emission.

In this article, we present new results for the GW signature of
rotating core collapse and first results on its early postbounce
neutrino signature obtained from multi-D simulations.  We study the
dependence of collapse, bounce, and early postbounce evolution, GW and
neutrino signal on precollapse rotation, progenitor structure and
thermodynamics, and on the inclusion of deleptonization, neutrino
cooling and heating in the early postbounce phase.

Our new results are based on a set of full GR simulations with our
\code{Zelmani} core-collapse simulation package, which is based on the
open-source \code{Einstein Toolkit} \cite{et:11,einsteintoolkitweb}
and includes deleptonization during collapse and neutrino cooling,
heating, and deleptonization after bounce via a three-species
energy-averaged neutrino leakage scheme. Our simulations use the
finite-temperature microphysical nuclear EOS of Lattimer \& Swesty
\cite{lseos:91} with a choice of the nuclear incompressibility $K =
220\,\mathrm{MeV}$, which leads to cold neutron stars broadly
consistent with current theoretical and observational neutron star
mass and radius constraints (see, e.g.,
\cite{ott:12hanse,steiner:10,hebeler:10}).

We focus on the early postbounce evolution up to $25\,\mathrm{ms}$
after bounce which has been shown to proceed essentially
axisymmetrically in rotating progenitors
\cite{ott:07prl,scheidegger:08,scheidegger:10b}. The \code{Zelmani}
simulations are carried out in 3D, but only in an octant of the 3D
cube, using periodic (``rotating'' or ``$\pi/2$-symmetry'') boundary
conditions on the $x-z$ and $y-z$ planes and a reflective boundary
condition on the $x-y$ plane.  This choice of ``octant-3D'' is
appropriate for the collapse and essentially axisymmetric early
postbounce phases (see \cite{ott:07prl,ott:07cqg,scheidegger:10b}),
allows for the development of 3D convective flows, but excludes the
development of global low-$m$-mode nonaxisymmetric rotational
dynamics.  The latter could lead to strong elliptically-polarized GW
emission, possibly exceeding the linearly-polarized signal from core
bounce in emitted energy
\cite{ott:09,ott:07prl,scheidegger:10b,shibata:05}.

We simulate core collapse and early postbounce evolutions in a
$12$-$M_\odot$ (at ZAMS) and in a $40$-$M_\odot$ (at ZAMS)
presupernova stellar model, both drawn from the single-star
solar-metallicity stellar evolutionary calculations of Woosley \&
Heger \cite{woosley:07}. For each model, we carry out and compare
results of six simulations, varying the precollapse rotation rate from
zero to very rapid rotation. Every simulation is carried out once with
postbounce neutrino leakage and once without for comparison. Rotation
is set up by a rotation law that enforces near-uniform angular
velocity in the inner core, as motivated by stellar evolutionary
studies~(e.g., \cite{heger:05}). In contrast to previous work that
studied the progenitor dependence of collapse dynamics and GW signal
\cite{ott:04,dimmelmeier:08,ott:06spin}, we set up our models to have
approximately the same precollapse inner core angular momentum
distribution as a function of enclosed mass
coordinate\footnote{In or near spherical symmetry, any
  quantity can be expressed as a function of radius $r$ or enclosed
  baryonic mass $M$ and the two are related by $d M(r) = 4\pi r^2
  \rho(r)dr$.}, rather than the same precollapse angular velocity or
$T/|W|$, since the former is a conserved quantity while the latter two
are useful parameters only when core structures (i.e., central
density, compactness) at the onset of collapse are very similar, which
generally is not the case and is certainly not the case for the
$12$-$M_\odot$ and $40$-$M_\odot$ progenitors considered here.

Our simulations confirm previous results on the dependence of the GW
signal on precollapse rotation (e.g.,
\cite{scheidegger:10b,dimmelmeier:08,dimmelmeier:07,takiwaki:11}), but
we identify the excitation of non-linear quadrupolar PNS pulsations at
bounce in rapidly rotating cores. These pulsations occur around
$700-800\,\mathrm{Hz}$, last for many cycles and \emph{our simulations
  show that they are clearly manifest in both the GW signal and the
  neutrino luminosities}.

We demonstrate that the dynamical effects of neutrino cooling and
deleptonization and their impact on the GW signal of bounce and the
very early postbounce evolution are tremendous at low rotation rates,
but are strongly reduced with increasing rotation. This lends credence
to previous work
\cite{scheidegger:10b,dimmelmeier:08,dimmelmeier:07,takiwaki:11} that
focused on the bounce and early postbounce GW signal of rapidly
rotating core collapse.

We furthermore show that different rapidly rotating progenitors, if
endowed with (approximately) the same angular momentum distribution in
their inner $0.5-1 M_\odot$, will lead to the same late collapse,
bounce, and early postbounce dynamics and associated GW signal,
essentially independent of their precollapse structures and
thermodynamics.

This paper is structured as follows: In Section~\ref{sec:methods}, we
lay out our computational methods, input physics and GW extraction
technique. In Section \ref{sec:initial_models}, we discuss our model
set and rotational setup. Section~\ref{sec:results1} discusses our new
result on correlated GW and neutrino signals from rapidly rotating
collapse.  We go into detail on the nature of the PNS 
pulsations, effects of neutrino leakage, and differences between
simulations using the $12$-$M_\odot$ and the $40$-$M_\odot$ progenitor
model in Section~\ref{sec:res2_modes}, \ref{sec:res2_leakage}, and
\ref{sec:progcomp}, respectively.  Section \ref{sec:res2_rotrate}
discusses the evolution of the rotation rate and prospects for
nonaxisymmetric rotational instabilities in our simulated models. The
detectability of the GW signals and of the high-frequency variations
of the neutrino luminosity are assessed in
Section~\ref{sec:detect}. Finally, we summarize and conclude in
Section~\ref{sec:summary}.


\section{Methods}
\label{sec:methods}

We use the {\tt Zelmani} GR core collapse simulation package that is
based on the open-source {\tt Cactus Computational Toolkit}
\cite{Goodale2002,cactusweb}, the open-source {\tt Carpet} adaptive
mesh refinement driver, and the open-source {\tt Einstein Toolkit}
\cite{einsteintoolkitweb}.  In the following, we describe the various
parts of {\tt Zelmani} that we use in this study.

\subsection{Spacetime Evolution and General-Relativistic
  Hydrodynamics}

We evolve the full Einstein equations in a $3+1$ split (a Cauchy
initial boundary value problem), using the BSSN
formulation~\cite{baumgarte:99,shibata:95}, a $1+\log$ slicing
condition \cite{Alcubierre2000} controlling the lapse function
$\alpha$, and a modified $\Gamma$-driver condition
\cite{Alcubierre2003b} for the coordinate shift vector $\beta^i$. The
BSSN equations and the shift conditions are implemented in the code
{\tt McLachlan} \cite{ES-Brown2007b}, which is part of the {\tt
  Einstein Toolkit}.  The implementation makes use of fourth-order
accurate finite differencing and details are given in
\cite{reisswig:11ccwave,et:11}.

For the evolution of the equations of GR hydrodynamics, we use the
{\tt Einstein Toolkit} GR hydrodynamics code~{\tt GRHydro}
\cite{et:11}, which implements the Valencia
flux-conservative form of GR hydrodynamics~(e.g., \cite{font:08})
in the standard finite-volume high-resolution shock capturing
approach. The equations are kept in semi-discrete form and first-order
(in time) Riemann problems are solved at cell interfaces with the
approximate HLLE solver~\cite{HLLE:88} on the basis of states
reconstructed with the PPM \cite{colella:84} algorithm. High accuracy
in time and coupling with the spacetime evolution is achieved via the
Method of Lines~\cite{Hyman-1976-Courant-MOL-report} and fourth-order
Runge-Kutta time integration. The time step is limited by the
light crossing time and we use a Courant-Friedrichs-Levy factor
of $0.4$.

In simulations employing hot nuclear matter in nuclear statistical
equilibrium, the electron fraction $Y_e$ is the only compositional
variable that must be tracked and the EOS is a function of rest-mass
density $\rho$, temperature $T$ (or specific internal energy
$\epsilon$), and $Y_e$. {\tt GRHydro} solves an additional equation to
advect the conserved scalar $\sqrt{\gamma} \rho Y_e$ \cite{ott:07cqg},
where $\gamma$ is the determinant of the 3-metric.

{\tt GRHydro} has undergone a rigorous set of standard
tests~\cite{et:11}.  A previous version of {\tt
  GRHydro} has already been applied to stellar collapse with a
finite-temperature nuclear EOS~\cite{ott:07prl,ott:07cqg}
and we have verified that the version used here reproduces previous
results for the same set of input physics.

\subsection{Adaptive Mesh Refinement and Grid Setup}

We employ the {\tt Carpet} adaptive-mesh refinement (AMR) driver
\cite{ES-Schnetter2003b} and perform our simulations in an octant of a
3D Cartesian cube, using periodic boundary conditions on two of the
inner faces of the octant (implementing a $\pi/2$ rotational symmetry)
and reflective boundary conditions on the equatorial plane. This
limits 3D structure to spherical harmonic modes of even $\ell$ and $m$
that are multiples of $4$, which is not a significant limitation for
the current study, since rotating core collapse and the very early
postbounce evolution are likely to proceed nearly
axisymmetrically~\cite{ott:07prl,scheidegger:08,scheidegger:10b}

{\tt Carpet} implements Berger-Oliger style AMR \cite{Berger1984}, where the
fine grids are aligned with coarse grids, refined by factors of two.
{\tt Carpet} also implements subcycling in time, where finer grids take two
time steps for every coarse grid step. The latter greatly improves
efficiency, but also introduces significant complexity into the time
evolution method. The refined regions can be chosen and modified
arbitrarily.

In our baseline resolution, we employ 8 levels of AMR in a box-in-box
grid hierarchy, refining central regions where the highest resolution
is required. The basegrid outer boundary is at $3400\,\mathrm{km}$ and
at the onset of collapse, four levels of AMR (including the basegrid)
are active. During collapse, we dynamically add additional finer
levels, starting when the central density reaches
$5\times10^{10}\,\mathrm{g\,cm}^{-3}$ and thereafter whenever the
central density increases by a factor of $4$. The finest refinement
level has a linear resolution of $\sim 450\,\mathrm{m}$ and is set up
to encompass the PNS after bounce, while we dynamically adjust the
extent of the second-finest level (linear resolution $\sim
900\,\mathrm{m}$) to
always fully enclose the shock and provide constant high resolution
for tracking the postshock flow. In addition to the baseline
resolution, we perform higher-resolution simulations with
$\sim 370\,\mathrm{m}$ linear resolution on the finest level for select
models to test the resolution dependence of our results.

\subsection{Equation of State}
\label{sec:eos}

We employ the tabulated finite-temperature nuclear EOS by Lattimer \&
Swesty \cite{lseos:91}. The Lattimer-Swesty (LS) EOS is based on the
compressible liquid-droplet model. The transition from 
  inhomogeneous to homogeneous nuclear matter is
controlled by a Maxwell construction, and the nucleon-nucleon
interactions are expressed in terms of a Skyrme model potential. The
LS EOS comes with a nuclear symmetry energy of $29.3\,\mathrm{MeV}$
and has variants with an incompressibility of nuclear matter $K$ of
$180\,\mathrm{MeV}$, $220\,\mathrm{MeV}$, $375\,\mathrm{MeV}$. The
variant with $K=180\,\mathrm{MeV}$ yields a maximum cold neutron star
mass below that of the currently most massive known neutron star of
$(1.97\pm 0.04) M_\odot$ \cite{demorest:10} and thus has been ruled
out.  We choose the variant with $K=220\,\mathrm{MeV}$ (LS220), since
it is the EOS that is in best agreement with current theoretical and
observational neutron star mass and radius constraints (e.g.,
\cite{hebeler:10,steiner:10}).

More details on the EOS table and on the implementation
of the contribution of electrons, positrons, and photons, as well as
other details of the construction of this EOS table are described in
O'Connor \& Ott~\cite{oconnor:10}.  Note that for the LS220 EOS, the
Tolman-Oppenheimer-Volkoff (TOV) solution for a neutron star (NS) with
$T=0.1$ MeV in neutrinoless $\beta$-equilibrium yields a maximum
gravitational (baryonic) NS mass of $\sim 2.04 M_\odot$ ($\sim 2.41
M_\odot$).

We point out that, although the properties of the EOS at supranuclear
densities are poorly known and are model dependent, they are expected
to have only a quantitative, but not a qualitative impact on rotating
collapse, bounce, the early postbounce phase and on the associated
GW signal \cite{dimmelmeier:08}. Therefore, in this study, we limit
the dimensionality of the explored input parameter space by using
a single EOS.

\subsection{Neutrino Treatment}
\label{sec:neutrinos}

We employ the neutrino microphysics provided by the open-source code
{\tt GR1D} \cite{oconnor:10,oconnor:11,oconnor:11b}, which is
available for download from \cite{stellarweb}. This includes a
parameterized $Y_e(\rho)$ deleptonization prescription proposed by
Liebend\"orfer \cite{liebendoerfer:05fakenu} during the collapse
phase, a neutrino leakage scheme that approximates deleptonization,
cooling, and heating in the postbounce phase, and contributions to the
pressure from the presence of trapped neutrinos.

Liebend\"{o}rfer's parameterized $Y_e(\rho)$ deleptonization
prescription follows from the observation that the electron fraction
during the collapse phase predicted from Boltzmann neutrino transport
calculations is very well approximated as a function of density and
shows only small variations with progenitor and EOS that have little
impact on the dynamics
\cite{liebendoerfer:05fakenu,dimmelmeier:08}. More details on this
scheme and the particular implementation we use can be found in
\cite{oconnor:10,ott:07cqg}. Following \cite{liebendoerfer:05fakenu},
we employ an analytic fit, based on the $Y_e(\rho)$ profile previously
used in \cite{ott:07prl,ott:07cqg}, which was originally obtained for
a $20$-$M_\odot$ progenitor with the neutrino radiation-hydrodynamics
transport code and neutrino microphysics of \cite{buras:06a}.  Details
on this fit are available on-line~\cite{ccleakweb}.  We employ it for
both our $12$-$M_\odot$ and $40$-$M_\odot$ progenitor models (see
\S\ref{sec:initial_models}).

In the collapse phase of our simulations, we update the electron
fraction in each zone according to the $Y_e(\rho)$ profile after each
spacetime/hydro timestep in an operator-split manner. We follow the
prescription laid out in \cite{liebendoerfer:05fakenu} for 
updating the specific entropy to account for thermalization of neutrinos
at intermediate densities and to enforce constant entropy under $Y_e$
changes at densities above $2\times10^{12}\,\mathrm{g}\,\mathrm{cm}^{-3}$,
where we assume neutrinos to be trapped.

\begin{figure}
 \includegraphics[angle=0,width=0.8\columnwidth,  clip=false]{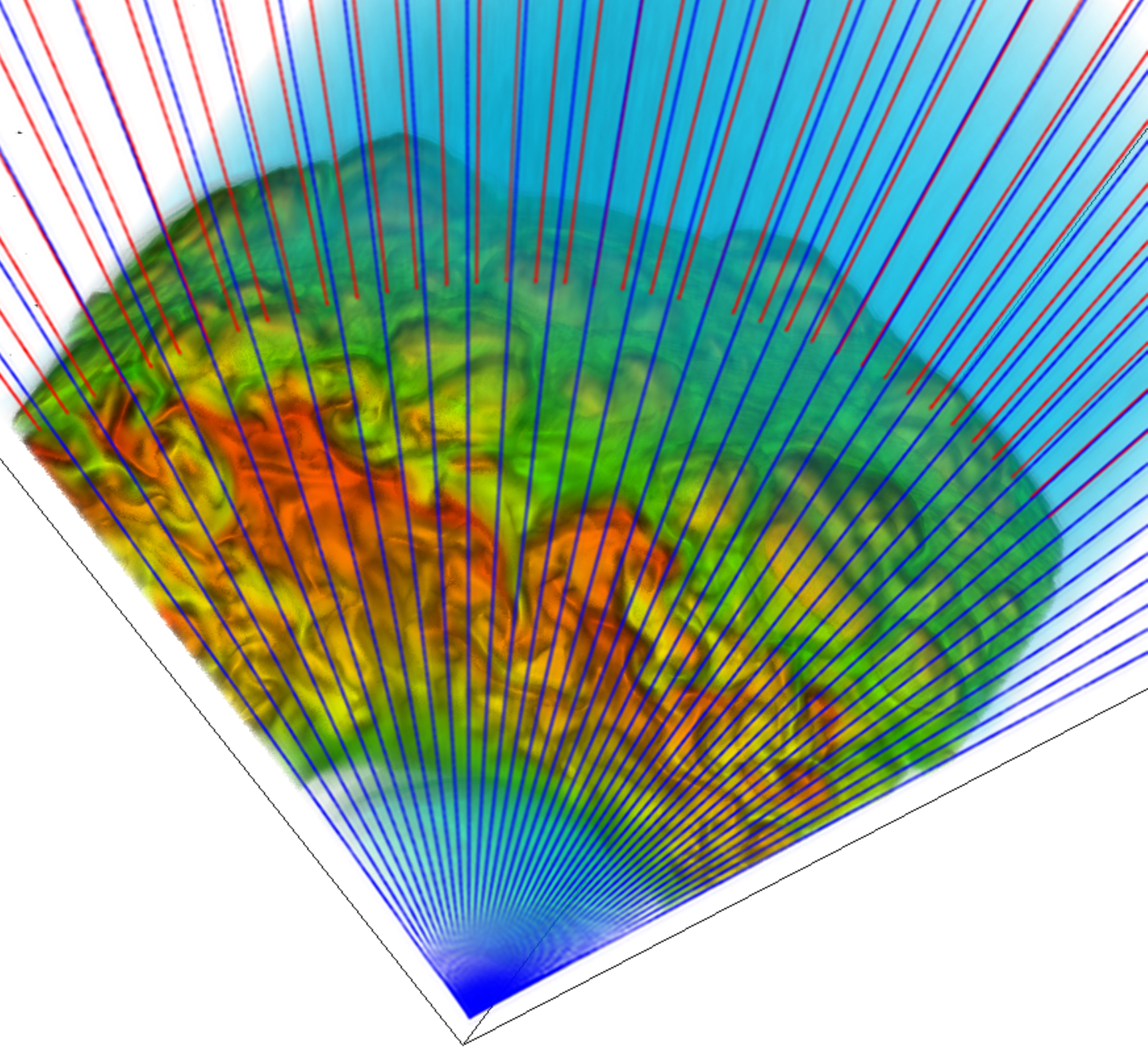}
  \caption{Illustration of the rays along which calculations 
  are performed for the neutrino leakage scheme.  Shown here are two layers
  of rays overlaid on a volumetric visualization of a snapshot of the 
  specific entropy distribution 
  taken in the postbounce phase of model s12WH07j0. Light blue corresponds
  to low entropy ($0.6\,k_\mathrm{B}\,\mathrm{baryon}^{-1}$) and
  red corresponds to high entropy ($15\,k_\mathrm{B}\,\mathrm{baryon}^{-1}$).
  \label{fig:leakrays}}
\end{figure}

During the later stages of the collapse phase and throughout the
postbounce phase, the trapped neutrinos exert a stress on the
surrounding material. The effect of this neutrino stress on the
hydrodynamics of core-collapse supernovae naturally arises when
coupling the evolution of the neutrino fields to the matter; however,
in our treatment of neutrinos we must explicitly take this term into
account.  At matter densities higher than the assumed trapping density,
$2\times10^{12}\,$g cm$^{-3}$, we assume that the neutrinos behave as a
relativistic ideal Fermi gas and approximate the stress based on the
neutrino Fermi pressure gradient. The neutrino stress is then included
as a source term in the GR hydrodynamics equations at each time-integration
substep as detailed in \cite{ott:07cqg} and the neutrino Fermi pressure 
is included in the stress-energy tensor.

Following core bounce, which we define as the time at which the
entropy at the edge of the inner core reaches
$3\,k_\mathrm{B}\,\mathrm{baryon}^{-1}$, the simple $Y_e(\rho)$ prescription
becomes inaccurate and cannot capture the effects of neutrino cooling,
deleptonization, and neutrino heating. Hence, we switch to the leakage
scheme of \cite{oconnor:10}, which incorporates elements of previous
work by \cite{rosswog:03b,ruffert:96}.  We consider three neutrino
species, $\nu_e$, $\bar{\nu}_e$ and $\nu_x =
\{\nu_\mu,\bar{\nu}_\mu,\nu_\tau,\bar{\nu}_\tau\}$.  Neutrino pairs of
all species are made in thermal processes of which we include
electron-positron pair annihilation, plasmon decay \citep{ruffert:96},
and nucleon-nucleon Bremsstrahlung~\cite{oconnor:11b}.  In addition,
charged-current processes lead to the emission of $\nu_e$s and
$\bar{\nu}_e$s.

The leakage scheme provides approximate energy and number emission rates
based on local thermodynamics and an estimate of the energy-averaged
optical depth $\tau_{\nu_i}$ for each neutrino species. The latter requires
a non-local calculation, which we implement in a ray-by-ray approach
as schematically depicted by Fig.~\ref{fig:leakrays} and compute
\begin{equation}
\tau_{\nu_i}(r,\theta_l,\varphi_m) = \int_r^{R_\mathrm{max}}
\sqrt{\gamma_{rr}}\, \overline{\kappa}_{\nu_i}\, dr'\bigg|_{\text{ray}\, l,m} \,\,,
\end{equation}
where $\overline{\kappa}_{\nu_i}$ is the sum of the energy-averaged
absorptive and scattering opacities for each neutrino species and
$\gamma_{rr}$ is the $rr$-component of the 3-metric. A second such
integral is computed with integrand $\chi_{\nu_i} = \kappa_{\nu_i} /
\epsilon_{\nu_i}^2$, in which $\kappa_{\nu_i}$ is the sum of the
energy-dependent absorption and scattering opacities. It is divided by
its energy dependent term, the square of the neutrino energy
$\epsilon_{\nu_i}^2$. $\chi_{\nu_i}$ is needed for computing the
neutrino energy and number loss rates following \cite{rosswog:03b}.

The ray-by-ray approach, pioneered by \cite{bhf:95} for flux-limited
neutrino diffusion, has recently been employed in modern 2D
neutrino radiation-hydrodynamics calculations
(e.g., \cite{marek:09b,mueller:12a,bruenn:09}) and in the 2D GR leakage scheme
of \cite{sekiguchi:11a}. It works well in situations that are spherical at
lowest order like the one considered here. However, it is
not appropriate for globally asymmetric systems, such as
accretion disks around merged compact objects. In such situations it
is necessary to evaluate the optical depth by integrating rays in
various directions from each local emission region.

In our simulations, we use 40 rays in latitude $\theta$ and 20
azimuthal rays in $\varphi$, covering $[0,\pi/2]$ in both angles as
appropriate for the octant in which our simulations are performed. The
radial resolution of the rays is $500\,\mathrm{m}$ out to
$300\,\mathrm{km}$ and is logarithmically decreased further
out. The outer radius for the ray calculation is set to
$3000\,\mathrm{km}$. We have studied variations in angular and radial
resolution and find that the choices given here yield converged
results.

Our 3D neutrino leakage calculation proceeds as follows: We
interpolate density, temperature, $Y_e$, and the 3-metric from the 3D
Cartesian grid onto the leakage rays and carry out a 1D leakage
calculation as described in \cite{oconnor:10} along each ray and for
each neutrino species. This involves finding the optical depth
$\tau_{\nu_i}$ and the quantity $\chi_{\nu_i}$ and, for each zone of
the rays interior to the shock, computing the instantaneous number and
energy leakage rates to obtain an approximate ``neutrino luminosity''
and mean energy as a function of radius and angle for each species.

The results of this calculation are then interpolated back onto our 3D
Cartesian grid, where we locally compute the instantaneous number and
energy leakage rates as well as the number and energy
  absorption rates computed on the basis of the interpolated
  approximate neutrino luminosity coming radially from below. In this
approximation, we must attenuate the absorption rates with a term
$\propto \exp(-\tau_{\nu_i})$ to avoid incorrect spurious heating at
high optical depths \cite{oconnor:10}.

All leakage calculations are carried out in operator-split fashion
after the spacetime/hydro update and are first order in time, which we
find to be sufficiently accurate given the small spacetime/hydro
timestep size. Updates are applied to the fluid rest-frame quantities,
ignoring velocity dependence and other relativistic effects in
consideration of the general approximate nature of the leakage scheme.
In the highly dynamic first few milliseconds after bounce, we carry
out all leakage calculations at every timestep. Subsequently we switch
to calculating the ray-by-ray calculations only every 8 fine-grid
timesteps (corresponding to every $5\times10^{-6}\,\mathrm{s}$) while
continuing to evaluate the local expressions at every timestep.

We have tested our 3D leakage scheme against the implementation in
\code{GR1D} using a nonrotating model and found excellent agreement in
the emitted luminosities, deleptonization, and cooling/heating rates
in the early postbounce phase. The leakage scheme in \code{GR1D}
itself has been compared with full multi-energy neutrino
radiation-hydrodynamics simulations and found to produce qualitatively
robust results \cite{oconnor:10,oconnor:11} with quantitative differences
of order $\sim$$20\%$ across all transport variables. 

\subsection{Gravitational Wave Extraction and Analysis}
\label{sec:gws}

We employ the quadrupole formalism for extracting the GW signal from
our simulations.  We neglect higher spatial orders whose GW emission
is suppressed by factors of $c^{-\alpha}$, $\alpha \ge 1$ and GW
emission by anisotropic neutrino emission whose reliable prediction is
not possible with the neutrino leakage scheme employed here.

The quadrupole formalism estimates the GW signal from the dynamical
quadrupolar matter distribution alone.  It does not take into account
non-linear curvature effects and is strictly valid only when the
fields are weak $\frac{G}{c^2}\frac{R}{M}\ll1$ and in slow motion $v/c
\ll1$.

Even though simple, comparison with curvature-based methods has
proven the quadrupole formalism to be sufficiently accurate for the
core collapse scenario~\cite{reisswig:11ccwave}. Most notably, it
yields waveforms which are in good agreement (within a few percent)
with those extracted at future null infinity $\scri^+$ using the
technique of Cauchy-characteristic extraction (e.g.,
\cite{Reisswig:2009rx}).  It must be noted, however, that the
quadrupole formalism breaks down in the context of black hole
formation \cite{ott:11a}, when the matter has fallen inside the black
hole and the GW signal is entirely due to (vacuum) black hole
ringdown.  This is not the case in our present study.

We estimate the GW strain
by
\begin{equation} \label{eq:quad-strain}
  h_{jk}^{TT}(t,\mathbf{x})=\frac{2}{c^4}\frac{G}{D}\left[\frac{d^2}{dt^2}I_{jk}(t-D/c)
  \right]^{TT}\,,
\end{equation}
where $D$ is the distance to the source and
\begin{equation} \label{eq:quad-mass}
  I_{jk}=\int\tilde{\rho}(t,\mathbf{x})\left[x_j x_k -
    \frac{1}{3}x^2\delta_{jk}\right]d^3x
\end{equation}
is the reduced mass-quadrupole tensor and the superscript $TT$
denotes projection into the transverse-traceless gauge~(see, e.g.,
\cite{schutz:85}).  The mass quadrupole is not uniquely defined in GR and
the choice of density variable is ambiguous.  Following previous work
(e.g.,
\cite{dimmelmeier:02,shibata:04,dimmelmeier:08,ott:07cqg}),
we set $\tilde{\rho} = \sqrt{\gamma}\, W \rho = \hat{D}$, because,
(i), this is the conserved density variable in our code, and (ii),
$\sqrt{\gamma}\, d^3 x$ is the natural volume element. 

The reduced mass-quadrupole tensor can be computed directly from the
computed distribution $\hat{D}(t,\mathbf{x})$.  
Numerical noise, introduced by the second time derivative
of Eq.~(\ref{eq:quad-mass}), may limit the accuracy of the result.
We can circumvent this by making use of the continuity
equation to obtain the first time derivative of
Eq.~(\ref{eq:quad-mass}) without numerical differentiation
\cite{finnevans:90,blanchet:90},
\begin{equation}
  \frac{d}{dt}I_{jk}=\int \hat{D}(t,\mathbf{x})\left[\tilde{v}^j x^k + \tilde{v}^k x^j - \frac{2}{3}(x^l\tilde{v}^l)\delta^{jk}\right]d^3x\,\,,
\label{eq:dtquad}
\end{equation}
where we follow \cite{dimmelmeier:02a} and employ physical velocity
components $\tilde{v}^i \equiv \{\tilde{v}_x, \tilde{v}_y,
\tilde{v}_z\} \approx \{\sqrt{\gamma_{11}} v^1, \sqrt{\gamma_{22}}
v^2, \sqrt{\gamma_{33}} v^3\}$ that are individually bound to $v <
c$. This assumes that the 3-metric is nearly diagonal (which is the
case in our gauge; see \cite{reisswig:11ccwave}).  Also note that we
have switched to contravariant variables in the integrand as these are
the ones present in the code. This is possible since in the
weak-field slow-motion approximation the placement of indices is
arbitrary.

The two dimensionless independent GW strain
polarizations $h_+$ and $h_\times$ incident on a detector located at
distance $D$ and at angular coordinate $(\theta,\phi)$ in source
coordinates are given by
\begin{equation}
  \label{eq:hDecomposition}
  h_+ - ih_\times =
  \frac{1}{D}\sum_{\ell=2}^{\infty}\sum_{m=-\ell}^\ell H_{\ell m}(t)\,
  \Ytwo_{\ell m}(\theta,\phi)\,\,,
\end{equation}
where $\Ytwo_{\ell m}$ are the spin-weighted spherical harmonics of
weight $-2$ \cite{thorne:80} and the $H_{\ell m}$ are
expansion coefficients, which, in the quadrupole case, are related to
the second time derivative of the mass-quadrupole tensor by
\begin{eqnarray}
  H^\mathrm{quad}_{20}    &=& \sqrt{\frac{32\pi}{15}} \frac{G}{c^4} 
  \left(\ddot{I}_{zz} 
    - \frac{1}{2}(\ddot{I}_{xx}
    +\ddot{I}_{yy})\right)\,\,,\\
  H^\mathrm{quad}_{2\pm1} &=& \sqrt{\frac{16\pi}{5}} \frac{G}{c^4}
  \left(\mp \ddot{I}_{xz} 
    + i \ddot{I}_{yz}\right)\,\,,\\
  H^\mathrm{quad}_{2\pm2} &=& \sqrt{\frac{4\pi}{5}} \frac{G}{c^4} 
  \left(\ddot{I}_{xx} 
    - \ddot{I}_{yy} \mp 
    2 i \ddot{I}_{xy} \right)\,\,.
\end{eqnarray}

The rotating core collapse models considered in this study stay almost
perfectly axisymmetric in the collapse and early postbounce phases. In
axisymmetry about the $z$-axis, $I_{xx} = I_{yy} = -\frac{1}{2}
I_{zz}$ and $I_{xy} = I_{xz} = I_{yz} = 0$.  $h_\times$ vanishes and
$h_+$ becomes
\begin{equation}
h_+ = \frac{G}{c^4} \frac{1}{D} \frac{3}{2} \ddot{I}_{zz} \sin^2\theta\,.
\end{equation}
We will generally plot $h_+ D$ in units of centimeters
when displaying gravitational waveforms.

The energy emitted in gravitational waves is given by
\begin{equationarray}
E_\text{GW}& = &\frac{1}{5}\frac{G}{c^5} \int^{\infty}_{-\infty}\, dt\, 
{\Ibardotdotdot_{ij}}\, {\Ibardotdotdot_{ij}}\, dt\, \nonumber \\
&=& \frac{1}{5}\frac{G}{c^5} \int^{\infty}_{-\infty}\,dt\ \bigg[
\Ibardotdotdot_{xx}^{\,2} + \Ibardotdotdot_{yy}^{\,2} + \Ibardotdotdot_{zz}^{\,2}\nonumber\\
&&\hspace{2cm} + 2 \big(\Ibardotdotdot_{xy}^{\,2} + \Ibardotdotdot_{xz}^{\,2} + \Ibardotdotdot_{yz}^{\,2}
\big)
\bigg]\,\,.
\end{equationarray}

In the special case of axisymmetry and in terms of $h_{+,e} = h_+
/ \sin^2 \theta$, this becomes 
\begin{equation}
  E_\mathrm{GW}^{\mathrm{axi}} = \frac{2}{15} \frac{c^3}{G^5}\, D^2 
 \int^{\infty}_{0}\,dt \left(\frac{d}{dt} h_{+,e} \right)^2\,\,.
\label{eq:egw2}
\end{equation}

The spectral GW energy density is given by
\begin{equation}
\frac{dE_{\text{GW}}}{df} = \frac{2}{5}\frac{G}{c^5}  (2 \pi f)^2
\left|\tilde{\Ibardotdot}_{ij}\right|^2 \,\,,
\end{equation}
so that 
\begin{equation}
E_\text{GW} = \int_0^\infty df\, \frac{dE_{\text{GW}}}{df}\,\,.
\end{equation}
In the above, we have introduced the Fourier transform of the
mass-quadrupole tensor, $\Ibardotdottilde_{ij}(f)$, and and denoted it
with a tilde accent.

In axisymmetry, the spectral GW energy density is related to $h_{+,e}$
by
\begin{equation}
\frac{dE_{\text{GW}}^{\mathrm{axi}}}{df} = \frac{4}{15} \frac{c^3}{G}\,D^2\, (2\pi f)^2 \left| \tilde{h}_{+,e} \right|^2\,\,.
\label{eq:degwdf}
\end{equation}
When showing the spectral energy density, we will plot
the dimensionless characteristic strain \cite{flanhughes:98},
\begin{equation}
\label{eq:hchar}
h_{\text{char}} (f) = 
\sqrt{\frac{2}{\pi^2} \frac{G}{c^3} \frac{1}{D^2} \frac{dE_{\text{GW}}(f)}{df}}\,\,,
\end{equation}
which can be compared to the GW detector root-mean-squared noise,
\begin{equation}
h_{\mathrm{rms}}(f) = \sqrt{f\,S(f)}\,\,,
\end{equation}
where $\sqrt{S(f)}$ is the one-sided detector noise amplitude spectral
density in units of $(\mathrm{Hz})^{-1/2}$. For making rough
statements about detectability, we use the single-detector
optimal-orientation signal-to-noise ratio, which is given by
\begin{equation}
\label{eq:snr}
(\mathrm{SNR})^2 = \int_0^\infty d \ln{f}\,
  \frac{h^2_\mathrm{char}}{h^2_\mathrm{rms}}\,\,.
\end{equation}

Note that we cut the calculation of integrals in the Fourier domain at
$3000\,\mathrm{Hz}$ to filter out numerical high-frequency
noise. Wherever we need $\sqrt{S(f)}$, we employ the projected broadband
Advanced LIGO noise curve (the so-called zero-detuning, high-power
configuration [ZD-HP]), available as file {\tt ZERO\_DET\_high\_P.txt} from
\cite{LIGO-sens-2010}.

For quantifying the difference between two gravitational waveforms
$h_1(t)$ and $h_2(t)$, we introduce the mismatch \cite{owen:96,damour:98},
\begin{equation} 
\mathcal{M}_{\rm mis}=1-\mathcal{M}\,\,,
\label{eq:mismatch}
\end{equation} where $\mathcal{M}$ is
the match, defined in terms of a maximization over time of arrival
$t_0$ and GW phases $\phi_i$ of the two waveforms as
\begin{equation}
\mathcal{M}=\max_{t_0}\max_{\phi_1}\max_{\phi_2}\mathcal{O}[h_1,h_2]\,,
\end{equation}
where $\mathcal{O}[h_1,h_2]$ is the overlap, given by
\begin{equation}
\mathcal{O}[h_1,h_2]:=\frac{\langle h_1 | h_2 \rangle}
{\sqrt{\langle h_1 | h_1 \rangle\langle h_2 | h_2 \rangle}}\,
\end{equation}
and where
\begin{equation}
\langle h_1 | h_2 \rangle = 4\, \mathrm{Re} \int_0^\infty df \frac{\tilde{h}_1(f)\tilde{h}_2^*(f)}{S_h(f)}\,
\end{equation}
is the noise-weighted inner product. A mismatch of $\mathcal{M}_{\rm
  mis}=0$ indicates that the two given waveforms are
identical. Conversely, a mismatch of $\mathcal{M}_{\rm mis}=1$
indicates that both waveforms are completely different.  For
calculating $\mathcal{M}_{\rm mis}$, we employ the
\texttt{pyGWDataAnalysis} package \cite{Reisswig:2010di}, which is part
of the \texttt{Einstein Toolkit}.

\section{Initial Models and Setup}
\label{sec:initial_models}

\begin{figure}
\centering
 \includegraphics[width=1.0\linewidth]{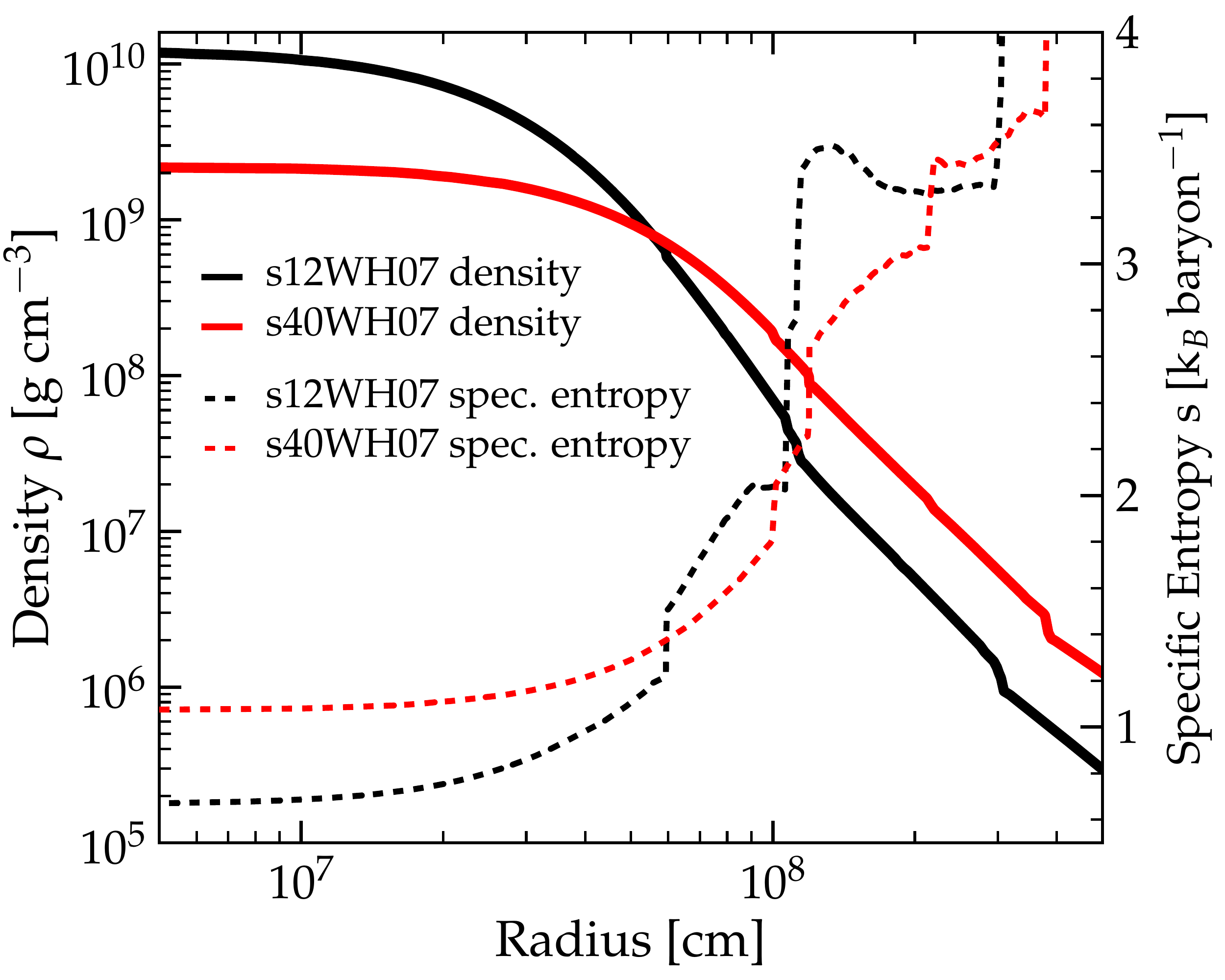}
  \caption{Precollapse structure of the $12$-$M_\odot$ (black graphs)
    and $40$-$M_\odot$ (red graphs) progenitor models taken from
    \cite{woosley:07}. Shown in solid lines is the radial density
    profile and dashed lines denote the radial entropy distribution as derived
    from our EOS.
    Note that, as expected from stellar evolution theory (e.g.,
    \cite{whw:02}), the more massive progenitor has, at the onset of
    collapse, a lower central density, a higher central entropy, and a
    more extended core than its lower-mass counterpart. }
  \label{fig:initial}
\end{figure}

\subsection{Progenitor Structure}
\label{sec:prog}

We perform our collapse simulations with $12$-$M_\odot$ and
$40$-$M_\odot$ (both at ZAMS) solar-metallicity presupernova stellar
models evolved and provided to us by Woosley~\&~Heger
(2007)~\cite{woosley:07}. We refer to them as s12WH07 and s40WH07,
respectively. Stellar winds lead to mass loss and at the onset of core
collapse, defined as the time when the peak negative radial velocity
exceeds $1000\,\mathrm{km\,s^{-1}}$, model s12WH07 is a red supergiant
with a final mass of $\sim$$10.9\,M_\odot$ and a radius of
$\sim$$4.4\times10^{13}\,\mathrm{cm}$ ($\sim$$630\,R_\odot$). s40WH07
is a much more compact Wolf-Rayet star and has a
radius of $\sim$$8\times10^{11}\,\mathrm{cm}$ ($\sim$$10\,R_\odot$) at
a final mass of $\sim$$15.3\,M_\odot$. Dynamically relevant during
collapse, though, are only the central regions and at the precollapse
stage, s12WH07's central density, temperature, and entropy are
$\sim$$1.2\times10^{10}\,\mathrm{g\,cm}^{-3}$,
$\sim$$7\times10^{9}\,\mathrm{K}$, and
$0.67\,k_\mathrm{B}\,\mathrm{baryon}^{-1}$, respectively. Its iron
core has an approximate mass of $1.3\,M_\odot$ and a radius of
$\sim$$1200\,\mathrm{km}$. The spherical radius that encloses
$1\,M_\odot$ is $\sim$$634\,\mathrm{km}$.

Model s40WH07's central structure is quite different at the
precollapse stage. Its central density, temperature, and entropy are
$\sim$$2.2\times10^{9}\,\mathrm{g\,cm}^{-3}$,
$\sim$$8\times10^{9}\,\mathrm{K}$, and
$1.08\,k_\mathrm{B}\,\mathrm{baryon}^{-1}$, respectively. Its iron
core has a mass of $\sim$$1.8\,M_\odot$ and an approximate radius of
$\sim$$2200\,\mathrm{km}$. The spherical radius that encloses the
innermost $1\,M_\odot$ of material is $\sim$$945\,\mathrm{km}$.

Picking these two thermodynamically and structurally rather different
models (cf. also Fig.~\ref{fig:initial}) allows us to test if the
systematics in the collapse dynamics, GW and neutrino signals that our
simulations uncover are independent of thermodynamical and structural
differences at the precollapse stage.

We point out that both progenitor models considered here do not have
the observational characteristics associated with long-GRB progenitor
stars, which are expected to be rapidly spinning, low-metallicity,
stripped-envelope or chemically homogeneous stars (e.g.,
\cite{woosley:06}). However, stellar evolution theory has not yet
converged on a robust prediction for the core structure and rotational
configuration of long-GRB progenitors (see the discussion in
\cite{dessart:12a}).  Since the electron degenerate iron cores of our
two progenitor models have rather similar structure as many of the
proposed long-GRB progenitors of \cite{woosley:06}, the results found
for the current set of models are very likely to transfer to these
models.

\begin{table*}[t]
\begin{center}
  \caption{Key initial model parameters for our simulations using the
    s12WH07 and s40WH07 progenitors.  Models using the former
    progenitor are set up according to the rotation law specified in
    Eq.~\ref{eq:rotlaw}. Rotating s40WH07 models are set up to have
    the same specific angular momentum (in enclosed-mass coordinate)
    on the equator. Specific angular momentum is constant on
    cylindrical shells in both sets of initial models.  $j_{\infty}$
    is the specific angular momentum at infinity.
    $\Omega_{\mathrm{c,initial}}$ is the initial central angular
    velocity. $J_{0.5 M_\odot}$, $J_{0.6 M_\odot}$, $J_{0.7 M_\odot}$,
    and $J_{1 M_\odot}$ are the total angular momentum within a
    enclosed-mass coordinate of $(0.5,0.6,0.7,1.0)\,M_\odot$,
    respectively. $T/|W|_\mathrm{initial}$ is the initial ratio of
    rotational kinetic energy to gravitational energy as measured at
    a radius of $3400\,\mathrm{km}$.}
\begin{tabular}{lcccccccc}
  \hline \hline
  \multicolumn{1}{c}{Model} & $j_{\infty}$           & $\Omega_{c,\mathrm{initial}}$     & $J_{0.5 M_\odot}$ &$J_{0.6 M_\odot}$ &$J_{0.7 M_\odot}$ & $J_{1 M_\odot}$ & $T/|W|_\mathrm{initial}$          \\  
  \multicolumn{1}{c}{Name}  & ($10^{16} \mathrm{cm}^2 \mathrm{s}^{-1}$) & (rad s$^{-1}$) & ($10^{48} \mathrm{g\, cm}^2\,s^{-1}$) & ($10^{48} \mathrm{g\, cm}^2\,s^{-1}$) & ($10^{48} \mathrm{g\, cm}^2\,s^{-1}$) & ($10^{48} \mathrm{g\, cm}^2\,s^{-1}$) & (\%)                \\  
  \hline
  s12WH07j0 & $0.000$            & 0\phantom{.000}  & 0.00 & 0.00& 0.00& \phantom{0}0.00     & 0\phantom{.000}            \\
  s12WH07j1 & $0.404$            & 1\phantom{.000}  & 0.35 & 0.49& 0.65& \phantom{0}1.31      & 0.018 \\
  s12WH07j2 & $0.809$            & 2\phantom{.000}  & 0.70 & 0.98& 1.31& \phantom{0}2.62      & 0.072 \\
  s12WH07j3 & $1.620$            & 4\phantom{.000}  & 1.40 & 1.96& 2.62& \phantom{0}5.25      & 0.289 \\
  s12WH07j4 & $2.430$            & 6\phantom{.000}  & 2.10 & 2.95& 3.93& \phantom{0}7.88      & 0.650 \\ 
  s12WH07j5 & $3.240$            & 8\phantom{.000}  & 2.80 & 3.93& 5.23& 10.51                & 1.150 \\
  \hline                                                    
  s40WH07j0 & $0.000$            & 0.000            & 0.00 & 0.00& 0.00& \phantom{0}0.00    & 0\phantom{.000}  \\
  s40WH07j1 & $0.404$            & 0.453            & 0.31 & 0.43& 0.58& \phantom{0}1.16    & 0.010 \\
  s40WH07j2 & $0.809$            & 0.906            & 0.62 & 0.86& 1.15& \phantom{0}2.31    & 0.041 \\
  s40WH07j3 & $1.620$            & 1.810            & 1.23 & 1.73& 2.30& \phantom{0}4.62    & 0.165 \\
  s40WH07j4 & $2.430$            & 2.720            & 1.85 & 2.59& 3.46& \phantom{0}6.94    & 0.373 \\ 
  s40WH07j5 & $3.240$            & 3.620            & 2.46 & 3.45& 4.62& \phantom{0}9.25    & 0.656 \\

\hline\hline
\end{tabular}
\label{table:initial}
\end{center}
\end{table*}

\subsection{Rotational Setup}
\label{sec:rotsetup}

Both model s12WH07 and model s40WH07 were evolved in a
spherically-symmetric stellar evolution code without
rotation \cite{woosley:07}. While a limited set of presupernova models
with rotation is available~(e.g.,
\cite{heger:00,heger:05,woosley:06}), we choose to use nonrotating
models and impose precollapse rotation by hand to study its effects
during the collapse, bounce, and early postbounce phases in a
controlled fashion.

We are interested in variations of the dynamics and the associated GW
and neutrino signals that (\emph{1}) arise from changes in the
precollapse rotational configuration and (\emph{2}) are a consequence
of differences in the progenitor structure and thermodynamics.  In the
following, we describe the rotational setup that allows us to
differentiate between the two.

According to the Poincar\'e-Wavre theorem (e.g., \cite{tassoul:78}) we
expect the electron-degenerate iron cores (which are essentially
barotropic, $P \approx P(\rho)$) to have constant specific angular
momentum $j = \Omega \varpi^2$ on cylindrical shells of distance
$\varpi$ from the rotation axis. We impose rotation using the
rotation law \cite{zwerger:97,ott:04}
\begin{equation}
\Omega(\varpi) = \Omega_{c,\text{initial}} \left[ 1 + \left(\frac{\varpi}{A} \right)^2 \right]^{-1}\,\,,
\label{eq:rotlaw}
\end{equation}
where $\Omega_{c,\text{initial}}$ is the initial central angular
  velocity and the parameter $A$ controls the cylindrical radius at
  which differential rotation becomes significant. This rotation law
  leads to constant specific angular momentum at $\varpi \gg A$.

We first set up rotation for model s12WH07 and choose $A =
634\,\mathrm{km}$, which corresponds to the spherical radius that
encloses $1 M_\odot$ at the precollapse stage. This yields an angular
velocity profile that is consistent with results from stellar
evolution calculations that include rotation \cite{heger:00,heger:05}
and ensures that the inner core is essentially uniformly spinning.
Homologous collapse will preserve its nearly uniform rotation
\cite{ott:06spin,dessart:12a}.

We vary the initial rotation rate by choosing the initial central
angular velocity $\Omega_0$ from the set
$\{0,1,2,4,6,8\}\,\mathrm{rad}\,\mathrm{s}^{-1}$, which correspond to
specific angular momenta $j_{\infty}$ at infinity  ranging from $0$ to
$3.2\times 10^{16}\,\mathrm{cm}^2\,\mathrm{s}^{-1}$. We enumerate
models as s12WH07j$n$, where $n$ ranges from $0$ to $5$, corresponding
to the 6 choices of initial central angular velocity.

In order to isolate the influence of the progenitor star, we must
compare simulations using the s12WH07 and s40WH07 progenitors that
have identical or at least nearly identical angular momentum
distributions. Since angular momentum is conserved per unit comoving
mass, we compute the specific angular momentum on the equator (on the
equator, the spherical radius $r$ is equal to the cylindrical radius
$\varpi$) of a given rotating s12WH07j$n$ model as a function of
enclosed mass $M(r=\varpi)$. We then take this specific angular
momentum distribution and map it in mass coordinate to the equator of
the corresponding s40WH07j$n$ model and enforce constant angular
velocity on cylinders $\Omega(\varpi) = j(M(r=\varpi))/\varpi^2$. With
this approach, the equatorial regions (where $r \approx \varpi$) of
both models have identical specific angular momentum as a function of
enclosed mass. This mapping becomes, of course, more approximate at
regions away from the equator. In principle, it would be possible to
introduce an exact mapping $j(M,\theta)$ for enclosed mass at lateral
angle $\theta$, but this would violate the constraint of constant
specific angular momentum on cylindrical shells, which we intend to
adhere to in this study.

In Tab.~\ref{table:initial}, we summarize key properties of our
initial configurations. Since model s40WH07 is initially much more
extended than model s12WH07 (see Fig.~\ref{fig:initial}) it is
initially much more slowly spinning. The rotational setup described in
the above nevertheless yields integrated s12WH07 and s40WH07 angular
momenta of the innermost $0.5 M_\odot$ that agree within $\sim$$12\%$
at a given j$n$. The s40WH07j$n$ models have systematically slightly
less total angular momentum than their s12WH07j$n$ counterparts at the
same enclosed mass.

In addition to performing 6 simulations with varying initial rotation
for each progenitor, we also repeat each simulation without neutrino
leakage after core bounce. We append ``nl'' to these models (denoting
``no leakage'') and use them for comparison with our runs with leakage
to assess the impact of neutrino leakage on the postbounce dynamics
and the resulting GW signal. This allows us to quantify the systematic
error in the waveform predictions of previous GR core collapse
simulations without postbounce neutrino leakage (e.g.,
\cite{dimmelmeier:08,abdikamalov:10}).

\subsection{Mapping and Initial Spacetime Setup}

We map the 1D progenitor models from 1D to 3D using linear
interpolation and set up rotation by imposing constant angular
velocity on cylindrical shells as discussed in the previous section
\ref{sec:rotsetup}. We keep the mass distribution spherically
symmetric when mapping and do not solve for rotational equilibrium,
since this can be found consistently only for models with constant
entropy and $Y_e$. The error introduced
by not using models in rotational equilibrium is small, since
collapse proceeds initially slowly, the region in sonic contact
is large, and the core has sufficient time to adjust to the appropriate
angular density stratification before the dynamical phase of collapse
is reached \cite{zwerger:97,ott:04}.

We set up the initial $3-$metric and the lapse function using the
quasi-Newtonian spherically symmetric line element
\cite{schutz:85}. The extrinsic curvature and shift are set to zero on
the initial slice. Since our precollapse configurations are
essentially Newtonian, this yields accurate results and the 2-norm of
the initial Hamiltonian (Momentum) constraint violation is only of
order $10^{-10}$ ($10^{-11}$).

\section{Results: Correlations Between Core Dynamics, Gravitational Waves,
and Neutrino Signals}
\label{sec:results1}

\begin{figure*}[t]
\centering
\vspace*{-0.6cm}
\begin{minipage}{\linewidth}
\includegraphics[width=0.38\linewidth]{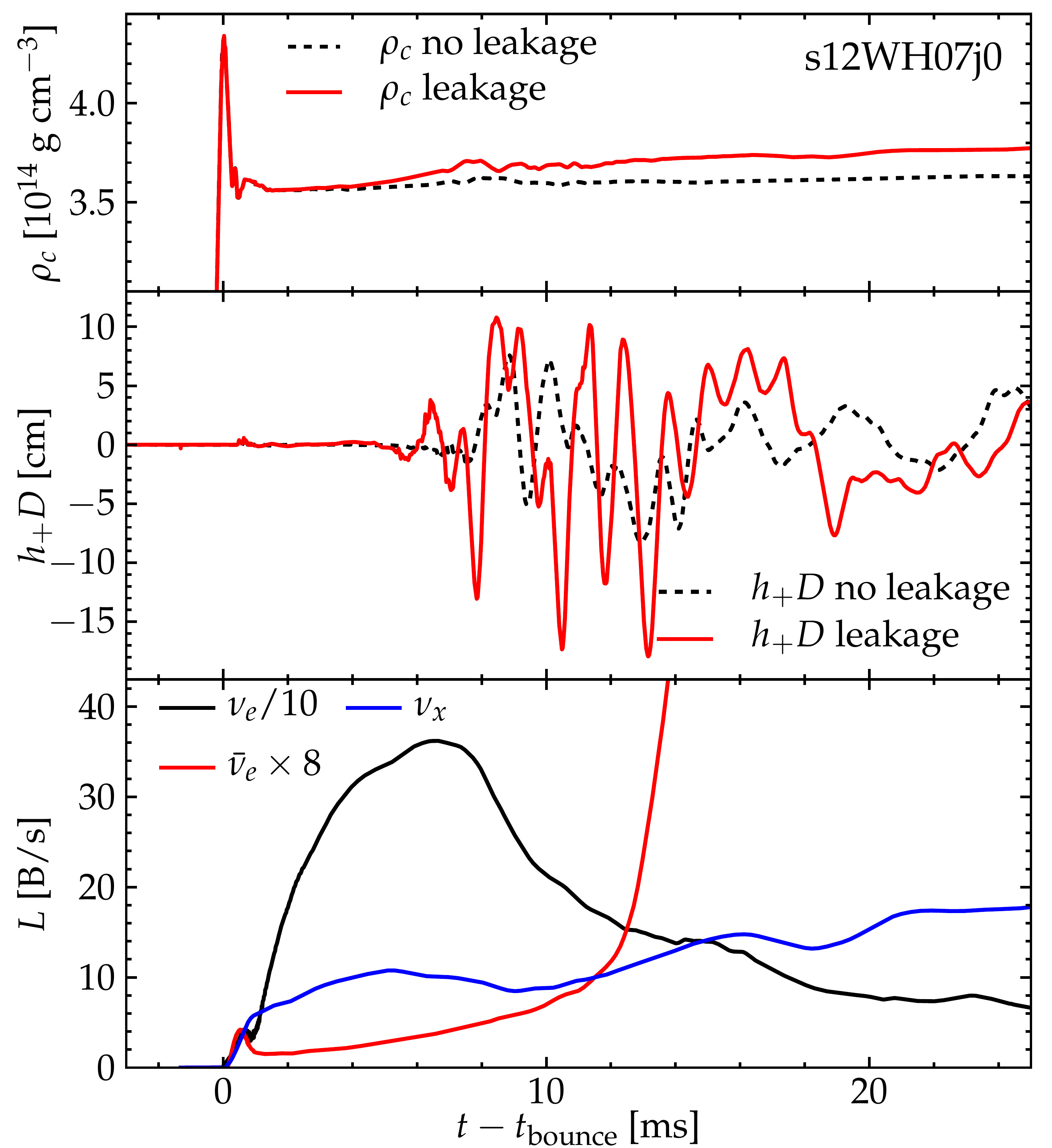}
\includegraphics[width=0.38\linewidth]{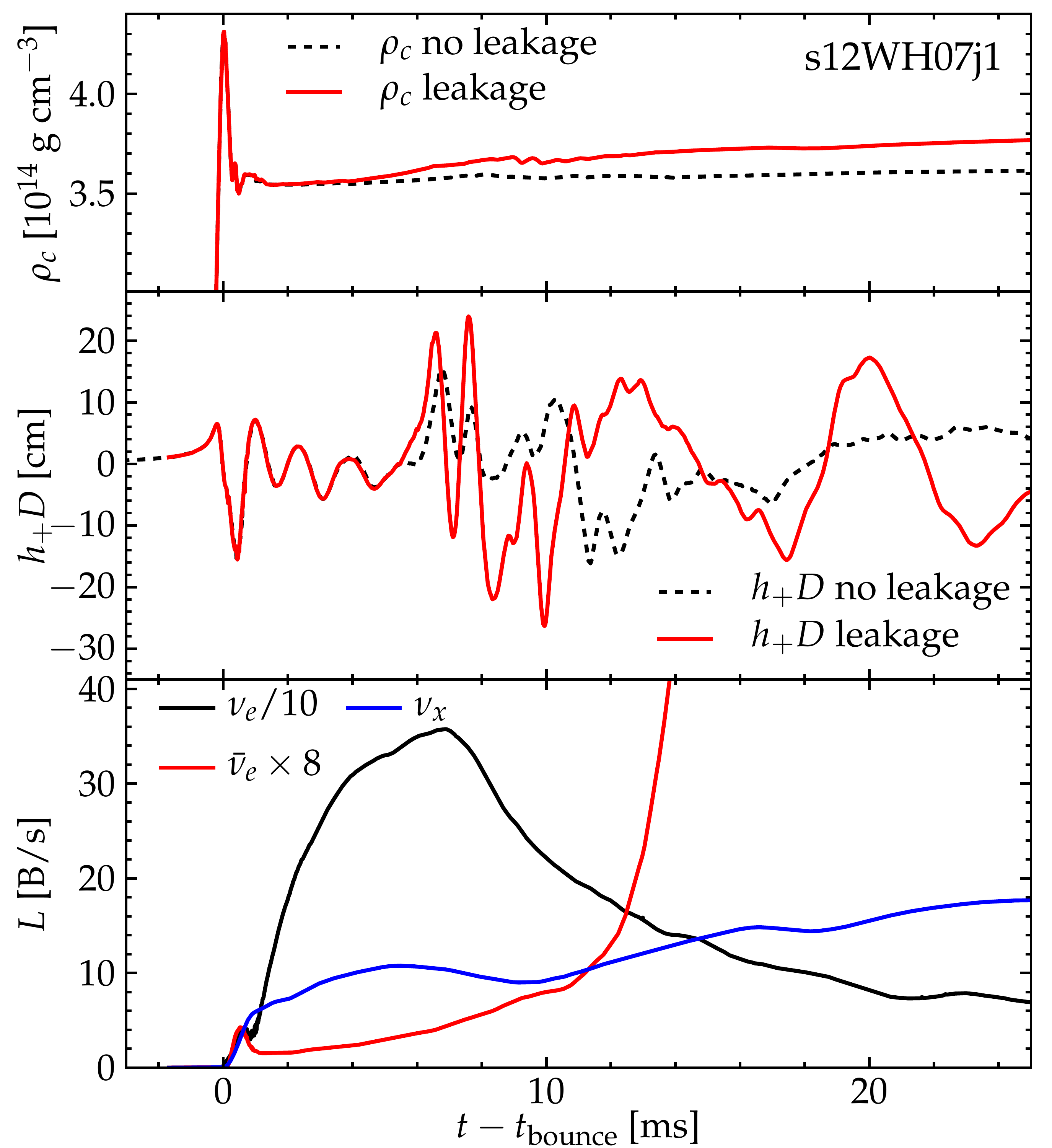}
\end{minipage}
\begin{minipage}{\linewidth}
\vspace*{-0.0cm}
\includegraphics[width=0.38\linewidth]{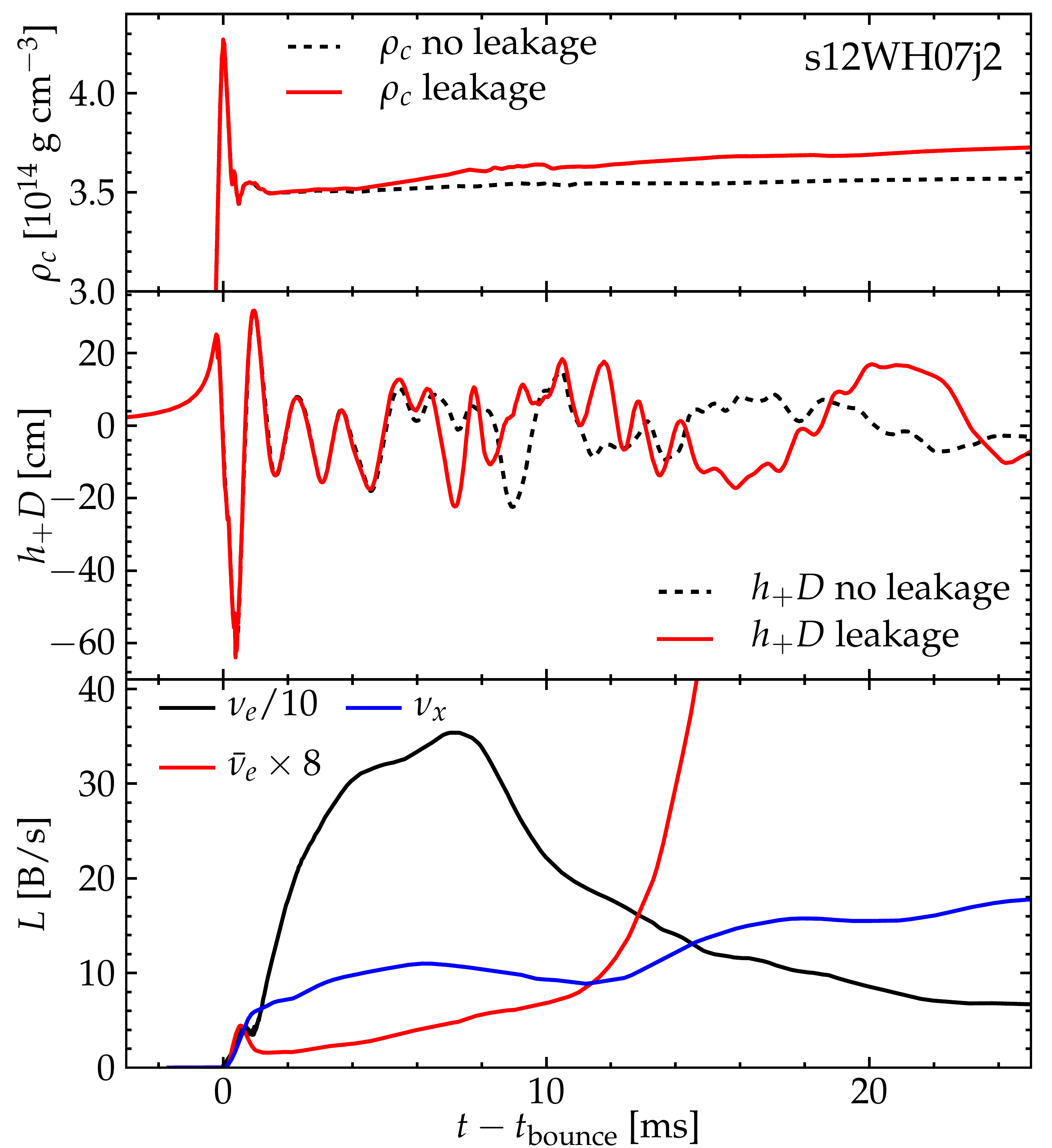}
\includegraphics[width=0.38\linewidth]{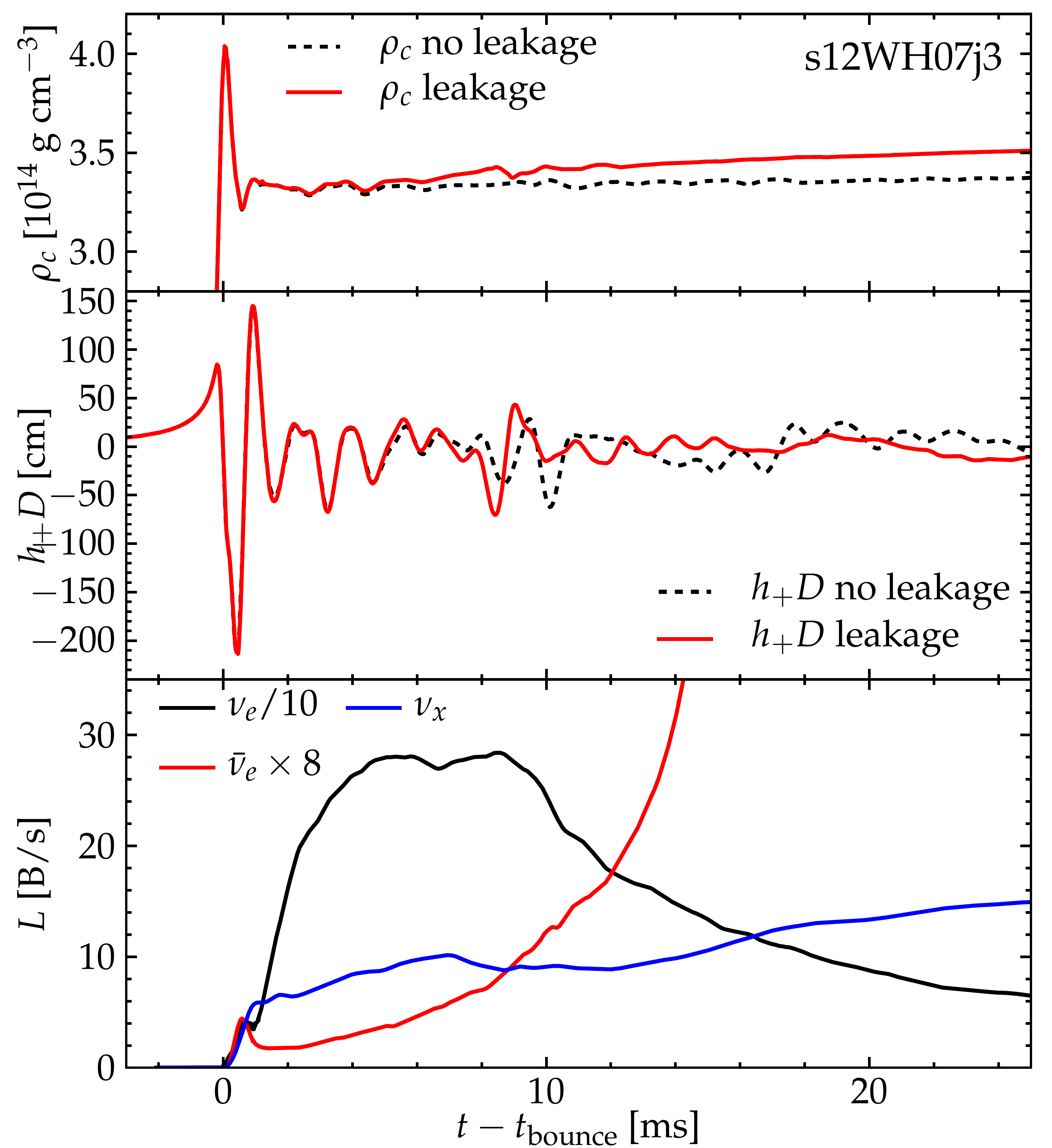}
\end{minipage}
\begin{minipage}{\linewidth}
\vspace*{-0.0cm}
\includegraphics[width=0.38\linewidth]{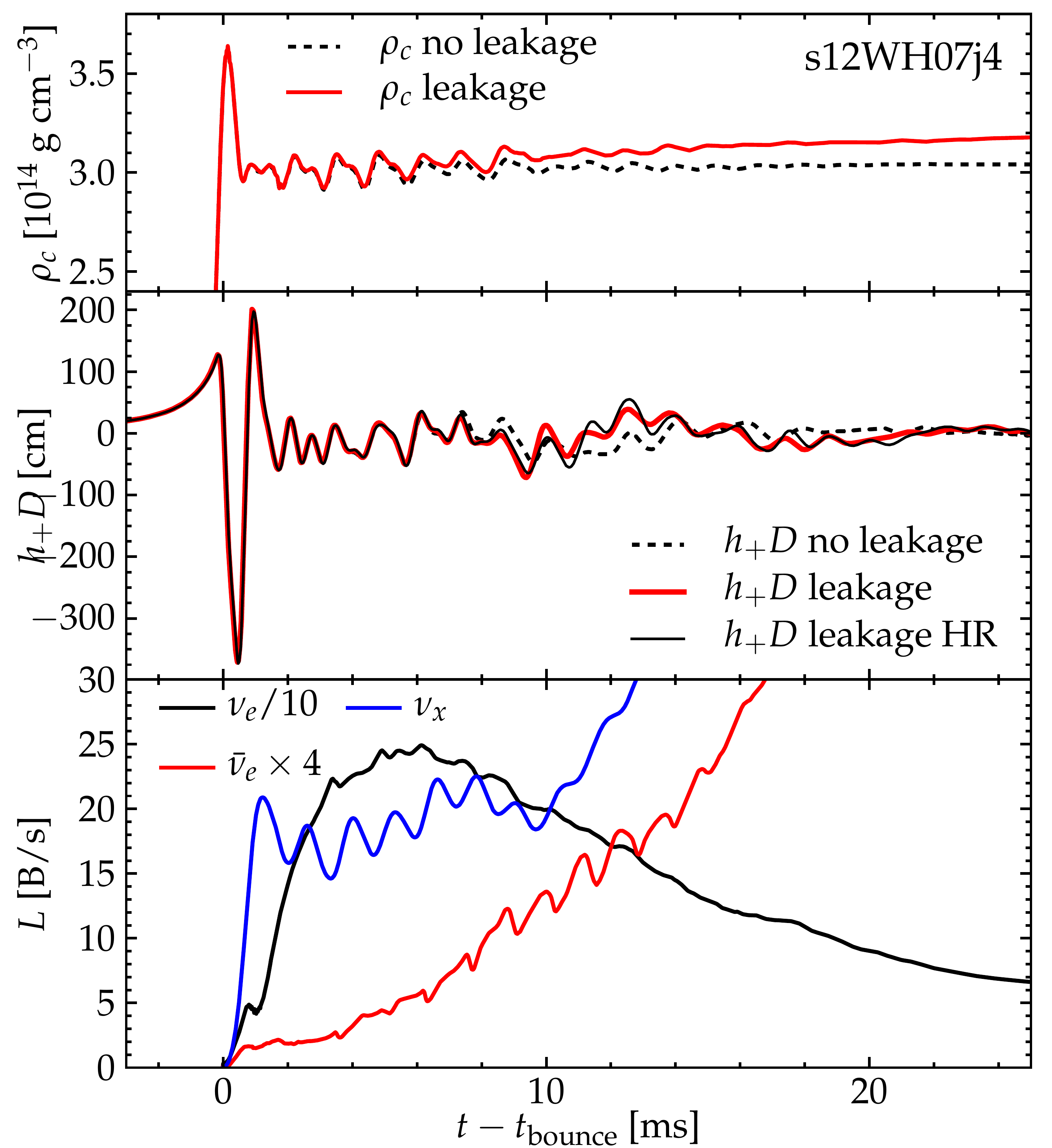}
\includegraphics[width=0.38\linewidth]{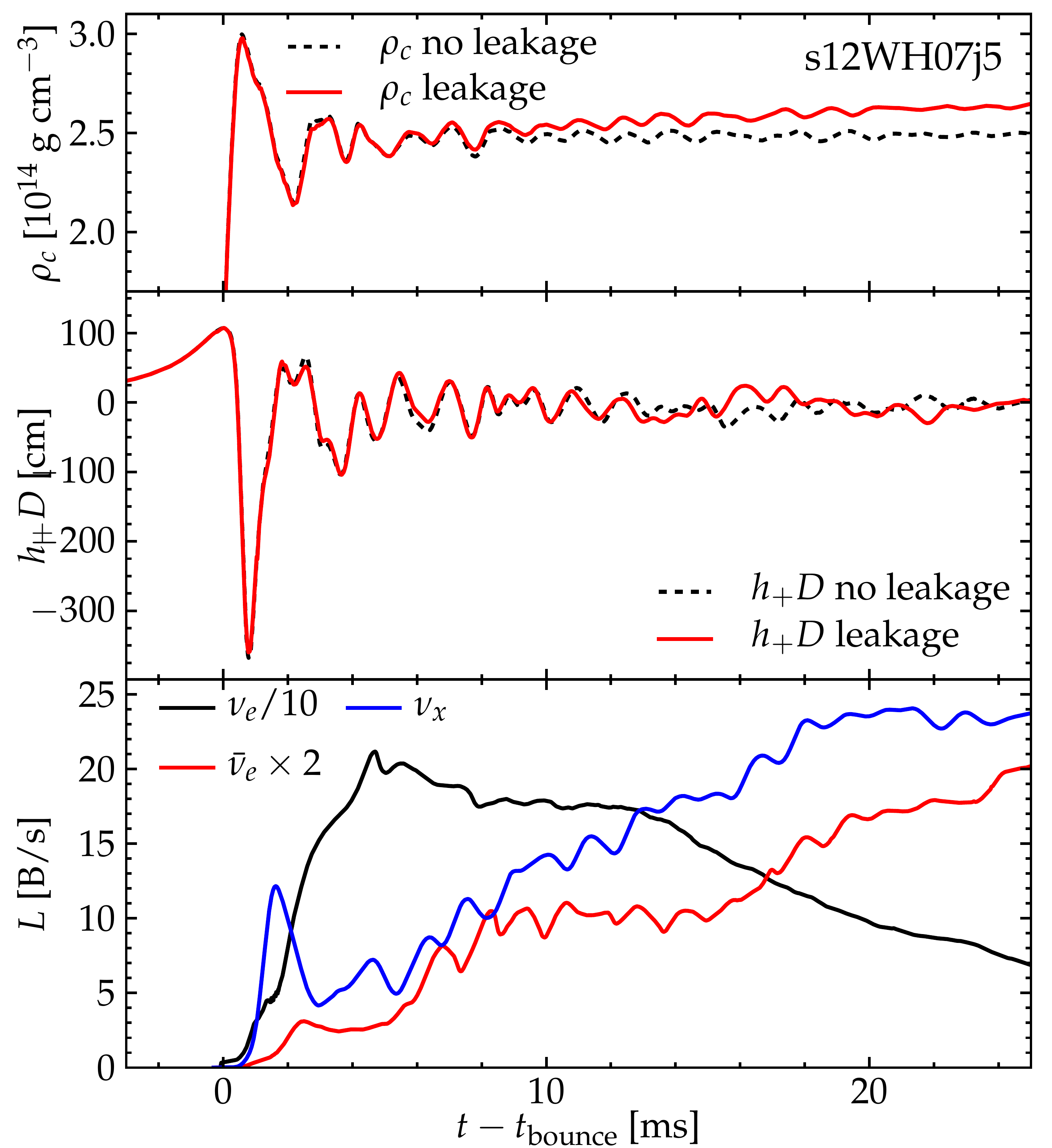}
\end{minipage}
\caption{Evolution of the central density ($\rho_c$, top sub-panel),
  GW signal ($h_+ D$, rescaled by distance $D$) seen by an equatorial
  observer (center sub-panel), and $\nu_e$, $\bar{\nu}_e$, $\nu_x$
  neutrino luminosities $L_{\nu_i}$ (bottom sub-panel) from shortly
  before core bounce to $25\,\mathrm{ms}$ after bounce in the s12WH07
  models ordered according to their precollapse rotation rate (left to
  right, top to bottom). The $L_{\nu_i}$ are scaled to emphasize the
  very early postbounce phase.  Also shown, in dashed lines, are the
  $\rho_c$ evolutions and GW signals of the simulations without
  neutrino leakage. For the j4 model, we also plot the GW signal as
  obtained from a simulation with 20\% higher resolution (HR). The
  effects of neutrino leakage on the dynamics and GW emission are
  strongest in nonrotating or slowly rotating models (j\{0-2\}).
  Note the prominent correlated oscillations in $\rho_c$,
  GWs, and neutrino signals starting, at low amplitude, in the j3
  model and becoming quite prominent in the j4 and j5 models.}
\label{fig:s12_rho_h_nu}
\end{figure*}

Our focus in this section is on the primary and most important new
result of this study: the correlation of GW and neutrino signals in
the very early postbounce evolution of rapidly rotating core-collapse
supernovae. We use the s12WH07 model set with neutrino leakage as the
basis of our discussion and postpone a detailed quantitative
analysis of the effects of including neutrino leakage, differences due
to progenitor structure, and various other aspects of our simulations
to Section~\ref{sec:results2}.

Figure~\ref{fig:s12_rho_h_nu} depicts, for the s12WH07 model set, the
central density ($\rho_c$) evolution, the GW signal, and the total
approximate luminosities of the $\nu_e$, $\bar{\nu}_e$, and $\nu_x$
neutrinos as predicted by our leakage scheme. The panels are ordered
according to the initial rotation rate of each model from left to
right and top to bottom.  The times are given with respect to the
times of core bounce in each model, which corresponds almost exactly
to the time of the pronounced global maximum in $\rho_c$ seen in the
top sub-panels of Fig.~\ref{fig:s12_rho_h_nu} in each model.

Neutrino leakage is activated at core bounce and we plot the resulting
neutrino luminosities in the bottom sub-panels of
Fig.~\ref{fig:s12_rho_h_nu}.  The leakage scheme captures the large
neutronization burst in the $\nu_e$ luminosity ($L_{\nu_e}$),
which reaches its peak within $5-8\,\mathrm{ms}$ of bounce. This burst
sets in when the shock breaks through the $\nu_e$ neutrinosphere (the
location where the optical depth $\tau_{\nu_e} \approx 2/3$) and electrons
capture on protons freshly liberated from shock-dissociated heavy
nuclei, producing electron neutrinos. The typical, much slower rise of
the $\bar{\nu}_e$ and $\nu_x$ luminosities (e.g., \cite{buras:06a}) is
also present. The luminosities shown here are the instantaneous
integral luminosities and we do not consider neutrino
oscillations. Due to the approximate and energy-averaged nature of the
leakage scheme, our quantitative predictions for the $L_{\nu_i}$
should be taken with a grain of salt, while their qualitative features
are most likely robust \cite{oconnor:10,oconnor:11}. We focus our
discussion on the qualitative aspects of the neutrino signal.

The GW signals are shown in the center sub-panels of
Fig.~\ref{fig:s12_rho_h_nu}. They are the waveforms observed by an
observer located in the equatorial plane and their
amplitudes have been rescaled by source distance $D$ and are given in
units of centimeters.  In the nonrotating model s12WH07j0, GW emission
is entirely due to prompt postbounce convection, which is driven
initially by the negative entropy gradient that naturally develops
behind the stalling shock, immediately after bounce. Prompt convection
is seeded by perturbations coming from numerical noise in our
simulations, which is, due to our Cartesian grid, of $\ell = 4$ (and
even $m$) character. Hence, prompt convection has initially $\ell = 4$
structure and becomes visible in quadrupole GWs
only\footnote{Accelerated $\ell = 4$ dynamics will also emit GWs, but
  their amplitudes are suppressed by $c^{-2}$ compared to quadrupole
  ($\ell = 2$) waves \cite{schutz:85}.}  once $\ell = 2$ components
develop in its non-linear phase. GW emission sets in around
$8\,\mathrm{ms}$ after bounce in model s12WH07j0, reaches peak
amplitudes of $\sim$$17 \,\mathrm{cm}$ (corresponding to a
dimensionless strain of $\sim$$5.5\times 10^{-22}$ at
$D=10\,\mathrm{kpc}$) and its GW spectral energy density peaks at
$\sim$$800\,\mathrm{Hz}$.  The onset of GW emission and the waveform
characteristics sensitively depend on the shock dynamics, the
thermodynamics of the region behind the shock, and on the distribution
and magnitude of the seed perturbations (i.e., numerical resolution in
our case). Hence, the GW signal observed for model s12WH07j0 is just
one possible realization and should not be taken as being
representative for prompt convection (see the discussion in
\cite{marek:09b,ott:09}). The neutrino emission in the nonrotating
model shows the global features generally seen in 1D (spherically
symmetric) simulations (e.g., \cite{buras:06a,oconnor:10}). Besides
that the GW and neutrino signals set in within milliseconds of each
other, there is no other obvious correlation between them in model
s12WH07j0.

Rotation leads to a centrifugal flattening of the inner
core into oblate shape, which results in a prominent spike in the GW
signal at core bounce, when the inner core experiences extreme
acceleration, essentially reverting its infall velocity on a timescale
of $\mathcal{O}(1)\,\,\mathrm{ms}$.  This bounce spike is clearly
visible, though dwarfed by the signal from prompt
convection, in the most slowly spinning model s12WH07j1. It starts out
with an initial period $P_0 = 2\pi\,\mathrm{s}$ and makes a
protoneutron star with a spin period of $\sim$$10.2\,\mathrm{ms}$ and
a ratio of rotational kinetic energy to gravitational energy ($T/|W|$)
of $\sim$0.4\%, which is a very small rotation rate in
the context of rotating core collapse.  Consequently, the overall
collapse, bounce, and postbounce dynamics and the neutrino signals are
very similar to the nonrotating case. However, the onset of
appreciable GW emission already occurs robustly $5-8\,\mathrm{ms}$
before the peak of the $\nu_e$ neutronization burst and
$\gtrsim$$20\,\mathrm{ms}$ before the $\bar{\nu}_e$ and $\nu_x$
luminosities grow to their postbounce steady-state values.

Increasing precollapse core rotation results in a more oblate inner
core at bounce, which translates directly into a more prominent bounce
spike in the gravitational waveform. Model s12WH07j2 starts out with
a central precollapse period of $P_0 = \pi\,\mathrm{s}$ and is spun up
by collapse to a postbounce period of $\sim$$5.2\,\mathrm{ms}$ and a
modest $T/|W|$ of $\sim$1.3\%. Its inner-core dynamics,
diagnosed by $\rho_c$, and its neutrino luminosities are still very
similar to the nonrotating case.  Due to the increased oblateness of
the inner core, the GW signal now has a peak amplitude of
$\sim$$60\,\mathrm{cm}$ (corresponding to $\sim$$1.9\times10^{-21}$ at
$10\,\mathrm{kpc}$) and its $dE_\mathrm{GW}/df$ peaks at
$\sim$$750\,\mathrm{Hz}$.

Model s12WH07j3 starts out with $P_o = \pi/2\,\mathrm{s}$ and achieves
a postbounce period of $\sim$$2.4\,\mathrm{ms}$ and $T/|W|$ of
$\sim$4.6\%.  It is the first model in this sequence that is
appreciably affected by rotation: its maximum density at bounce is
reduced by $\sim$$10\%$ due to the stabilizing centrifugal force and
its peak GW amplitude is a factor of $\sim$$3.5$ greater than in model
s12WH07j2. The maximum GW amplitude reached is
$\sim$$210\,\mathrm{cm}$ ($\sim$$6.8\times10^{-21}$ at
$10\,\mathrm{kpc}$) and $dE_\mathrm{GW}/df$ peaks at
$\sim$$716\,\mathrm{Hz}$. On the neutrino side, the rotationally
deformed, less compact core, leads to a $L_{\nu_e}$ maximum that is
significantly broadened, but reaches a lower maximum and settles at
smaller postbounce values, which is consistent with what was found by
Ott~et~al.~\cite{ott:08} in 2D rotating core-collapse supernova
simulations with more accurate neutrino transport.  The GW signal
exhibits a number of pronounced postbounce oscillations due to
pulsations of the PNS core whose radial components are reflected in
small variations of $\rho_c$. The neutrino signals also exhibit slight
traces of oscillations, though without clear correlation with $\rho_c$
and GW signal.

Model s12WH07j4 (bottom-left panel of Fig.~\ref{fig:s12_rho_h_nu}) has
a central precollapse spin period $P_0 = \pi/3\,\mathrm{s}$, a PNS
spin period of $\sim$$2\,\mathrm{ms}$ and an early postbounce $T/|W|$
of $\sim$$8.9\%$. Its late collapse and bounce dynamics is affected by
rotation, lowering the peak density at bounce by $\sim$$20\%$ compared
to the nonrotating value, but still not significantly widening the
bounce spike in the GW signal, which peaks at an amplitude of
$\sim$$364\,\mathrm{cm}$ ($\sim$$1.2\times10^{-20}$ at
$10\,\mathrm{kpc}$). The spectral GW energy density
$dE_\mathrm{GW}/df$ peaks at $\sim$$787\,\mathrm{Hz}$. This upward
shift from model s12WH07j3 is due to the more pronounced and higher
frequency early-postbounce oscillations of the GW signal in this
model. These oscillations in the GW signal are also prominently
visible in its early-postbounce central density evolution. This
suggest that a PNS pulsation mode has been excited and leads to global
oscillations of the PNS core that persist at appreciable amplitude for
$\gtrsim$$20\,\mathrm{ms}$ after bounce (see
Section~\ref{sec:res2_modes} for further analysis of the nature of
this pulsation mode).  The periodic contraction and expansion of the
PNS core seen in model s12WH07j4 has a significant impact on the
neutrino signal and leads to a characteristic modulation that is
strongest in the $\bar{\nu}_e$ and $\nu_x$ luminosities and weaker in
the $\nu_e$ luminosity.

To check the sensitivity of our results to resolution, we carry out
another simulation of model s12WH07j4 with 20\% higher resolution
(HR). We find that the HR and standard-resolution GW signals (see
Fig.~\ref{fig:s12_rho_h_nu}) and all other quantities match almost
perfectly until $\sim$10\,ms after bounce. At later times, the damping
of the PNS pulsations is slightly stronger in the lower-resolution
calculation and resolution-dependent convective dynamics starts to
play a role in the GW emission, thus leading to slight differences
between the HR and standard-resolution simulations.

\begin{figure}[t]
\centering
\includegraphics[width=\columnwidth]{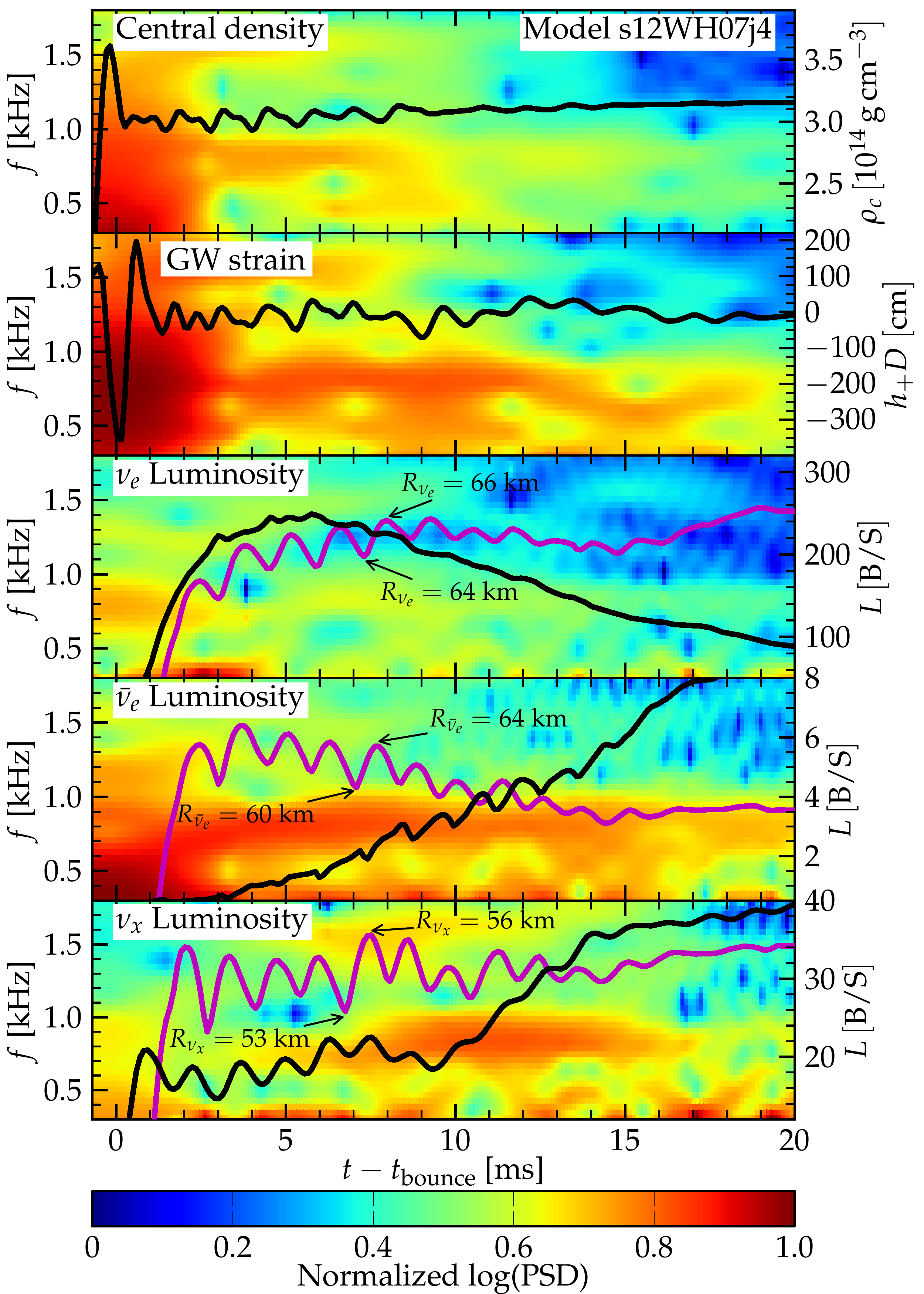}
\caption{Spectrograms of the power spectral densities computed via
  short-time Fourier transforms of the central density evolution (top
  panel), GW signal (second panel), $L_{\nu_e}$ (third panel),
  $L_{\bar{\nu}_e}$ (fourth panel), and $L_{{\nu}_x}$ (bottom panel).
  In each panel, the time-domain evolution of the respective quantity
  is superposed with the scales set by the right ordinates. In the
  panels showing neutrino luminosity data, we also provide graphs of
  the time evolution of the corresponding energy and angle averaged
  neutrinosphere radii. All quantities show significant correlated
  excess power in their power spectral densities (PSDs) around
  $\sim$$700-800\,\mathrm{Hz}$. The correlation in the $\nu_e$
  luminosity is weak, since it is primarily sourced by electron
  capture outside the $\nu_e$ neutrinosphere.  The spectrograms are
  obtained by sliding an $8$-$\mathrm{ms}$ Hann window over the data
  with a $0.1\,\mathrm{ms}$ timestep. The scales on which the time
  series data are shown have been chosen to highlight the correlated
  multi-messenger signal oscillations. }
\label{fig:s12j4_all}
\end{figure}

\begin{table*}[t]
\begin{center}
  \caption[Summary of Simulation Results]{Summary of key simulation
    results for the s12WH07j\{0-5\} and s40WH07j\{0-5\} model sets.
    $t_\mathrm{b}$ is the time of core bounce, defined as the time at
    which the entropy reaches $3\,k_\mathrm{B}\,\mathrm{baryon}^{-1}$
    at the edge of the inner core, $\rho_\mathrm{max,b}$ is the
    maximum density reached, $M_\mathrm{ic,b}$ is the mass of the
    inner core at the point of maximum compression, $J_\mathrm{ic,b}$
    is the angular momentum contained in the inner core,
    $T/|W|_\mathrm{ic,b}$ and $T/|W|_\mathrm{e}$ are the ratio of
    rotational kinetic energy to gravitational energy at bounce and at
    $25\,\mathrm{ms}$, respectively. The rightmost five columns
    contain key GW-related quantities: $|h_+|_\mathrm{max} D$ is the
    magnitude of the peak GW amplitude, $E_\mathrm{GW}$ is the emitted
    energy in GWs, $h_\mathrm{char,max}$ is the peak value of the
    characteristic GW strain spectrum (Eq.~\ref{eq:hchar}) and
    $f_\mathrm{char,max}$ is its location, and the optimal SNR in
    Advanced LIGO at $10\,\mathrm{kpc}$ is computed according to
    Eq.~\ref{eq:snr}.  }
\begin{tabular}{lccccccccccc}
 \hline \hline
Model&$t_\mathrm{b}$&$\rho_\mathrm{max,b}$&$M_\mathrm{ic,b}$&$J_\mathrm{ic,b}$ &$T/|W|_\mathrm{ic,b}$&$T/|W|_\mathrm{e}$          &$|h_\mathrm{+}|_\mathrm{max}| D$&$E_\mathrm{GW}$         &$h_\mathrm{char,max}$&$f_\mathrm{char,max}$& Optimal SNR \\
     &(ms) &($10^{14}$           &$(M_\odot)$      &$(10^{48}$ g      & (\%)                &(\%)          &                   &$ (10^{-9}$             &$(10^{-22})$         &(Hz)                 & in aLIGO    \\
     &     &g cm$^{-3}$)         &                 &cm$^{2}$ s$^{-1})$&                     &&                   &$\,\,\,\,\,M_\odot c^2)$&                     &                     & at $10$ kpc \\
\hline
s12WH07j0  &\lowentry{$178.52$}&\lowentry{${4.34}$}&\lowentry{$0.56$}&\phantom{0}\lowentry{$0.00$}&\lowentry{$0.0$} &0.0&$17.5$            &$\phantom{0}0.27$&$\phantom{0}9.0$  &$802$&$\phantom{0}6.02$ \\
s12WH07j0nl&                   &                   &                 &                            &                 &0.0&$\phantom{0}8.2$  &$\phantom{0}0.04$&$\phantom{0}3.8$  &$871$&$\phantom{0}3.07$ \\
\hline
s12WH07j1  &\lowentry{$179.60$}&\lowentry{${4.33}$}&\lowentry{$0.57$}&\phantom{0}\lowentry{$0.44$}&\lowentry{$0.4$} &0.4&$25.5$            &$\phantom{0}0.51$&$12.2$            &$730$&$12.10$ \\
s12WH07j1nl&                   &                   &                 &                            &                 &0.3&$15.6$            &$\phantom{0}0.17$&$\phantom{0}9.4$  &$738$&$\phantom{0}6.61$ \\ 
\hline
s12WH07j2  &\lowentry{$182.93$}&\lowentry{${4.28}$}&\lowentry{$0.57$}&\phantom{0}\lowentry{$0.91$}&\lowentry{$1.6$} &1.3&$60.3$            &$\phantom{0}1.58$&$28.5$            &$750$&$15.37$ \\
s12WH07j2nl&                   &                   &                 &                            &                 &1.3&$59.6$            &$\phantom{0}1.36$&$21.3$            &$736$&$12.97$ \\
\hline
s12WH07j3  &\lowentry{$198.40$}&\lowentry{${4.04}$}&\lowentry{$0.59$}&\phantom{0}\lowentry{$1.82$}&\lowentry{$5.1$} &4.9&$209.6\phantom{0}$&$20.08$          &$75.0$            &$716$&$40.75$ \\
s12WH07j3nl&                   &                   &                 &                            &                 &4.6&$209.0\phantom{0}$&$19.65$          &$83.3$            &$721$&$41.19$ \\
\hline
s12WH07j4  &\lowentry{$235.63$}&\lowentry{${3.64}$}&\lowentry{$0.62$}&\phantom{0}\lowentry{$2.88$}&\lowentry{$9.2$} &9.5&$363.5\phantom{0}$&$47.13$          &$121.6\phantom{0}$&$787$&$66.22$ \\
s12WH07j4nl&                   &                   &                 &                            &                 &8.9&$363.4\phantom{0}$&$46.47$          &$117.2\phantom{0}$&$771$&$65.14$ \\
\hline
s12WH07j5  &\lowentry{$383.36$}&\lowentry{${3.00}$}&\lowentry{$0.66$}&\phantom{0}\lowentry{$4.42$}&\lowentry{$13.6$}&13.8&$346.3\phantom{0}$&$23.03$         &$86.1$            &$754$&$73.42$ \\ 
s12WH07j5nl&                   &                   &                 &                            &                 &12.4&$353.3\phantom{0}$&$24.32$         &$92.0$            &$737$&$73.64$ \\ 
\hline \hline
s40WH07j0  &\lowentry{$418.22$}&\lowentry{${4.41}$}&\lowentry{$0.62$}&\phantom{0}\lowentry{$0.00$}&\lowentry{$0.0$} &0.0&$41.4$            &$\phantom{0}0.03$&$\phantom{0}7.1$  &$381$&$\phantom{0}9.81$ \\
s40WH07j0nl&                   &                   &                 &                            &                 &0.0&$16.8$            &$\phantom{0}0.07$&$\phantom{0}3.8$  &$711$&$\phantom{0}4.74$ \\
\hline
s40WH07j1  &\lowentry{$419.93$}&\lowentry{${4.40}$}&\lowentry{$0.62$}&\phantom{0}\lowentry{$0.54$}&\lowentry{$0.4$} &0.5&$41.5$            &$\phantom{0}0.53$&$14.4$            &$769$&$10.17$ \\
s40WH07j1nl&                   &                   &                 &                            &                 &0.5&$27.7$            &$\phantom{0}0.23$&$\phantom{0}8.2$  &$607$&$12.95$ \\
\hline
s40WH07j2  &\lowentry{$425.22$}&\lowentry{${4.32}$}&\lowentry{$0.62$}&\phantom{0}\lowentry{$1.06$}&\lowentry{$1.4$} &1.8&$67.9$            &$\phantom{0}1.64$&$26.4$            &$711$&$16.41$ \\
s40WH07j2nl&                   &                   &                 &                            &                 &1.7&$67.2$            &$\phantom{0}1.60$&$23.0$            &$677$&$16.15$ \\
\hline
s40WH07j3  &\lowentry{$448.23$}&\lowentry{${4.10}$}&\lowentry{$0.67$}&\phantom{0}\lowentry{$2.34$}&\lowentry{$5.0$} &6.2&$237.2\phantom{0}$&$23.16$          &$93.4$            &$708$&$47.70$ \\
s40WH07j3nl&                   &                   &                 &                            &                 &6.2&$235.9\phantom{0}$&$22.41$          &$96.4$            &$689$&$46.83$ \\
\hline
s40WH07j4  &\lowentry{$495.68$}&\lowentry{${3.65}$}&\lowentry{$0.68$}&\phantom{0}\lowentry{$3.38$}&\lowentry{$8.9$} &11.6&$367.0\phantom{0}$&$45.94$         &$131.9\phantom{0}$&$766$&$75.35$ \\ 
s40WH07j4nl&                   &                   &                 &                            &                 &11.1&$366.8\phantom{0}$&$45.09$         &$137.2\phantom{0}$&$757$&$73.78$ \\ 
\hline
s40WH07j5  &\lowentry{$591.19$}&\lowentry{${3.00}$}&\lowentry{$0.72$}&\phantom{0}\lowentry{$4.89$}&\lowentry{$12.6$}&14.7&$402.8\phantom{0}$&$30.20$         &$87.3$            &$743$&$86.24$ \\
s40WH07j5nl&                   &                   &                 &                            &                 &13.1&$383.4\phantom{0}$&$26.07$         &$85.7$            &$716$&$85.12$ \\
\hline\hline
\end{tabular}
\label{table:results}
\end{center}
\end{table*}

In Fig.~\ref{fig:s12j4_all}, we present spectrograms obtained via
short-time Fourier transforms of the central density ($\rho_c$), GW
signal and $\nu_e$, $\bar{\nu}_e$, and $\nu_x$ luminosities for model
s12WH07j4.  To guide the eye, we also plot the time series data for
each spectrogram panel and add graphs of the corresponding
energy/angle-averaged neutrinosphere radii to the panels depicting
neutrino luminosity data. There is a clear correlation between
dynamics ($\rho_c$), GW signal, and neutrino signal: their power
spectra all exhibit significant excess power at
$\sim$$700-800\,\mathrm{Hz}$ between $\sim$$2\,\mathrm{ms}$ and
$\sim$$20\,\mathrm{ms}$ after bounce.

The $\bar{\nu}_e$ and $\nu_x$ neutrino luminosities exhibit variations
of order $\lesssim$$20\%$. Their emission occurs primarily below or
near their neutrinospheres, so $L_{\bar{\nu}_e}$ and $L_{\nu_x}$ are
particularly sensitive to their respective neutrinosphere radii
$(R_{\nu_i})$.  As demonstrated by Fig.~\ref{fig:s12j4_all}, the
variations in $R_{\nu_i}$ due to the oscillations of the PNS core are
sizable and have relative amplitudes of $\Delta R_{\nu_i} / R_{\nu_i}
\sim 5\%$. They lead to a change of the neutrino-emitting volume of
the same order and temporarily uncover regions of higher
emissivity. This drives the variations in the $\bar{\nu}_e$ and
$\nu_x$ luminosities. The $\nu_e$ luminosity, on the other hand, is
due primarily to electron capture on free protons occurring above
the neutrinosphere. Hence, $L_{\nu_e}$ is less affected by changes of
its neutrinosphere's radius.  Note that the neutrinospheres are
centrifugally deformed and since we are plotting their angle-averaged
radii, we do not expect phase coherence between the $L_{\nu_i}$ and
$R_{\nu_i}$ curves shown in Fig.~\ref{fig:s12j4_all}. 

\begin{figure*}[t]
\centering
 \includegraphics[width=0.95\textwidth]{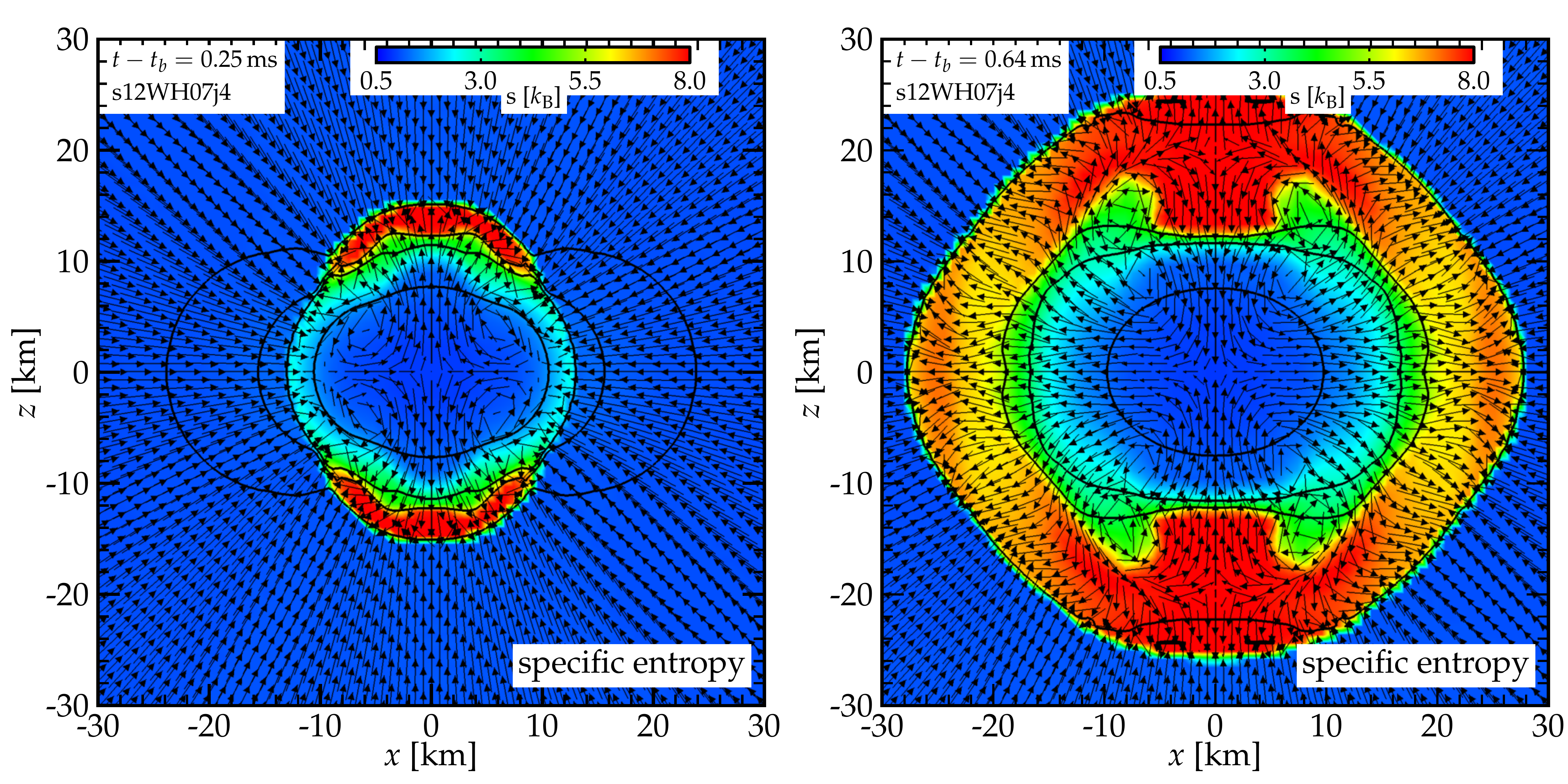}
 \caption{Colormaps of the specific entropy in the meridional plane
   of the rapidly rotating model s12WH07j4. The contour lines mark
   isocontours of rest-mass density at $\{10^{14},10^{13},5\times
   10^{12},10^{12}\}\,\mathrm{g}\,\mathrm{cm}^{-3}$ (in this order
   from the origin).  Velocity vectors are superposed and their length
   is saturated at $0.01\,c$. The left panel shows a
   snapshot at $0.25\,\mathrm{ms}$ after bounce while the right panel
   shows the situation at $0.64$ ms. Due to rapid rotation and thus
   strong centrifugal support at low latitudes, bounce occurs earlier
   and is stronger (leading to higher specific entropies) near the
   poles. As shown in the left panel, at $0.25$ ms after bounce, the
   inner core is re-expanding along the polar direction while most of
   the material near the equatorial plane is still falling in. At the
   slightly later time shown in the right panel, matter along the
   rotation axis is receding, while the inner core is expanding
   on the equator. The flow pattern in the inner
   core has strong quadrupolar features that excite the fundamental
   quadrupole oscillation mode.}
  \label{fig:colormapbounceframe}
\end{figure*}

The bottom right panel of Fig.~\ref{fig:s12_rho_h_nu} shows $\rho_c$,
$h_+ D$, and $L_{\nu_i}$ for model s12WH07j5, our most rapidly
spinning model. Its precollapse central spin period is $P_0 =
\pi/4\,\mathrm{s}$, which translates to an early postbounce spin
period of $\sim 1.5\,\mathrm{ms}$ and a $T/|W|$ of $\sim$$12.5\%$.
The late collapse and bounce phase of this model is strongly affected
by rotation and bounce occurs partially centrifugally at a central
density of $\sim$$3\times 10^{14}\,\mathrm{g}\,\mathrm{cm}^{-3}$. The
strong influence of centrifugal effects slows down the late collapse
and bounce of the inner core, which is reflected in the widening
of the pronounced bounce peaks in $\rho_c$ and in the GW signal. Model
s12WH07j5's inner-core quadrupole deformation is significantly greater
than that of s12WH07j4's inner core. However, the rotationally-slowed
bounce dynamics and the resulting smaller net acceleration of the
inner core leads to a peak GW amplitude of
$\sim$$346\,\mathrm{cm}$ ($\sim$$1.17\times 10^{-20}$ at
$10\,\mathrm{kpc}$), which is slightly smaller than in model
s12WH07j4. Model s12WH07j5's GW spectral energy density
$dE_\mathrm{GW}/df$ has a broad peak at $754\,\mathrm{Hz}$ with a FWHM
of $60\,\mathrm{Hz}$ in the $720-780\,\mathrm{Hz}$ interval.

The central density evolution, GW signal, and neutrino luminosities of
model s12WH07j5 show similar correlated behavior as in the less
rapidly spinning model s12WH07j4: The postbounce central density
evolution exhibits quasi-periodic oscillations that have appreciable
amplitudes for many cycles, indicative of a PNS pulsation mode that is
excited at core bounce and subsequently rings down. This dynamics is
imprinted on GW and neutrino signals, which exhibit variations
that correlate with the oscillations of the central density.

The correlated oscillations in central density, GW signal, and
neutrino luminosities driven by rotationally excited 
pulsations are strong indicators of rapid rotation (at both
precollapse and PNS stages) and the combined GW and neutrino data from
the next galactic core-collapse event may thus allow one to constrain
the rotation rate of its core. Provided the next galactic
core-collapse supernova occurs sufficiently nearby, it will not only
be possible to reconstruct its complete two-polarization GW signal but
also to deduce detailed neutrino ``lightcurves'' and spectra from the
events observed by neutrino detectors. Our present energy-averaged
neutrino leakage scheme is too crude to make robust quantitative
predictions on the expected neutrino signals. Qualitatively, however,
we expect it to correctly capture the most important features and
systematics. This includes the strong oscillations in the early
$L_{\bar{\nu}_e}$ and $L_{\nu_x}$ evolutions going along with the PNS
oscillations, but also the global trends with increasing rotation that one
notes from Fig.~\ref{fig:s12_rho_h_nu} for all
$L_{\nu_i}$. Universally, increasing rotation leads to a decrease of
$L_{\nu_e}$ and to a broadening of its early-postbounce peak. This
change is small for models s12WH07j0 to s12WH07j2, but already
significant in model s12WH07j3 and strong in models s12WH07j4 and
s12WH07j5.  The same systematics holds for the $\nu_x$ luminosities in
our models. The $\bar{\nu}_e$ luminosity, on the other hand, exhibits
an increase with rotation in the first $12-15\,\mathrm{ms}$ after bounce,
which subsequently turns into a decrease with rotation at later times.
This behavior is most likely due to the fact that the
rotationally-flattened neutrinospheres allow for greater early
emission of $\bar{\nu}_e$ from small-radius polar regions at early times
while the subsequent rise due to emission at greater radii is weaker
in rapidly spinning models, since rapid rotation leads to uniformly lower
temperatures (and entropies) \cite{ott:06spin,ott:08}.

\section{Results: Detailed Discussion}
\label{sec:results2}

\subsection{Analysis of Postbounce Core Oscillations}
\label{sec:res2_modes}

The spectrograms of central density, GW signal, and neutrino
luminosities of model s12WH07j4 shown in Fig.~\ref{fig:s12j4_all} all
exhibit a strong long-lasting contribution at $f \sim
700-800\,\mathrm{Hz}$.  The extreme, quasi-discontinuous dynamics at
core bounce inject power at a broad range of frequencies of $\sim$$100
-$$\sim$$2000\,\mathrm{Hz}$, but within $2-3\,\mathrm{ms}$ after bounce the
clear and persistent peak of the PSD at $\sim 700-800\,\mathrm{Hz}$
emerges and lasts for $\gtrsim$$20\,\mathrm{ms}$. 

The pulsation is clearly of global nature, has significant amplitude
out to $\gtrsim 60\,\mathrm{km}$, and its oscillation period is of the
same order as the dynamical time, $t_\mathrm{dyn} \sim (\bar{\rho}
G)^{-1/2}$, of this region, suggesting that a pulsational eigenmode of
the PNS has been excited. Moreover, we find a clear trend with
increasing rotation (as is apparent even from the time-series data
shown in Fig.~\ref{fig:s12_rho_h_nu}). No oscillations are present in
nonrotating or slowly rotating models (j0, j1, j2), but the moderately
rapidly rotating j3 model and the rapid rotators j4 and j5 all exhibit
the prominent oscillation mode at $700-800\,\mathrm{Hz}$. This trend
is independent of progenitor model (see
Section~\ref{sec:progcomp}). 

Rotation leads to a quadrupole deformation of the collapsing and
bouncing inner core and the close connection of the appearance of
the oscillation mode with rapid rotation strongly suggests that
a quadrupole eigenmode is excited. 

In nonrotating or only slowly rotating models, bounce is nearly
spherical and most of the kinetic energy of the inner core is
transferred to the nearly spherical hydrodynamic shock. In the case of
rapid rotation, core bounce is non-uniform. In
Fig.~\ref{fig:colormapbounceframe}, we show colormaps of the specific
entropy in the meridional plane of model s12WH07j4 at
$0.25\,\mathrm{ms}$ and $0.64\,\mathrm{ms}$ after the entropy has
first reached $3\,k_\mathrm{b}/\text{baryon}$, which is our criterion
for core bounce. Velocity vectors and representative density
isocontours are superposed. Due to the centrifugal
deformation of the inner core, bounce occurs earlier and is stronger
along the polar direction than in the equatorial plane. When the inner
core begins to re-expand along the equatorial direction, it is already
contracting again at the poles. This large-scale non-uniform dynamics
acts as the initial perturbation exciting the pulsation of the newborn
PNS.

Figure~\ref{fig:colormapbounceframe} also demonstrates that the
velocity field of the inner core is primarily of quadrupole character
and similar to the eigenfunction of the fundamental quadrupole mode
(typically denoted by ${}^2 f$; c.f., Fig.~4.7 of \cite{kastaun:07}).
This perturbation should predominantly excite the fundamental
quadrupole mode of the star, which is then responsible for the
dominant power at $\sim$$700-800\,\mathrm{Hz}$ observed in our rapidly
spinning models (Fig.~\ref{fig:s12j4_all}). To gain further confidence
in this conclusion, we study a time sequence of the 2D meridional
velocity field of model s12WH07j4. We remove the contribution from all
frequencies by applying a bandpass filter on $(800 \pm
75)\,\mathrm{Hz}$ and analyze the result. We find strong evidence for
a velocity field akin to what has been found for the ${}^2 f$ mode of
nonrotating Newtonian stars (e.g., \cite{unno:89}) and rotating
relativistic stars in the Cowling
approximation~\cite{kastaun:07,font:01}: (\emph{i}) The radial
velocity $v_r$ does not exhibit any nodes in the PNS. (\emph{ii})
$|v_r| \propto r$ in the same region, but with lateral dependence due
to the oblate shape of the PNS. (\emph{iii}) $v_r$ along the polar
axis and in the equatorial plane are out of phase by half a
cycle. (\emph{iv}) $v_\theta \propto \sin(2\theta) $ in the PNS.  All
of these features together are specific only to the fundamental
quadrupole mode.

\begin{figure}
 \includegraphics[width=0.95\linewidth]{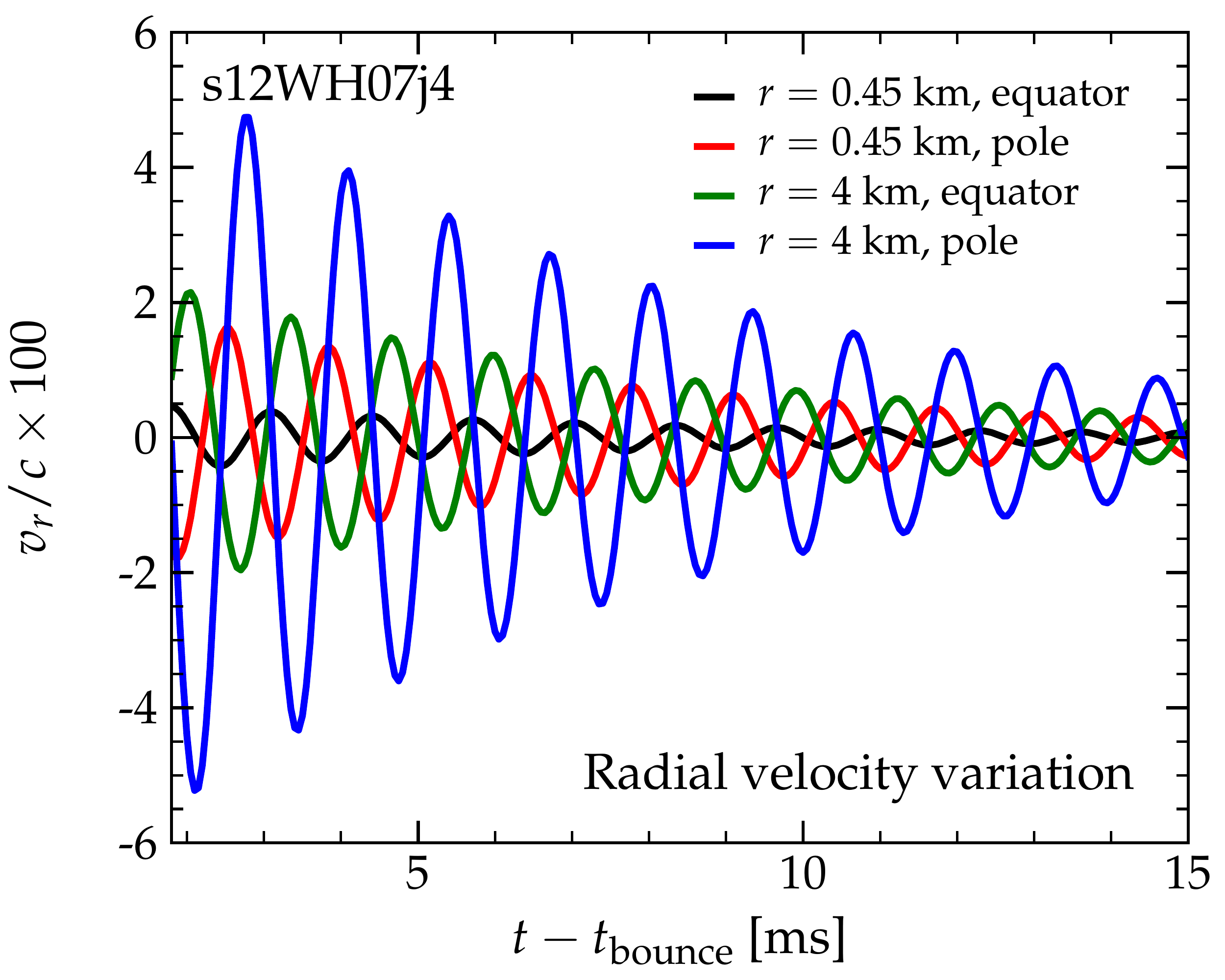}
  \caption{The time-evolution of the radial velocity of the PNS along
    polar and equatorial directions, band-passed around the frequency
    of the leading peak of its oscillation spectrum, for model
    s12WH07j4. The radial velocity along polar and equatorial
    directions are out of phase by half a cycle at any given radius in
    the PNS interior. Such behavior is also exhibited by the ${}^2 f$
    mode of a non-rotating Newtonian star.
  \label{fig:v_1_vs_t}}
\end{figure}

In Fig.~\ref{fig:v_1_vs_t} we show the time evolution of the filtered
radial velocity along polar and equatorial directions at
$0.45\,\mathrm{km}$ and $4\,\mathrm{km}$ from the origin. $v_r$ along
polar and equatorial directions at a given radius is out of phase by
half a cycle, as expected for a ${}^2 f$ oscillation. The variations
in $v_r$ are of order $\sim$5\% of the speed of light and, when
compared with the typical speed of sound in the PNS ($\sim$0.15\,c),
clearly represent non-linear pulsations. They are exponentially
damped, most likely by the emission of sound waves.

The oscillations exhibited by our rapidly rotating models have
manifestly non-linear amplitudes and the excitation at bounce, while
being primarily quadrupolar, involves large radial components and does
not exactly match the eigenfunction of the pure ${}^2 f$ mode. This
leads to the excitation of a broad spectrum of other modes (see, e.g.,
\cite{dimmelmeier:06}) as seen in the spectrograms of
Fig.~\ref{fig:s12j4_all}.  In the linear regime, the ${}^2 f$ mode
does not modify the central density and should not lead to the
pronounced variations in the central density observed in our
models. We attribute the presence of these variations to non-linear
driving of a quasi-radial pulsation at the same frequency as the ${}^2
f$ mode.

Finally, we note that the PNS oscillations in rapidly rotating models
observed and investigated here are present also in the central density
evolutions and GW signals in many models of previous parameter studies
of rotating iron core collapse
\cite{zwerger:97,dimmelmeier:02,ott:04,dimmelmeier:08} and
accretion-induced collapse of massive white
dwarfs~\cite{abdikamalov:10}. However, only the latter work made an
effort to study these pulsations and first pointed out that
fundamental quadrupole mode is likely to be the dominant mode of
pulsation.

\subsection{Leakage vs.\ No Leakage}
\label{sec:res2_leakage}

\begin{figure}
 \includegraphics[width=0.95\linewidth]{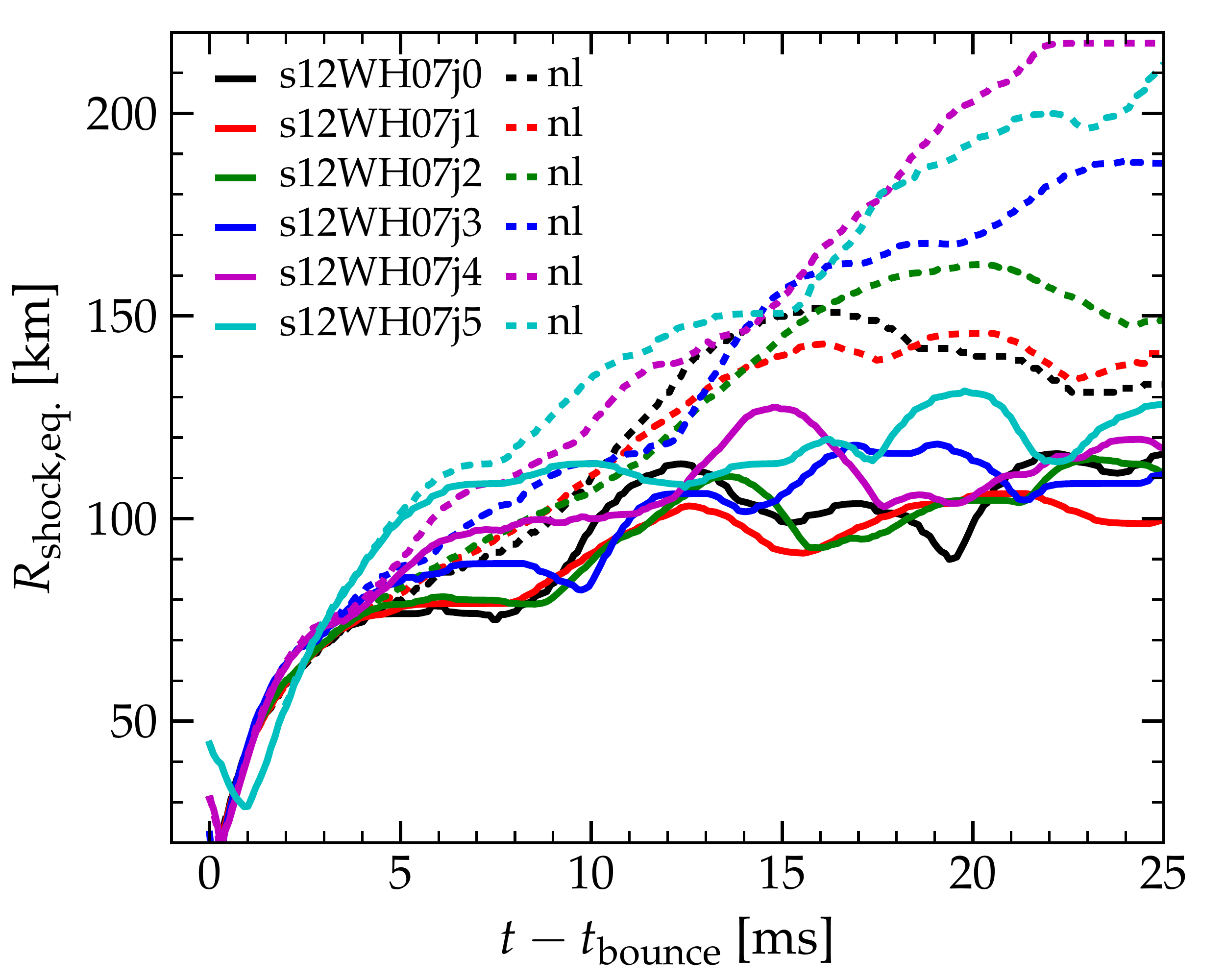}\\
  \caption{Shock position on the equator (along the positive $x$-axis)
    as a function of time after core bounce for the s12WH07j\{0-5\}
    model set. Models without neutrino leakage (``nl'', dashed lines)
    reach significantly larger shock radii, since they are, unlike
    the models that include neutrino leakage (solid lines), not affected
    by neutrino cooling and deleptonization. The rapidly spinning
    models s12WH07j4 and s12WH07j5 show an initial transient decrease
    in their equatorial shock radii. This is due to shock formation occurring
    first at the pole while the equatorial regions of the inner core
    are still contracting. Note that the discrete shock radius data has
    been spline-interpolated to produce this plot.
    \label{fig:s12shockrad}}
\end{figure}

\begin{figure*}[t]
\vspace{-0.35cm}
\centering
 \includegraphics[width=0.83\textwidth]{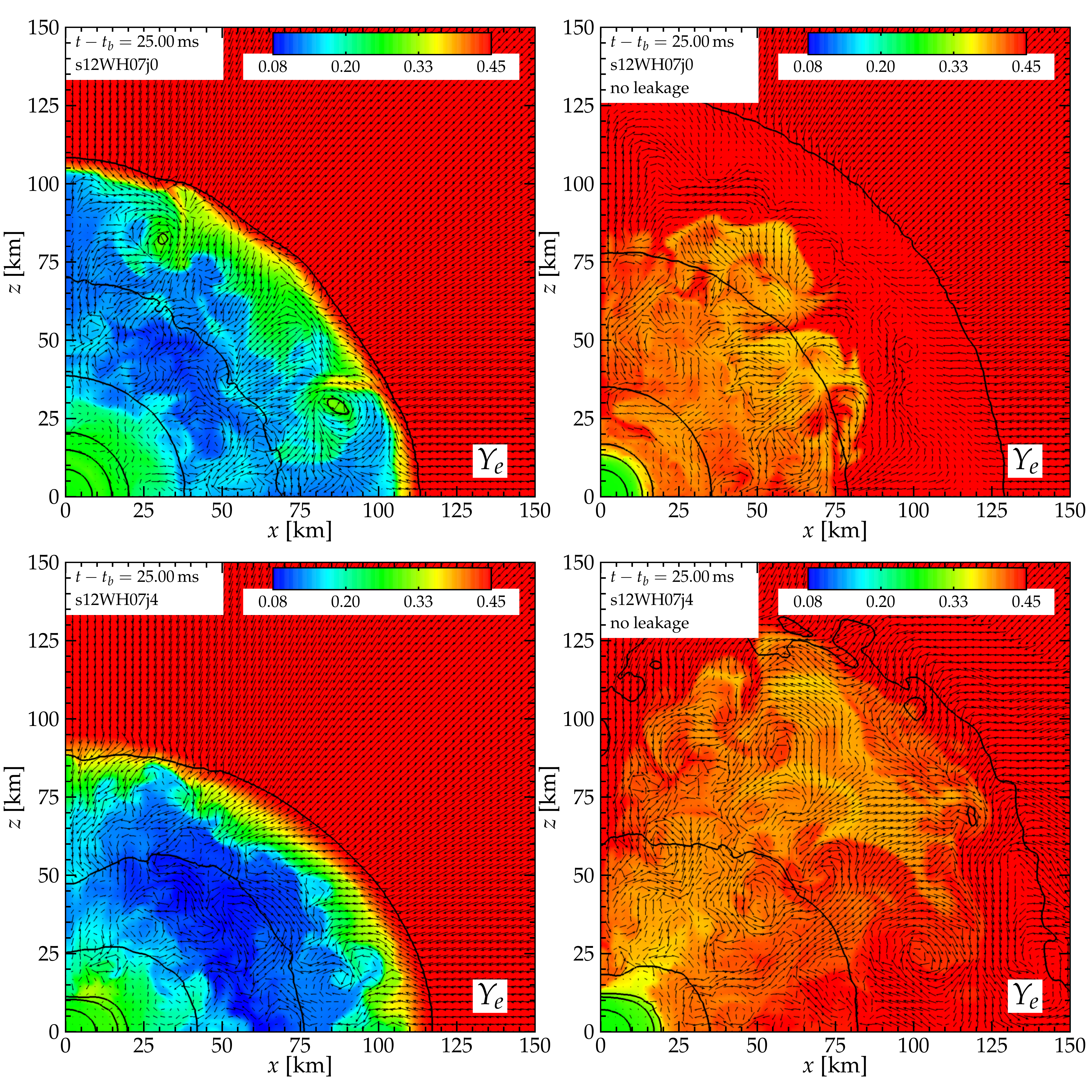}
 \caption{Colormaps of the electron fraction $Y_e$ in the meridional
   plane at $25\,\mathrm{ms}$ after bounce. The contour lines mark
   isocontours of rest-mass density (in this order from the origin) at
   $\{10^{14},10^{13},5\times
   10^{12},10^{12},10^{11},10^{10}\}\,\mathrm{g}\,\mathrm{cm}^{-3}$.
   Velocity vectors are superposed and their length is saturated at
   $0.0125\,c$ to prevent the high collapse velocities
     of the outer core outside of the shock from cluttering the plot. 
   The top panels show the nonrotating model s12WH07j0
   evolved with postbounce neutrino leakage (left panel) and without
   (right panel).  The bottom panels show the rapidly spinning model
   s12WH07j4 also with (left panel) and without neutrino leakage
   (right panel).  The simulations with neutrino leakage show the
   characteristic depletion of $Y_e$ due to electron capture in the
   region behind the shock. The shock is located at radii of
   $\sim$110-120\,km in the simulations with neutrino leakage, at
   $\sim$125\,km in the j0 model without leakage, and at $\sim$160\,km
   ($\sim$220\,km) at the pole (equator) of the j4 model without
   leakage. In the PNS core, neutrinos are trapped and $Y_e$ is at
   essentially its bounce value and only there the simulations with
   and without leakage yield the same $Y_e$ and structure. The
   deleptonization and neutrino cooling in the runs with leakage
   significantly reduce the pressure behind the shock, lead to much
   smaller shock radii, and allow more material to pile up on the
   outer PNS core.}
  \label{fig:colormapleak}
\end{figure*}

All simulated models in this study are run twice: once with postbounce
neutrino leakage and once without. In the collapse phase, the leakage
and no-leakage (suffix ``nl'' in the model identifier) simulations are
identical. After bounce, significant qualitative and quantitative
differences develop, which we discuss in the following to gain an
understanding of their relevance for early postbounce dynamics and the
associated GW signals. This will also allow us to assess the error
introduced by neglecting neutrino emission and deleptonization in the
extensive previous parameter study of Dimmelmeier~\emph{et
  al.}~\cite{dimmelmeier:08}. We focus our discussion on the
s12WH07j\{0-5\} model set. The effects of neutrino leakage are
identical in models s40WH07j\{0-5\}.

The hydrodynamic shock wave, launched at core bounce, moves out in
radius (and enclosed-mass coordinate) and dissociates infalling
iron-group nuclei of the outer core into neutrons and protons. The
degenerate electrons in the region behind the shock rapidly capture
onto the liberated protons, dramatically decreasing $Y_e$ and leading
to the characteristic neutronization $\nu_e$ burst when the shock
breaks out of the $\nu_e$ neutrinosphere. Nuclear dissociation and
neutrino losses reduce the pressure behind the shock and decelerate
its initially rapid expansion, eventually causing the shock to stall
and turn into an accretion shock within tens of milliseconds of
bounce. When neutrino leakage is not included, the shock still does
work to break up heavy nuclei, but neutronization does not occur, the
pressure behind the shock stays higher and the shock will stall at
much larger radii. This is what we observe in our simulations. In
Fig.~\ref{fig:s12shockrad}, we present the time evolution of the shock
position along the positive $x$-axis for all models of the
s12WH07j\{0-5\} set. In models with neutrino leakage, the shock
reaches $90-130\,\,\mathrm{km}$ at $25\,\mathrm{ms}$ after bounce,
while reaching $130$-$220\,\mathrm{km}$ in models run without neutrino
leakage. We also note from Fig.~\ref{fig:s12shockrad} that rapidly
rotating models tend to have larger equatorial shock radii (i.e.,
along the $x$-axis in the figure) than their slowly and nonrotating
counterparts. This is consistent with what was found by
\cite{ott:06spin,ott:08} and simply due to the rotational flattening
of rapidly spinning cores. The polar shock radii of rapidly rotating
models (not shown in Fig.~\ref{fig:s12shockrad}) are correspondingly
lower.

The top panels of Fig.~\ref{fig:colormapleak} show colormaps of the
$Y_e$ distribution in the meridional plane at $25\,\mathrm{ms}$ after
core bounce in the nonrotating model s12WH07j0 with neutrino leakage
(left panel) and without (right panel).  Fluid velocity vectors and
isocontours of rest-mass density are superposed. In the simulation
with neutrino leakage, the characteristic $Y_e$ trough behind the
shock and above the PNS core has clearly developed. In the PNS, at
densities of $\gtrsim 5\times 10^{12}\,\mathrm{g}\,\mathrm{cm}^{-3}$,
neutrinos are trapped and $Y_e$ is essentially at its bounce value. It
is only in this high-density region ($r \lesssim 20\,\mathrm{km}$)
where the simulations with and without leakage agree in their $Y_e$
distribution and structure. In the simulation with leakage, material
is able to deleptonize, cool, and sink onto the PNS core, which makes
it more extended, as shown by the $5\times
10^{12}\,\mathrm{g}\,\mathrm{cm}^{-3}$ and
$10^{12}\,\mathrm{g}\,\mathrm{cm}^{-3}$ density isocontours that are
at $3-5\,\mathrm{km}$ greater radii than in the simulation without
leakage. At the same time, however, the overall region behind the
shock is much more compact in the simulation with neutrino leakage:
The $10^{11}\,\mathrm{g}\,\mathrm{cm}^{-3}$ isocontour and the shock,
marked by the $10^{10}\,\mathrm{g}\,\mathrm{cm}^{-3}$ isocontour, are
at $\sim$$10-15\,\mathrm{km}$ smaller radii.

\begin{table*}[t]
\begin{center}
  \caption[Mismatch of GW signal]{Mismatch of GW signals of
    simulations with and without postbounce neutrino leakage computed
    according to Eq.~\ref{eq:mismatch} for an event at
    $10\,\mathrm{kpc}$. The mismatch is provided for the signal up to
    $5\,\mathrm{ms}$, $10\,\mathrm{ms}$, and $25\,\mathrm{ms}$ after
    bounce.}
\begin{tabular}{l@{~~~}c@{~~~}c@{~~~}c}
 \hline \hline
Model       &          Mismatch $<5$ ms &Mismatch $<10$ ms    &Mismatch $<25$ ms \\
\hline
s12WH07j0nl/s12WH07j0 & $\scis{3.7}{-1}$ & $\scis{3.5}{-1}$ & $\scis{5.1}{-1}$\\
s12WH07j1nl/s12WH07j1 & $\scis{1.3}{-3}$ & $\scis{5.2}{-1}$ & $\scis{6.6}{-1}$\\ 
s12WH07j2nl/s12WH07j2 & $\scis{1.6}{-3}$ & $\scis{2.1}{-1}$ & $\scis{5.3}{-1}$\\
s12WH07j3nl/s12WH07j3 & $\scis{4.6}{-4}$ & $\scis{5.8}{-2}$ & $\scis{1.3}{-1}$\\
s12WH07j4nl/s12WH07j4 & $\scis{1.3}{-4}$ & $\scis{1.4}{-2}$ & $\scis{5.9}{-2}$\\
s12WH07j5nl/s12WH07j5 & $\scis{1.0}{-3}$ & $\scis{2.4}{-3}$ & $\scis{2.8}{-2}$\\ 
\hline 
s40WH07j0nl/s40WH07j0 & $\scis{3.8}{-1}$ & $\scis{1.4}{-1}$ & $\scis{3.9}{-1}$\\
s40WH07j1nl/s40WH07j1 & $\scis{3.6}{-4}$ & $\scis{4.6}{-2}$ & $\scis{5.8}{-1}$\\
s40WH07j2nl/s40WH07j2 & $\scis{3.2}{-4}$ & $\scis{4.2}{-2}$ & $\scis{4.7}{-1}$\\
s40WH07j3nl/s40WH07j3 & $\scis{9.9}{-5}$ & $\scis{6.3}{-3}$ & $\scis{5.3}{-2}$\\
s40WH07j4nl/s40WH07j4 & $\scis{1.6}{-4}$ & $\scis{4.0}{-3}$ & $\scis{5.9}{-2}$\\ 
s40WH07j5nl/s40WH07j5 & $\scis{2.3}{-3}$ & $\scis{6.0}{-3}$ & $\scis{2.4}{-2}$\\
\hline\hline
\end{tabular}
\label{table:mismatch}
\end{center}
\end{table*}

The lower two panels of Fig.~\ref{fig:colormapleak} show colormaps of
the $Y_e$ distribution at $25\,\mathrm{ms}$ after bounce in the
meridional plane of the rapidly spinning model s12WH07j4 evolved with
(left panel) and without (right panel) neutrino leakage.  Comparing
leakage and no-leakage results, we find the same general trends
discussed for the nonrotating model in the above, but the differences
are increased by rotation. The shock in the simulation without
leakage has left the $150 \times 150\,\mathrm{km}$ region of the
frame, while it sits tight at small radii ($\sim$$115\,\mathrm{km}$ at
the equator and $\sim$$90\,\mathrm{km}$ at the pole) in the simulation
with neutrino leakage. As in the nonrotating model, neutrino leakage
allows material to cool, deleptonize, and sink to smaller radii,
leading to an extended high-density region at the edge of the
PNS core also in model s12WH07j4.

The top and center sub-panels of each model panel in 
  Fig.~\ref{fig:s12_rho_h_nu} depict the postbounce central density 
evolution and GW signal in each model, respectively. The results
obtained with neutrino leakage are shown in solid lines while the
no-leakage results are plotted as dashed curves. Independent of
rotation rate, the central density of models run with neutrino leakage
exhibits a secular upward trend due to the contraction of the PNS core
as more deleptonized material settles onto it. In the simulations
without leakage, material behind the shock cannot cool and deleptonize
and does not settle onto the PNS, whose postbounce central density
stays practically constant.

In the GW sector, the differences between simulations with and without
leakage vary strongly with increasing rotation. In order to quantify
the differences in the GW signals extracted from simulations with and
without neutrino leakage, we compute their mismatch as defined by
Eq.~(\ref{eq:mismatch}). The mismatch is zero for identical waveforms
and one if the waveforms have nothing in common. The mismatch measure
adopted here and discussed in Section~\ref{sec:gws} weights the
Fourier-space representations of the waveforms (for a source at
10\,kpc) with aLIGO detector sensitivity and thus emphasizes
differences relevant to aLIGO observations. We note that variations in
the waveforms due to different wave extraction methods alone lead to
mismatches of order $10^{3}-10^{-2}$
\cite{reisswig:11ccwave}. However, if all waveforms are extracted with
the same technique, as is the case here, the mismatch should be a
meaningful quantity.

In the nonrotating model s12WH07j0, the GW signal is due exclusively
to prompt convection. Prompt convection and the associated GW signal
are strongly modified by deleptonization and neutrino cooling, which
add lepton gradients and reduce the physical extent of the convection
zone. This leads to a stronger, higher-frequency GW signal from prompt
convection in the model evolved with neutrino leakage.  We compute
(and record in Tab.~\ref{table:mismatch}) a mismatch of $0.51$ between
the GW signals from the leakage and no-leakage simulations.  While
this mismatch is large, we point out that the GWs from prompt
convection seen in the simulations presented here are just example
realizations of GWs from prompt convection and are not representative
for the entire range of possible GW signals from this process.
Variations in progenitor core structure, thermodynamics, and
precollapse perturbations that seed convection may lead to large
qualitative and quantitative variations in the GW signal from prompt
convection (see, e.g., the discussion in \cite{ott:09}) that could
easily dwarf the difference between the leakage and no-leakage
variants that we see in our models.

With increasing rotation, the overall GW signature gradually becomes
dominated by the dynamics of the centrifugally flattened inner core,
which turns into the core of a nascent protoneutron star after bounce.
Neutrinos are trapped in the inner core and diffuse out on much longer
timescales than simulated here. Hence, the parts of the GW signal due
to inner core dynamics should be nearly identical in leakage and
no-leakage simulations. This is what we find already in the slowly
spinning model s12WH07j1 (top right panel of
Fig.~\ref{fig:s12j4_all}), whose GW signal is dominated by the inner
core in the first $\sim$$5\,\mathrm{ms}$ after bounce. Evaluated at
this time, the mismatch of leakage and no-leakage GW signal is only of
order $10^{-3}$. Subsequently, prompt convection takes over, leading
to a mismatch of $0.66$ at the end of the simulation. With further
increasing rotation, the signal due to the inner core's bounce becomes
globally dominant and the total (i.e., for the entire waveform)
mismatch decreases to 0.53 (s12WH07j2), 0.13 (s12WH07j3), 0.06
(s12WH07j4), and 0.03 (s12WH07j5). Most of the mismatch is built up at
times $\gtrsim$$5\,\mathrm{ms}$ after bounce, when the shock has
broken out of all neutrinospheres and when the large GW signal due to
inner core dynamics has sufficiently died down for the aspherical
flow in the semi-transparent region behind the shock to play a
appreciable role in the GW emission.

When considering the mismatch in the first $5\,\mathrm{ms}$ after
bounce only, we note somewhat different systematics. The mismatch is
large in the nonrotating model, small for the slowly to moderately
rapidly spinning models, and increases again for the most rapidly
spinning model s12WH07j5 (the same systematics holds for the
s40WH07j\{0-5\} models; see Tab.~\ref{table:mismatch}). We attribute
this to the strong global rotational deformation of the most
rapidly spinning model. It leads to shock formation, propagation,
and strong neutrino emission from polar regions milliseconds
before the shock forms at the equator. This results in an earlier
onset of differences between leakage and no-leakage simulations.

\subsection{Influence of Progenitor Structure}
\label{sec:progcomp}

\begin{figure*}
 \includegraphics[width=0.45\linewidth]{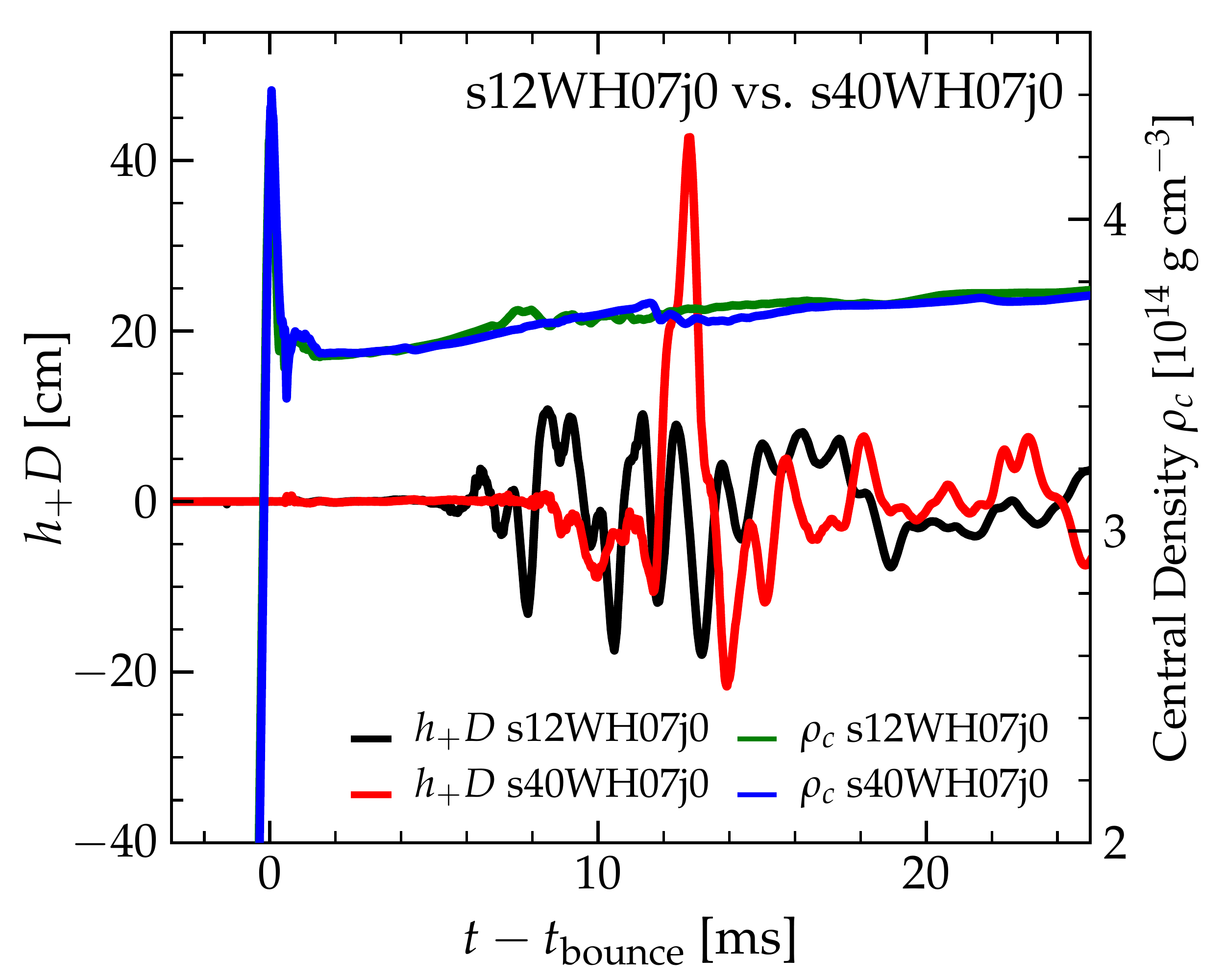}
 \includegraphics[width=0.45\linewidth]{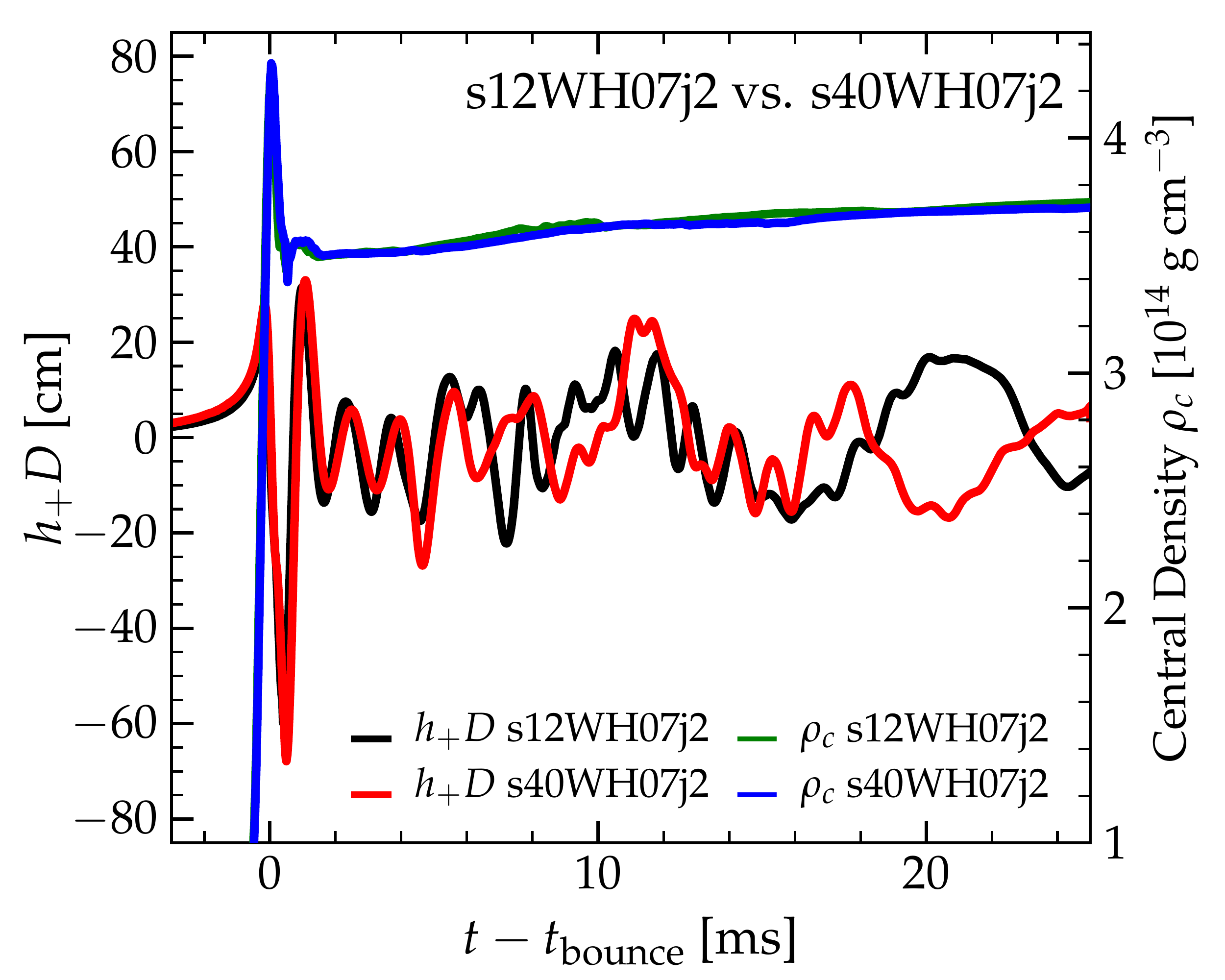}
 \includegraphics[width=0.45\linewidth]{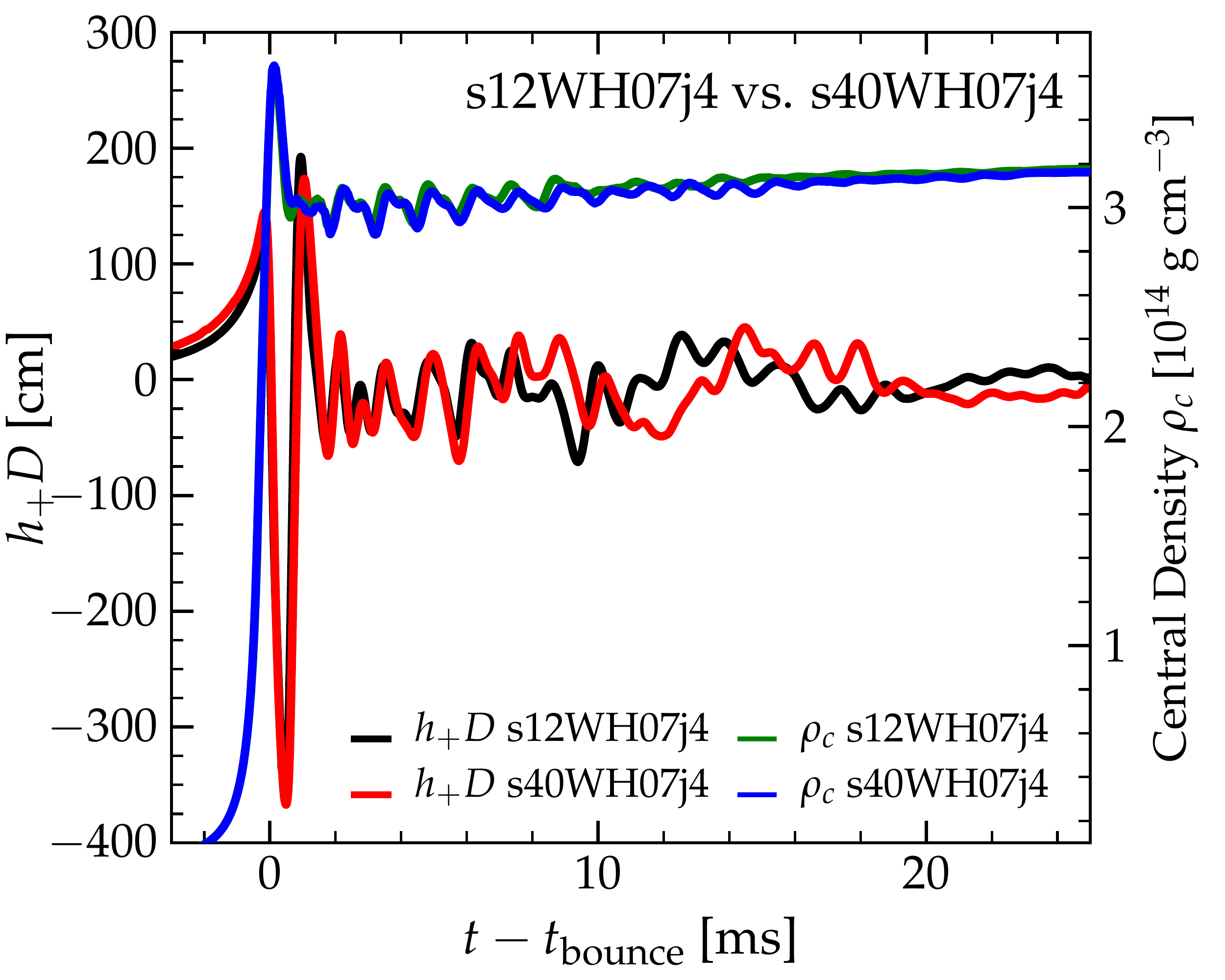}
 \includegraphics[width=0.45\linewidth]{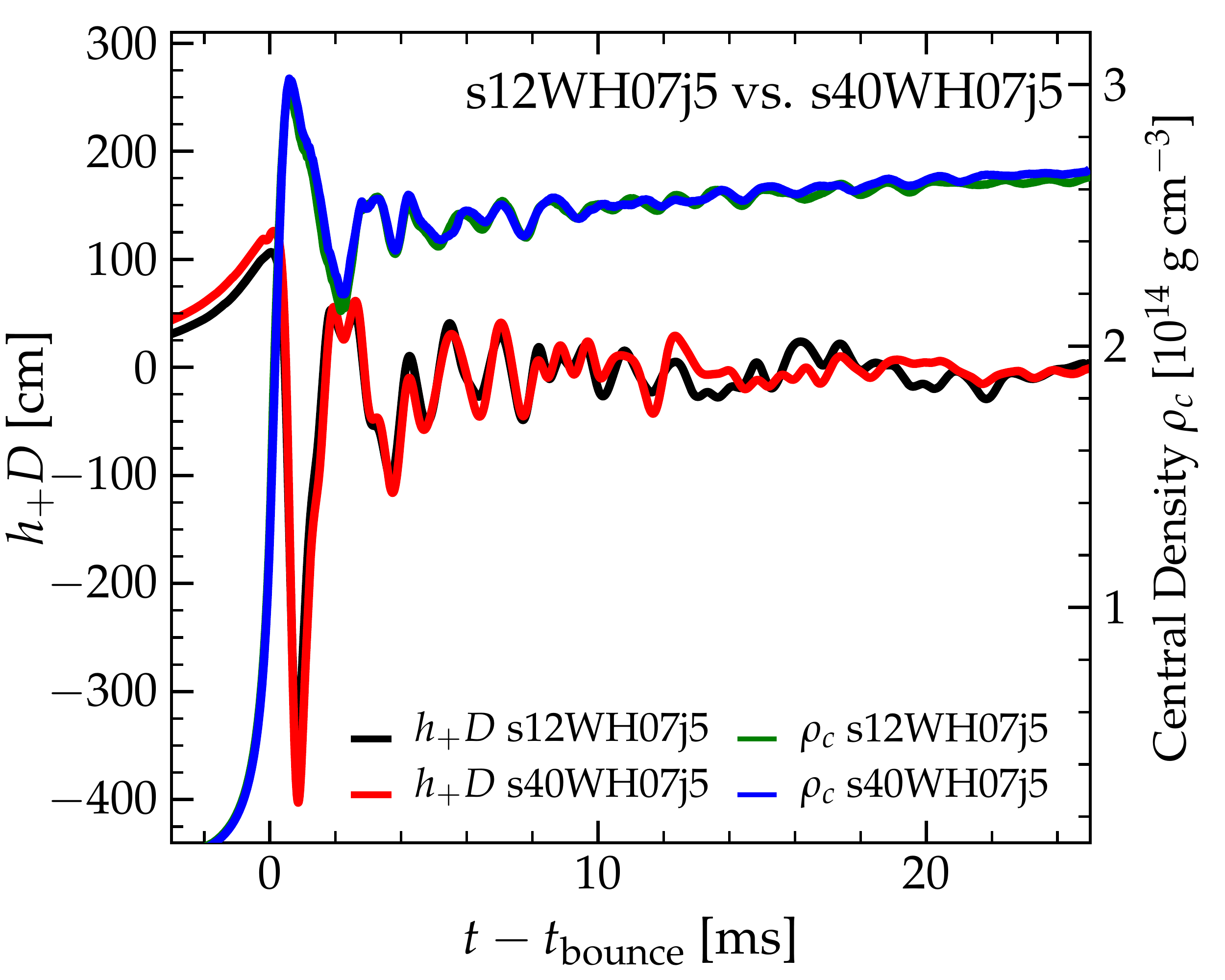}
  \vspace{-0.25cm}
  \caption{Comparison of select models of the s12WH07 model set (using
    the $12$-$M_\odot$ progenitor) with their s40WH07 counterparts
    (using the $40$-$M_\odot$ progenitor). Shown are the evolution of
    the GW signal seen by an equatorial observer (left ordinate) and
    of the central rest-mass density (right ordinate).  The GW signal
    $h_+$ is rescaled by source distance $D$ and plotted in units of
    centimeters, the central rest-mass density $\rho_c$ is given in
    units of $10^{14}\,\mathrm{g\,cm}^{-3}$, and the time is measured
    in milliseconds relative to the time of core bounce in each model.
    The central density evolutions in all s12WH07 and s40WH07 models
    are extremely close to one another. The GW signals of the
    nonrotating models (top left panel) differ significantly. However,
    with increasing rotation (from left to right, top to bottom), the
    dominant parts of the GW signal grow very similar.
    \label{fig:s12_s40_rho_gw_comp}}
\end{figure*}

\begin{table*}[t]
\begin{center}
  \caption{Mismatch of GW signals of simulations using the s12WH07 and
    the s40WH07 progenitor model computed according to
    Eq.~\ref{eq:mismatch} for an event at $10\,\mathrm{kpc}$. The
    mismatch is provided for the signals up to $5\,\mathrm{ms}$,
    $10\,\mathrm{ms}$, and $25\,\mathrm{ms}$ after bounce.}
\begin{tabular}{l@{~~~}c@{~~~}c@{~~~}c}
 \hline \hline
Model       &          Mismatch $<5$ ms &Mismatch $<10$ ms    &Mismatch $<25$ ms \\
\hline
s12WH07j0/s40WH07j0 & $\scis{4.1}{-1}$ & $\scis{3.2}{-1}$ & $\scis{5.8}{-1}$\\
s12WH07j1/s40WH07j1 & $\scis{1.9}{-2}$ & $\scis{5.6}{-1}$ & $\scis{6.9}{-1}$\\ 
s12WH07j2/s40WH07j2 & $\scis{3.1}{-2}$ & $\scis{2.4}{-1}$ & $\scis{6.5}{-1}$\\
s12WH07j3/s40WH07j3 & $\scis{1.1}{-2}$ & $\scis{8.8}{-2}$ & $\scis{1.6}{-1}$\\
s12WH07j4/s40WH07j4 & $\scis{1.9}{-2}$ & $\scis{7.6}{-2}$ & $\scis{2.0}{-1}$\\
s12WH07j5/s40WH07j5 & $\scis{1.2}{-2}$ & $\scis{1.7}{-2}$ & $\scis{5.2}{-2}$\\ 
\hline\hline
\end{tabular}
\label{table:mismatch_s12_s40}
\end{center}
\end{table*}

\begin{figure}
\centering
\includegraphics[width=\linewidth]{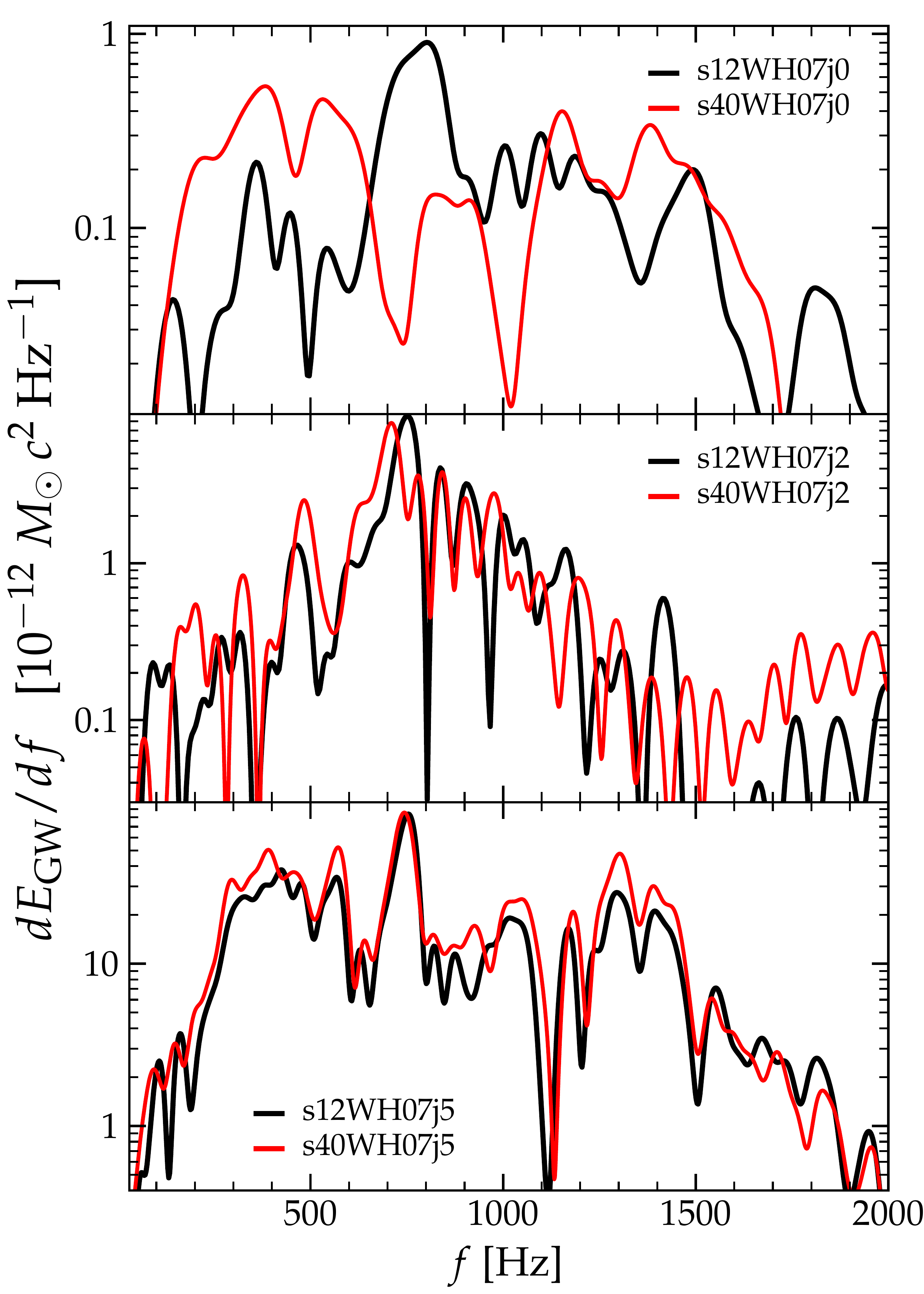}
\caption{Comparison of the spectral GW energy densities
  ($dE_\mathrm{GW}/df$; Eq.~\ref{eq:degwdf}) of select models using
  the $12$-$M_\odot$ and the $40$-$M_\odot$ progenitor at the same
  precollapse rotation rate. Top panel: Nonrotating case (j$0$); the
  spectra vary greatly. Center panel: Moderate rotation (j$2$);
  the spectra exhibit many similar features. Bottom panel:
  Rapid rotation (j$5$); the spectra are nearly identical.}
\label{fig:spect_s12_s40_comp}
\end{figure}

In the discussion up to now, we have relied exclusively on our
s12WH07j\{0-5\} model set that uses the $12$-$M_\odot$ (at ZAMS)
progenitor model of \cite{woosley:07}. We turn now to comparing these
simulations with the s40WH07j\{0-5\} model set that uses the
$40$-$M_\odot$ (at ZAMS) progenitor of the same study. As discussed in
Section~\ref{sec:rotsetup}, we have chosen a procedure for imposing rotation
that leads to approximately the same specific angular momentum as a function
of enclosed mass of inner core material for s12WH07 and s40WH07 models of the
same j$n$ parameter. In particular, our procedure ensures that at a
given rotational setup j$n$, the s40WH07j$n$ model will have
approximately the same or less, but never more angular
momentum than the s12WH07j$n$ model.

We first turn our attention to the nonrotating (j0) models.  Model
s12WH07j0 experiences bounce at $178.5\,\mathrm{ms}$ after the start
of the simulation, reaches a peak density of $4.3\times
10^{14}\,\mathrm{g}\,\mathrm{cm}^{-3}$, and has an inner core mass at
bounce ($M_\mathrm{ic,b}$) of $\sim$$0.56\,M_\odot$. The core of the
$40$-$M_\odot$ progenitor of model s40WH07j0 starts collapse at a
central density that is $\sim$$5.5$ times lower than s12WH07j0's and,
according to the free-fall estimate, we expect its collapse time to be
$\sim$$2.3$ times longer. Model s40WH07 indeed reaches core bounce at
$418.2\,\mathrm{ms}$, which is almost exactly what our estimate
predicts.  The central density at bounce is $4.4\times
10^{14}\,\mathrm{g}\,\mathrm{cm}^{-3}$. This is slightly higher than
the value of model s12WH07j0 and we attribute this to model
s40WH07j0's greater $M_\mathrm{ic,b}$ of $\sim$$0.62\,M_\odot$. The
difference in $M_\mathrm{ic,b}$ is due to s40WH07j0's higher
initial core entropy (see Section~\ref{sec:prog} and \cite{burrows:83}),
since both models are using the same $Y_e(\rho)$ prescription for
deleptonization in the collapse phase. In a full neutrino
radiation-hydrodynamics simulation, one would expect additional
differences due to progenitor-dependent variations in
deleptonization~\cite{dimmelmeier:08}.

In the top left panel of Fig.~\ref{fig:s12_s40_rho_gw_comp} we compare
the late collapse and early postbounce evolutions of the central
density and of the GW signal in model s12WH07j0 an s40WH07j0. Modulo
minute deviations and despite not insignificant differences in
precollapse structure and thermodynamics, the central density
evolutions of the two models practically lie on top of each other.
The situation is different for the GW signals. In both models, GW
emission is due to prompt convection, which is highly sensitive to
bounce dynamics, progenitor thermodynamics, and seed perturbations and
is generally of stochastic character (e.g.,
\cite{ott:09,marek:09b,kotake:09}).  Hence, given the differences in
inner core mass and core thermodynamics, the large difference that we
find in the GW signals from prompt convection in these models is not
surprising. Using our mismatch measure (Eq.~\ref{eq:mismatch}) we find
a mismatch of 0.58 for the GW signals of models s12WH07j0 and
s40WH07j0 in the first $25\,\mathrm{ms}$ after bounce. More data on
their mismatch are summarized in Tab.~\ref{table:mismatch_s12_s40}
while key quantitative results are provided in
Tab.~\ref{table:results}.

With increasing rotation, the dynamics of the centrifugally-deformed
inner core begins to dominate the GW signal and a prominent
characteristic peak develops whose shape and magnitude depends to very
good approximation only on mass and angular momentum of the inner
core \cite{dimmelmeier:08,abdikamalov:10}. This simplifies the
signal and drives the GW signals of the two model sets to congruence.

The top right panel and the two bottom panels of
Fig.~\ref{fig:s12_s40_rho_gw_comp} compare s12WH07 and s40WH07 models
with j$2$, $j4$, and $j5$ rotational setup. Models with $j1$ and $j3$
are omitted for clarity, but show no special behavior.  The close
qualitative and quantitative agreement in the central density
evolutions found for the nonrotating case is unchanged by rotation. On
the other hand, the effect of rotation on the differences between the
GW signals is remarkable, in particular when one focuses on the signal
up to $5\,\mathrm{ms}$ after bounce.  Taking the signals up to this
point and computing the mismatch, we find $0.41$ for the nonrotating
models, and $0.019$, $0.031$, $0.011$, $0.019$, and $0.012$ for j$1$,
j$2$, j$3$, j$4$, and j$5$, respectively. Thus even the slightest bit
of rotation already has a unifying effect and drives down the mismatch
to the percent level about which it hovers, affected primarily by
slight, but not negligible differences in the bounce peak and in the
first few postbounce oscillations. When considering the entire GW signal
out to the end of our simulations at $25\,\mathrm{ms}$ after bounce,
the mismatch is significantly larger (see
Tab.~\ref{table:mismatch_s12_s40}), since the GW emission begins to be
significantly affected by prompt convection outside the PNS core
starting at $5-10\,\mathrm{ms}$ after bounce.

It is interesting to compare some of the key representative quantities
summarized in Tab.~\ref{table:results} between the s12WH07j\{0-5\} and
s40WH07j\{0-5\} models. Rotation leads to an increase of the inner
core mass at bounce ($M_\mathrm{ic,b}$; e.g.,
\cite{dimmelmeier:08}). In the s12WH07 model set, $M_\mathrm{ic,b}$
increases from $\sim$$0.56\,M_\odot$ for j$0$ to $\sim$$0.66\,M_\odot$
for j$5$. The s40WH07 progenitor model's higher-entropy inner core is
more massive, its mass is $0.62\,M_\odot$ in j$0$ and increases to
$0.72\,M_\odot$ in j$5$. The s40WH07j\{1-5\} models have less angular
momentum as a function of enclosed mass (see
Tab.~\ref{table:initial}), but their inner cores, due to their greater
mass, contain up to $\sim$$15\%$ more angular momentum at bounce
($J_\mathrm{ic,b}$) than the inner cores of the s12WH07j\{1-5\}
models.  These differences in $M_\mathrm{ic,b}$ and $J_\mathrm{ic,b}$
lead to quantitative changes at the same or lower level in the maximum
density at bounce, the maximum GW signal amplitude, the peak of the GW
energy spectrum, and the values of $T/|W|$ obtained at bounce: s40WH07
models (\emph{i}) undergo bounce at $\lesssim$$3\%$ greater central
densities, (\emph{ii}) reach maximum GW signal amplitudes
$|h_+|_\mathrm{max} D$ that are $\lesssim$$15\%$ greater,
(\emph{iii}) have peaks of their $dE_\mathrm{GW}/df$ that are at
$\lesssim$$5\%$ smaller frequencies, and (\emph{iv}) assume maximum
values of $T/|W|$ that are $\lesssim$$10\%$ smaller than in their
s12WH07 counterparts.  These results are readily explained by
considering that the inner cores of the s40WH07 models are slightly
less compact (very similar $\rho_c$ at greater $M_\mathrm{ic,b}$ and
$J_\mathrm{ic,b}$) than their s12WH07 counterparts. The lower
compactness leads to a lower value of $T/|W|$ (at fixed $J$, $T
\propto R^{-2}$ and $|W| \propto R^{-1}$), but to a greater quadrupole
moment, varying at a just slightly slower rate, resulting in a
slightly higher $|h_+|_\mathrm{max} D$.

In Fig.~\ref{fig:spect_s12_s40_comp} we plot $dE_\mathrm{GW}/df$
spectra comparing results from s12WH07 and s40WH07 models with no
rotation (j$0$; top panel), moderate rotation (j$2$; center panel),
and rapid rotation (j$5$; bottom panel). As expected from the large
differences in the time domain seen in the top left panel of
Fig.~\ref{fig:s12_s40_rho_gw_comp}, the spectral GW energy densities
of models s12WH07j0 and s40WH07j0 are quite distinct. Slow rotation
already leads to many common features and the spectra of rapidly
spinning models are nearly identical.

\begin{figure}
\centering
 \includegraphics[width=\linewidth]{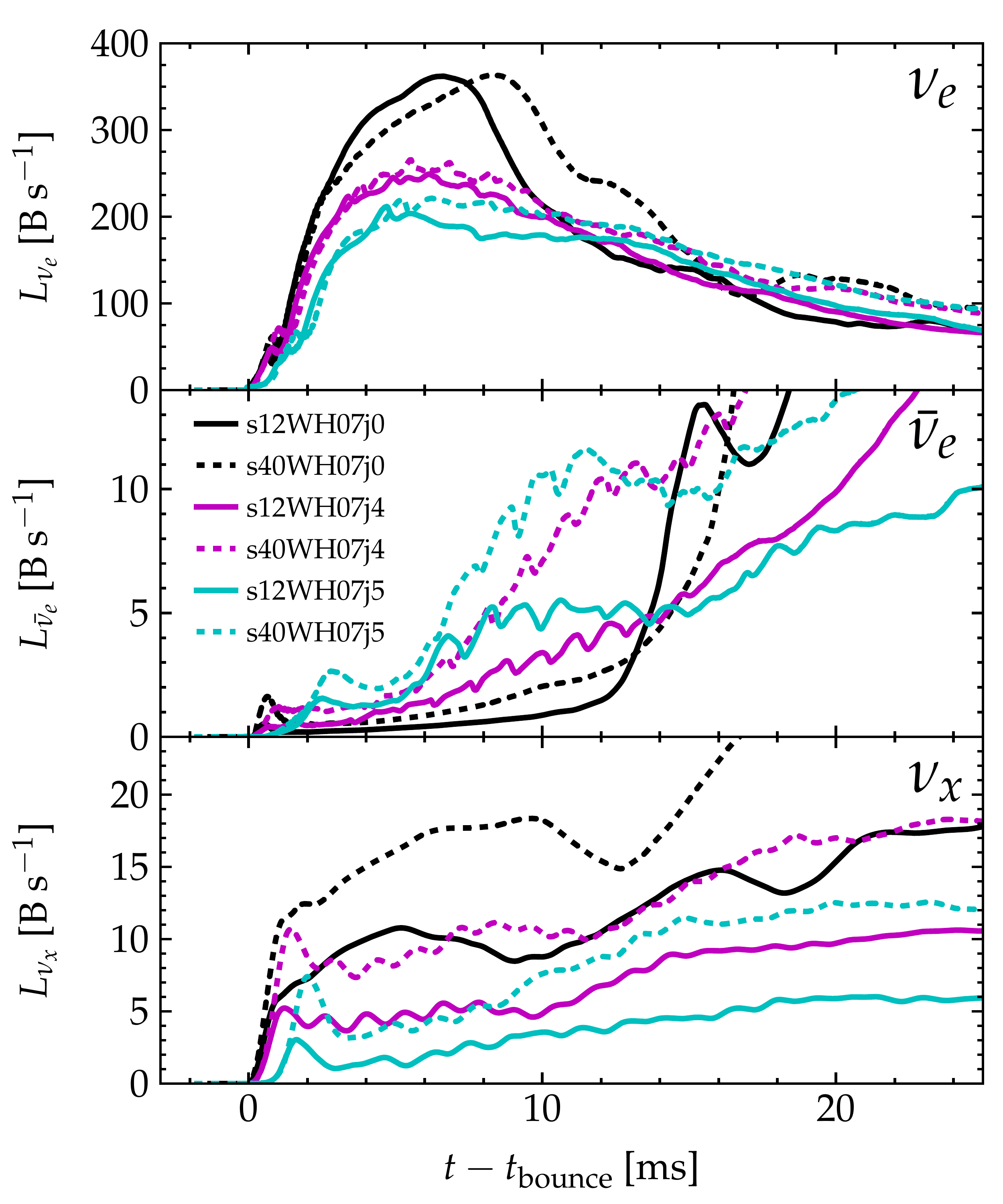}
  \caption{Comparison of the approximate $\nu_e$ (top panel),
    $\bar{\nu}_e$ (center panel), and $\nu_x$ (bottom) luminosities in
    models s12WH07j0, s12WH07j4, s12WH07j5 (all solid lines) with
    their s40WH07j0, s40WH07j4, s40WH07j5 counterparts (all dashed
    lines). The vertical scales have been chosen to emphasize the oscillations
    observed primarily in $L_{\bar{\nu}_e}$ and $L_{\nu_x}$ of the
      rapidly spinning j4 and j5 models. While quantitative aspects should be
      regarded as approximate, one notes that the s40WH07 models
      have systematically larger early postbounce luminosities than their
      s12WH07 counterparts.
    \label{fig:s12_s40_comp_nu}}
\end{figure}

The correlations between dynamics, gravitational waveforms, and
neutrino signals described in Section~\ref{sec:results1} for the
s12WH07j\{0-5\} model set also exist in very similar fashion in the
s40WH07j\{0-5\} models. For the central density evolution (a proxy for
the core dynamics) and GW signal, this is already apparent from the
two rapidly rotating cases shown in the bottom panels of
Fig.~\ref{fig:s12_s40_rho_gw_comp} in which the s40WH07 models
essentially mirror the evolution of their s12WH07 counterparts.
Fig.~\ref{fig:s12_s40_comp_nu} compares the approximate $\nu_e$ (top
panel), $\bar{\nu}_e$, and $\nu_x$ luminosity evolutions of j$0$
(black curves), j$4$ (magenta curves), and j$5$ (cyan curves) variants
of s12WH07 (solid lines) and s40WH07 (dashed lines). The
characteristic oscillations that go along with the quadrupole PNS
pulsations seen in models s12WH07j4/j5 also occur in their s40WH07j4/5
counterparts. While our leakage scheme permits us to make only
approximate quantitative predictions, we do find interesting
systematic trends when comparing the neutrino luminosities of s12WH07
and s40WH07 models. The rise times of the $\nu_e$ luminosity are
similar in both progenitors, but the peaks are broader in the s40WH07
models, most likely because bounce is stronger, leading to somewhat
larger shock radii and more dissociated material for electrons to
capture on at early times. The $\bar{\nu}_e$ emission is primarily
thermal at early times and the $\nu_x$ emission is completely thermal,
hence, for these neutrino species, the hotter cores of s40WH07 models
exhibit systematically higher luminosities than their cooler s12WH07
counterparts. We expect these more qualitative than quantitative
trends to carry over to future full energy-dependent neutrino
radiation-hydrodynamics simulations, which will be needed for reliable
quantitative estimates.

\begin{figure}
\centering
\includegraphics[width=\linewidth]{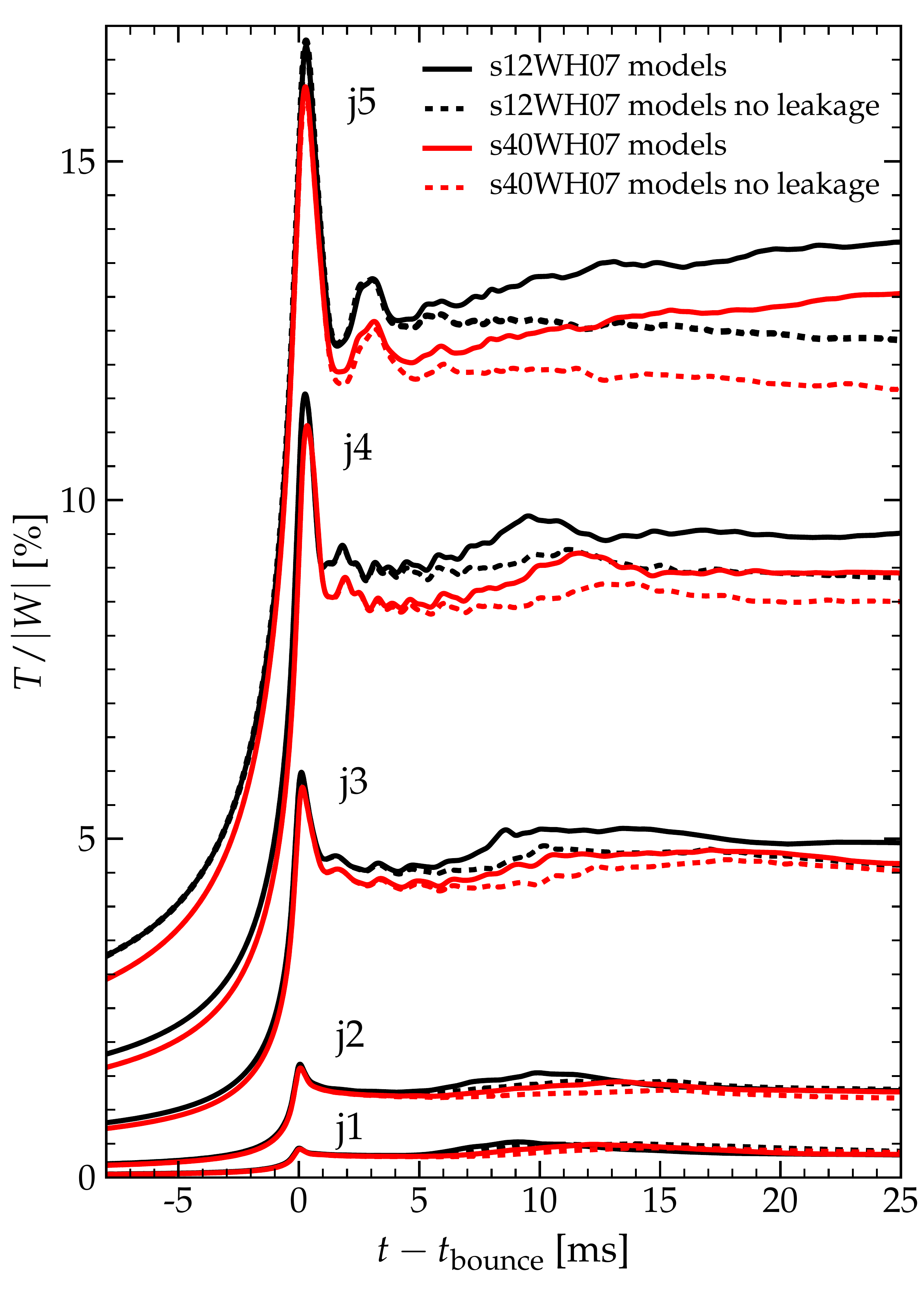}
  \caption{Rotation rate $T/|W|$ as a function of time after bounce
    for the rotating models of the s12WH07 (black lines) and s40WH07
    (red lines) model sets. Results for simulations with (solid lines)
    and without (dashed lines) neutrino leakage are shown.  s12WH07
    and s40WH07 models show qualitatively and quantitatively very
    similar $T/|W|$ evolutions.  s40WH07 models, which are endowed
    with slightly less angular momentum than their s12WH07
    counterparts, have slightly lower $T/|W|$ (see
    Tabs.~\ref{table:initial} and \ref{table:results}).  In rapidly
    rotating models, neutrino leakage leads to a more rapid
    contraction and spin-up after bounce than predicted by models
    without leakage. Nonaxisymmetric instability at low-$T/|W|$ is
    likely to occur in models with j3, j4, and j5 rotation. The j5
    models may even reach the threshold of $27\%$ for the high-$T/|W|$
    dynamical instability within a few hundred ms of bounce.
    \label{fig:s12_s40_t_over_w}}
\end{figure}

\subsection{Evolution of the Rotation Rate and Prospects for Rotational
Instability}
\label{sec:res2_rotrate}

Rotating core collapse proceeds axisymmetrically even in very rapidly
spinning cores (see, e.g.,
\cite{ott:07prl,ott:07cqg,scheidegger:10b}). After bounce,
nonaxisymmetric dynamics will naturally develop as a consequence of
convection in the region behind the shock and above the PNS
core. The latter, however, may become rotationally
  unstable to the growth of nonaxisymmetric modes that could lead to
  global deformation and extended quasi-periodic
  elliptically-polarized GW emission
  \cite{ott:09,ott:07prl,scheidegger:10b,shibata:05} that could
  surpass the linearly-polarized emission from core bounce in total
  emitted energy.  Such nonaxisymmetric rotational instability has
  been identified to come in three flavors:

(\emph{i}) At $T/|W| \gtrsim 27\%$, a dynamical purely
hydrodynamic instability will deform a spheroidal self-gravitating
fluid body into ellipsoidal ($m=2$, bar-like) shape (e.g.,
\cite{chandrasekhar69c}). 

(\emph{ii}) At $T/|W| \gtrsim
14\%$, the same kind of deformation may be driven by the presence of a
dissipative process (e.g., viscosity or GW back-reaction
\cite{friedman:78,chandrasekhar:70}).  The growth timescale for this
secular instability is likely to be of order seconds or longer
\cite{lai:01,gaertig:11a}. 

(\emph{iii}) A dynamical shear instability may operate on the free
energy stored in differential rotation, leading to nonaxisymmetric
spiral modes that redistribute angular momentum.  This kind of
instability is common in accretion disks (e.g., \cite{korobkin:11} and
references therein) and has been shown to occur also in differentially
rotating neutron star models already at values of $T/|W|$ of order
$\sim$1\%
\cite{centrella:01,shibata:03,watts:05,ou:06,corvino:10}. Hence, this
instability is commonly referred to as low-$T/|W|$ instability.  While
the inner PNS core is likely to be uniformly spinning, differential
rotation naturally develops in the outer core (e.g.,
\cite{dessart:12a,ott:06spin}) and full 3D simulations
\cite{ott:07prl, scheidegger:08,scheidegger:10b} have found the
instability to occur in PNSs with a $T/|W|$ in the range of
$\sim$5-13\%. The detailed conditions necessary for its
  development and its potential interaction with other processes,
  e.g., the magnetorotational instability \cite{balbus:91}, operating
  on differential rotation remain to be fully explored (but see
  \cite{fu:11}).

The simulations presented in this paper are carried out in an octant
of the 3D cube, thus do not allow the development of the typically
unstable $m = \{1,2,3\}$ nonaxisymmetric modes. While we cannot track
this low-order nonaxisymmetric dynamics, it is still interesting to
consider the rotational configuration of our postbounce models to
assess the potential for development of nonaxisymmetric rotation
dynamics in the postbounce epoch. Previous estimates have been based
on 1.5D (spherical symmetry plus angle-averaged rotation) GR
simulations \cite{oconnor:11}, 2D Newtonian simulations with
multi-group flux-limited diffusion neutrino transport
\cite{ott:06spin}, or on 2D GR simulations that did not include
deleptonization and neutrino cooling after bounce
\cite{dimmelmeier:08,abdikamalov:10}, so it is worthwhile to
reconsider the question at hand.

In Fig.~\ref{fig:s12_s40_t_over_w}, we show the evolution of $T/|W|$
from shortly before bounce to $25\,\mathrm{ms}$ after bounce for all
rotating models and for simulations with and without neutrino
leakage. As expected from the discussion in Section~\ref{sec:progcomp},
the only difference between models using the s12WH07 and the s40WH07
progenitor is that the latter models reach slightly lower peak and
postbounce $T/|W|$.

All models reach their peak $T/|W|$ at bounce and settle at a
$\sim$10-30\% lower early postbounce $T/|W|$ (see
Tab.~\ref{table:results} for quantitative details). The $T/|W|$
evolutions in initially slowly to moderately rapidly rotating models
(j1$-$j3) show little dependence on progenitor or neutrino leakage
and $T/|W|$ remains almost constant in their postbounce evolutions.
This is different in the rapidly spinning j4 and j5 models.
In simulations with neutrino leakage, accreting, high-angular momentum
material can efficiently cool, deleptonize, settle, and add spin
to the PNS. Hence, one observes a significant postbounce increase of $T/|W|$
from its value immediately after bounce to the end of the simulation.
This is not the case in simulations without leakage and $T/|W|$ may even
slightly decrease after bounce, since only low-angular momentum from
polar regions quickly reaches small radii.

The models with precollapse rotation choices j3, j4, and j5 have
early postbounce $T/|W|$ of $\sim$4.5\%, $\sim$9\%, and
$\sim$12.5\%. Consistent with previous results
\cite{ott:06spin,dimmelmeier:08,dessart:12a}, their PNS cores are
approximately uniformly spinning, but the angular velocity begins to
fall off roughly proportional to $r^{-\alpha}$ with $\alpha \sim$$1.3-1.7$
at radii $\gtrsim 10-20\,\mathrm{km}$. Hence, these models are likely
to be subject to the low-$T/|W|$ instability \cite{ott:07prl,scheidegger:08,
scheidegger:10b,dimmelmeier:08}.

When linearly extrapolating the postbounce $T/|W|$ growth in the j5
models under the simplifying assumption that the angular momentum of
the accreting material is approximately constant in time, we find that
a $T/|W|$ of $27\%$, the approximate threshold for the guaranteed
dynamical bar-mode instability, is reached at $\sim$$300\,\mathrm{ms}$
after bounce. Even if accretion stops, cooling and contraction of the
PNS to final NS form will likely lead to $T/|W|$ in excess of the
dynamical instability threshold in the j5, j4, and even in the
j3 models (see, e.g., the mapping of initial core spin to final NS
spin in \cite{ott:06spin}), unless angular momentum is being
redistributed or radiated by some other mechanism, e.g., the
low-$T/|W|$ instability, the secular instability, or MHD processes.

\subsection{Notes on Detectability}
\label{sec:detect}

\subsubsection{Gravitational Waves}
In the rightmost five columns of Tab~\ref{table:results}, we summarize
key quantities describing the GW emission characteristics of the
simulated models: the peak of the GW signal amplitude time series
($|h_{+}|_\mathrm{max} D$) as seen by an equatorial observer rescaled
by distance $D$, the emitted energy in GWs ($E_\mathrm{GW}$), the peak
value of the dimensionless characteristic strain
($h_\mathrm{char,max}(f)$; Eq.~\ref{eq:hchar}) in frequency space and
at an equatorial observer location of $10\,\mathrm{kpc}$, the
frequency $f_\mathrm{char,max}$ at which $h_\mathrm{char,max}$ is
located, and the single-detector Advanced LIGO optimal signal-to-noise
ratio (SNR) as calculated using Eq.~\ref{eq:snr} for a core collapse
event at $10\,\mathrm{kpc}$, the fiducial galactic distance scale.
Since we are considering linearly polarized signals in $h_+$, for
non-optimal orientation the SNR will scale $\propto \sin^2 \theta$,
where $\theta$ is the angle between observer line of sight an the
rotation axis of the collapsing star.

In the following, we focus exclusively on the physically more
realistic models that include neutrino leakage. Furthermore, as
discussed in Section~\ref{sec:progcomp}, the $12$-$M_\odot$ and the
$40$-$M_\odot$ progenitors lead to very similar GW emission in the
phases that we simulate and we do not discuss them separately.

\begin{figure}
\centering
\includegraphics[width=\linewidth]{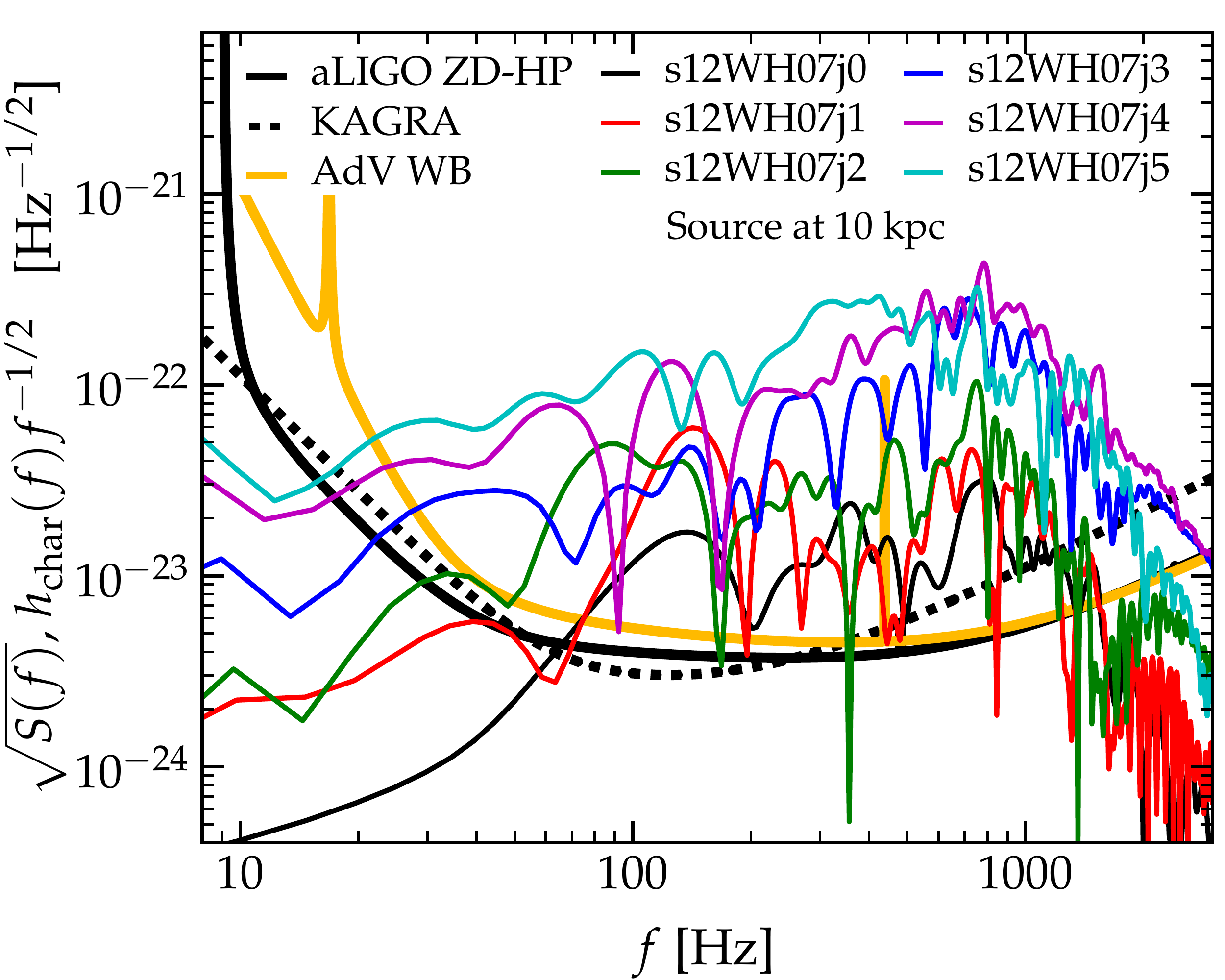}
\caption{Comparison of projected Advanced LIGO broadband (aLIGO ZD-HP --
  zero-detuning, high-power)
  \cite{LIGO-sens-2010}, KAGRA/LCGT
  \cite{lcgt:10}, and potential Advanced Virgo wide-band (AdV WB)
  \cite{AdV-sens-2010} sensitivity with the characteristic GW amplitudes
  $h_\mathrm{char}(f)f^{-1/2}$ of the s12WH07j\{0-5\} model set at a source
  location of $10\,\mathrm{kpc}$.
}
\label{fig:hchar_spect}
\end{figure}

The peak GW signal amplitudes of our models lie in the range $20
\,\mathrm{cm} \lesssim |h_{+}|_\mathrm{max} D \lesssim
400\,\mathrm{cm}$, which corresponds to $7\times 10^{-22} \lesssim
|h_{+}|_\mathrm{max} \lesssim 1.3 \times 10^{-20}$
at $10\,\mathrm{kpc}$ and is fully consistent with the results of
\cite{dimmelmeier:08}, who also focused on the linearly polarized GW
signal from core bounce and early postbounce evolution, but did not
include postbounce neutrino leakage. The lowest peak amplitudes are
reached in nonrotating (j0) or slowly rotating (j1) models, in which the
emission is primarily due to prompt convection. The highest amplitudes
are emitted by the most rapidly spinning models (j4 and j5). A further
increase of precollapse rotation would not result in significantly
higher peak amplitudes, since j5 models are already strongly affected
by centrifugal effects, which reduce the acceleration the inner core
is experiencing at bounce, thus lead to lower GW amplitudes when
rotation begins to dominate the dynamics.

The total energy emitted in GWs is in the range $2.7 \times
10^{-11} M_\odot c^2 \lesssim E_\mathrm{GW} \lesssim 4.7 \times
10^{-8} M_\odot c^2$. Again the nonrotating and slowly rotating models
mark the lower end of this range. The upper end is set by the j4
models, since the j5 models, due to the strong influence of rotation,
have more slowly varying waveforms and lower $E_\mathrm{GW}$
($E_\mathrm{GW} \propto \int (dh/dt)^2 dt$; Eq.~\ref{eq:egw2}).

Comparing our model predictions with GW detector sensitivity is 
done best in the frequency domain. In Fig.~\ref{fig:hchar_spect} we
contrast $h_\mathrm{char}(f)$ spectra of our s12WH07j\{0-5\} model set
with the projected noise levels in Advanced LIGO (in the
zero-detuning, high-power configuration \cite{LIGO-sens-2010}; aLIGO
ZD-HP), KAGRA/LCGT \cite{lcgt:10}, and Advanced Virgo (AdV) in a
potential wide-band configuration \cite{AdV-sens-2010}. Shown are the
one-sided detector noise amplitude spectral densities $\sqrt{S(f)}$ in
units of $\mathrm{Hz}^{-1/2}$ and $h_\mathrm{char} f^{-1/2}$ of our
models (the $f^{-1/2}$ rescaling is introduced to to conform to the
units of $\sqrt{S(f)}$), assuming a source distance of
$10\,\mathrm{kpc}$. $h_\mathrm{char}$ peaks in a narrow frequency
range of about $700 - 800\,\mathrm{Hz}$ for all rotating
models. Slowly spinning models typically have their $h_\mathrm{char}$
peak at the high end of this range and the frequencies of their
spectral peaks are influenced primarily by the properties of the
nuclear EOS (not studied in detail here; see
\cite{dimmelmeier:08}). Very rapidly spinning models tend towards the
lower end and develop strong low-frequency components, which almost
reach the level of the peak around $750\,\mathrm{Hz}$ in model
s12WH07j5.

The $h_\mathrm{char}$ spectra of all models shown in
Fig.~\ref{fig:hchar_spect} have large portions that lie above the
detector noise levels. By integrating the ratio $h^2_\mathrm{char}(f)
/ (f S(f))$ over frequency (Eq.~\ref{eq:snr}) and using $S(f)$ of
Advanced LIGO in ZD-HP mode \cite{LIGO-sens-2010}, we arrive at
single-detector optimal (i.e., most optimistic) SNRs at an assumed
distance of $10\,\mathrm{kpc}$ that range from $\sim$6 for the
nonrotating model j0 to $\sim$73 for the most rapidly rotating model
j5. While these numbers appear large and suggest that Advanced
LIGO-class detectors may be able to detect the emitted waveforms, we
emphasize their ``optimal'' nature: a real astrophysical core collapse
event is highly unlikely to be optimally oriented and optimally
located on the sky with respect to a detector or a network of
detectors on Earth (see, e.g., \cite{wen:10} for a discussion of sky
coverage for detector networks).  Furthermore, the noise in
interferometric GW detectors is typically non-Gaussian,
non-stationary, and detection may be further complicated by noise
artifacts (``glitches'') that were found in first-generation detectors
(e.g., \cite{ligo:09}) and may be present also in the detectors of the
advanced generation considered here. The SNR of a real signal above
which one may be confident of a detection is typically assumed to be
$\gtrsim 8$ for Gaussian noise and at at least two detectors observing
the event in coincidence \cite{abadie:10pop}. For real noise even
higher SNRs may be required and the exact threshold will depend on
the detector network, data quality, and search methodology.

In addition to the general question of detectability, one may ask:
(\emph{1}) Would advanced GW detectors be able to observe the
pronounced postbounce variations in the GW signal of rapidly
rotating models that are correlated with variations in the neutrino
luminosity?  (\emph{2}) Would advanced GW detectors be sensitive to
effects of neutrino leakage in the very early postbounce phase out to
$25\,\mathrm{ms}$ simulated for our models?  (\emph{3}) Will advanced
GW detectors be able to tell the difference between rotating core
collapse in a $12$-$M_\odot$ and in a $40$-$M_\odot$ progenitor with
the same angular momentum distribution?

A fully reliable answer to these questions would require a Bayesian
model selection / parameter estimation approach as taken by
\cite{logue:12,roever:09}. The answer to (\emph{1}) is most likely
``Yes'' for events taking place in the Milky Way, since the GW signal
of rapidly rotating models is so strong that the entire waveform may
be recovered for a galactic event.

For (\emph{2}) and (\emph{3}) we can attempt to give a heuristic
answer on the basis of the mismatches computed between waveforms and
the general simulation results: The short answer to both is likely
``No''. The long answer is slightly more involved: As we have
demonstrated in Sections~\ref{sec:res2_leakage} and
\ref{sec:progcomp}, both neutrino leakage and differences in
progenitor structure/thermodynamics have an effect primarily on
convective/turbulent dynamics driving GW emission in the region
outside the PNS core that has stochastic character. So, despite the
sizable mismatches listed in Tab.~\ref{table:mismatch} and
Tab.~\ref{table:mismatch_s12_s40}, it will not be possible to make a
case for telling $12$-$M_\odot$/$40$-$M_\odot$ or leakage/no-leakage
apart on the basis of the GW signal alone, since the shape (and
perhaps even the overall characteristics) of the parts of the signal
that differ is fundamentally unpredictable.

\subsubsection{High-Frequency Variations in the Neutrino Luminosities}

The dominant supernova neutrino reaction in water/ice-Cherenkov
detectors is electron antineutrino capture on protons (e.g.,
\cite{scholberg:12}). We consider in our analysis the current
detectors Super-Kamiokande \cite{ikeda:07} and IceCube
\cite{icecube:11sn}, and the proposed future Hyper-Kamiokande detector
\cite{hyperkamiokande:11}.  A core collapse event anywhere in the
Milky Way is likely to be detected by these detectors with high SNR
\cite{icecube:11sn,ikeda:07,scholberg:11}. In the rapidly spinning
protoneutron stars considered in this study, neutrino fluxes and
average energies will generally be higher along the polar and lower
along the equatorial direction as demonstrated by \cite{ott:08}, who
employed fully angle-dependent neutrino transport. The leakage scheme
employed in this study does not allow us to make robust predictions of
the flux asymmetry and we thus neglect it in the following discussion
and leave its inclusion to future work.

\begin{figure}
\centering
\includegraphics[width=\linewidth]{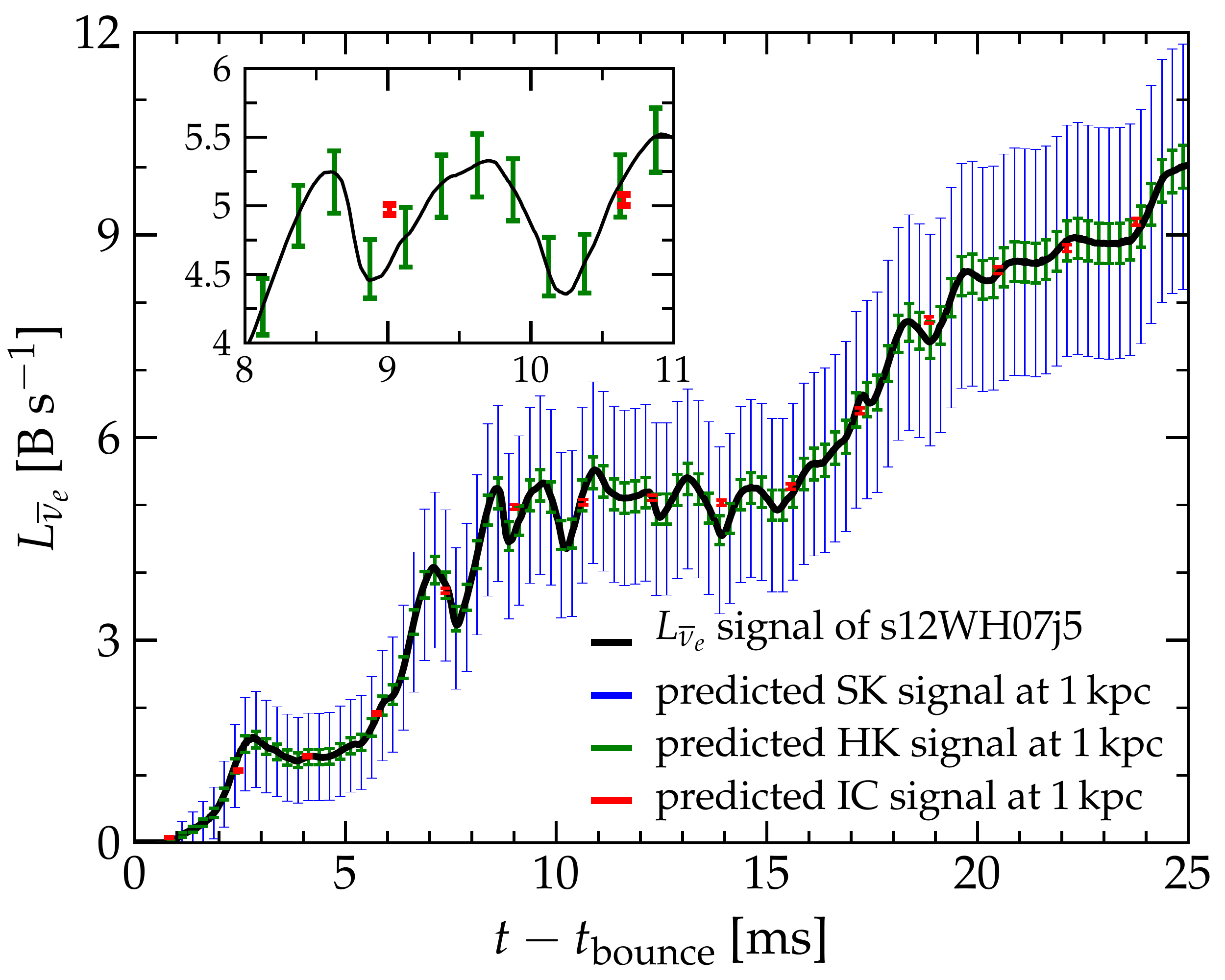}
\caption{Predicted neutrino signals and associated statistical error
  in Super-Kamiokande (SK, blue error bars), Hyper-Kamiokande (HK,
  green error bars), and IceCube (IC, red error bars) from model
  s12WH07j5 at 1\,kpc. The solid black line is the $\bar{\nu}_e$
  luminosity from our simulation.  In the inset we show only the HK
  and IC predicted signals and errors, zoomed in between 8 and 11\,ms.
  The error of the IC signal is small, but the imposed bin width of
  1.6384\,ms averages out variations on timescales smaller than
  this. If binned in 1.6384\,ms bins, the Hyper-Kamiokande SNR
  would be approximately half of the SNR in IceCube.}
\label{fig:nupredictions}
\end{figure}

Here we are interested in the question if the high-frequency
variations in the rising neutrino luminosity seen in
Fig.~\ref{fig:s12_rho_h_nu} are detectable. In consideration of the
smallness of the effect, we pick a fiducial distance of
$1\,\mathrm{kpc}$ for our estimates. More rigorous analyses of the
detectability of time variations in the supernova neutrino signal,
though for lower frequencies than considered here, were presented,
e.g., by Burrows~\emph{et al.}~\cite{burrows:92b}, Lund~\emph{et
  al.}~\cite{lund:10} and Brandt~\emph{et al.}~\cite{brandt:11}. For
simplicity, we ignore neutrino oscillations in this analysis,
but note that they may play an important role in the
  overall core-collapse supernova neutrino signature (e.g.,
  \cite{duan:10}).  Regarding the early postbounce
    signals considered here, one would expect some mixing of the
  $\bar{\nu}_e$ signal with the $\bar{\nu}_x$ signal \cite{halzen:09}.
  However, since we observe similar rotation-induced variations in the
  $\nu_x$ signal (Fig.~\ref{fig:s12_rho_h_nu}), we expect
  approximately the same results for a mixed $\bar{\nu}_e$ signal.

In Fig.~\ref{fig:nupredictions}, we plot the early $\bar{\nu}_e$
signal for model s12WH07j5 and superpose the error bars for the
statistical error of the estimated recovered luminosity in
Super-Kamiokande, Hyper-Kamiokande, and IceCube for a source distance
of $1\,\mathrm{kpc}$.  The results are very similar for model
s12WH07j4 and for the corresponding models using the s40WH07
progenitor.

For the electron antineutrino capture rate on protons in
water-Cherenkov detectors we obtain an estimate based on
\cite{vogel:99,horiuchi:09},
\begin{equation}
  R_{\bar{\nu}_e}^\text{SK/HK} \sim \frac{L_{\bar{\nu}_e}}{4\pi r^2}
  \frac{\sigma_0(1+3g_A^2)}{4m_e^2c^4} \frac{\langle
    E_{\bar{\nu}_e}^2\rangle}{\langle E_{\bar{\nu}_e} \rangle} \frac{M_\text{SK/HK}X_\text{p}}{m_\text{p}}\,\,,
\end{equation}
\noindent
where $\sigma_0 = 1.76\times10^{-44} \text{cm}^2$ is the reference
weak-interaction cross section, $g_A = -1.254$ is the axial coupling
constant, $M_\text{SK} = 22.4\,\text{kT}$, $M_\text{HK} =
740\,\text{kT}$ is the fiducial water mass of Super-Kamiokande and
Hyper-Kamiokande, respectively, $X_\text{p}$ is the number fraction of
protons (for H$_2$O, $X_\text{p} = 2/18$), $m_\text{p}$ is the proton
mass, and $\langle E^2\rangle/\langle E \rangle$ is the energy
averaged spectral factor.  We take $T_{\bar{\nu}_e}\,\sim$\,\,4\,MeV
and zero chemical potential as an approximation, giving $\langle
E^2\rangle/\langle E \rangle \sim 16.4\,$MeV.  With these numbers, for
Super-Kamiokande, $R_{\bar{\nu}_e}^\text{SK} \sim 12000\ \text{s}^{-1}
L_{\bar{\nu}_e, 51} / r_\text{1\,kpc}^2$ and for Hyper-Kamiokande,
$R_{\bar{\nu}_e}^\text{HK} \sim 400000\ \text{s}^{-1} L_{\bar{\nu}_e,
  51} / r_\text{1\,kpc}^2$, where $L_{\bar{\nu}_e, 51}$ is
  the electron antineutrino luminosity in units of
  $10^{51}\text{erg}\,\,\text{s}^{-1}$ and $r_\text{1\,kpc}$ is the
  distance in kpc.

In Super/Hyper-Kamiokande, electron antineutrino events are detected
individually, therefore, if the signal is strong enough, one can bin
the events in temporal bins of almost arbitrarily small width. To
capture the rotation-induced variations, we choose bin widths of
0.25\,ms. As seen in Fig.~\ref{fig:nupredictions}, within statistical
errors, Super-Kamiokande is insensitive to the variations, even at
1\,kpc for model s12WH07j5.  Hyper-Kamiokande, however, has $\sim$$33$
times the sensitivity of Super-Kamiokande and would be able to
marginally detected our predicted oscillations if the collapsing core
was rapidly rotating and within 1\,kpc. Super-Kamiokande and
Hyper-Kamiokande are sensitive to the total electron antineutrino
signal at 25\,ms with an estimate signal to noise ratio of 6 and 35,
respectively, at this distance.

For IceCube we use the rate prediction of \cite{halzen:09}. The
expected event rate per 1.6384\,ms timing sample in IceCube from
electron antineutrino interactions in the ice is,
\begin{equation}
R_{\bar{\nu}_e}^\text{IceCube} = 1860 \left(\frac{L_{\bar{\nu}_e}}{1\,\text{B}\
  \text{s}^{-1}}\right)\left(\frac{1\text{kpc}}{r}\right)^2\frac{\langle
  E_{15}^3\rangle}{\langle E_{15}\rangle}\, \text{bin}^{-1}
\end{equation}
\noindent
where, following \cite{halzen:09}, we assume $\langle E \rangle =
15$\,MeV and take the spectral factor $\langle E^3\rangle/\langle
E\rangle^3 \sim 2$, this gives $R^\text{IC}_{\bar{\nu}_e} \sim
2.3\times10^6\ \text{s}^{-1}L_{\bar{\nu}_e, 51} /
r_\text{1\,kpc}^2$. We estimate the errors based on IceCube's
background root-mean-square scatter of $\sim 47$ events
$\text{bin}^{-1}$ \cite{halzen:09} and the square root of the expected
count rate. IceCube is very sensitive to the overall $L_{\bar{\nu}_e}$
signal of model s12WH07j5 at a distance of 1\,kpc ($L_{\bar{\nu}_e} =
9.20\pm0.05\ \text{B}\ \text{s}^{-1}$ at 24\,ms after bounce). This is
due to both the large effective volume that IceCube covers,
$\sim$3.5\,Mt or $\sim 5 \times M_\text{HK}$, and the current relatively
large (compared to our choice for Super- and Hyper-Kamiokande) 
1.6384\,ms binning of the IceCube detector.  Unfortunately,
this large bin width makes it impossible to resolve the short-period
(1.2--1.3\,ms) variations in $L_{\bar{\nu}_e}$ in IceCube, even if one
applied a sub-sampling method as proposed by \cite{dasgupta:09qcd}.
We also note that at close distances ($\lesssim$1$-$few\,kpc),
detector saturation due to the high event rate may
prevent an accurate determination of the neutrino luminosity
\cite{icecube:11sn}.


\section{Summary and Conclusions}
\label{sec:summary}

We have presented results from a new set of (octant) 3D fully
general-relativistic (GR) rotating core collapse simulations. These
simulations are the first to be carried out with a
microphysical equation of state (EOS), progenitor models from stellar
evolution simulations, a parameterized treatment of deleptonization
during collapse and deleptonization and neutrino cooling/heating in
the postbounce phase via an approximate neutrino leakage scheme. Our
simulations track collapse, bounce, and the first $25\,\mathrm{ms}$ of
the postbounce phase. We considered precollapse rotation rates that lead
to early postbounce protoneutron star (PNS) spin periods as short as
$\sim$$1.5\,\mathrm{ms}$. Such rapid rotation is not expected for
garden-variety core-collapse supernovae (e.g.,
\cite{ott:06spin,heger:05}), but may be highly relevant for
potentially jet-driven explosions in the context of hyper-energetic
core-collapse supernovae and long gamma-ray bursts. Our simulations
are an extension of previous work by Dimmelmeier~\emph{et
  al.}~\cite{dimmelmeier:08}, who performed a large suite of 2D
conformally-flat GR rotating core collapse simulations with similar
microphysics and progenitor models, but neglected postbounce neutrino
leakage.

The major new result of this study is the discovery of a systematic
correlation between high-frequency ($\sim$$700-800\,\mathrm{Hz}$)
variations in the gravitational-wave (GW) signal and in the
luminosities of electron antineutrinos and heavy-lepton neutrinos.
These are induced by non-linear global oscillations of the PNS that
are excited at core bounce in rapidly rotating cores.  Detailed
analysis of the nature of the rotationally-induced PNS oscillation
indicates that the excited mode is the fundamental quadrupole
pulsation mode of the PNS. We find that an inner core ratio of
rotational kinetic energy to gravitational energy ($T/|W|$) at bounce
of $\gtrsim$$5\%$ (corresponding to a PNS with spin period
$\lesssim$$2.5\,\mathrm{ms}$ in our models) is required for the
oscillations to be excited at significant amplitudes. This requires an
inner core angular momentum $\gtrsim 2\times 10^{48}\,
\mathrm{g\,cm}^2\,\,{s}^{-1}$ at an inner core mass of $\sim
0.5-0.7\,M_\odot$ at bounce and is independent of other progenitor
properties.

Thus, neutrinos and gravitational waves can both be used to probe
rapid rotation in core collapse and combined can provide smoking-gun
evidence for core rotation at the level required to drive
magnetorotational explosions in the context of hyper-energetic
core-collapse supernovae (e.g., \cite{burrows:07b,takiwaki:11}).
However, while the GW signal is likely to be detectable throughout the
Milky Way by the advanced generation of laser interferometer GW
observatories, a water-Cherenkov detector would have to be of the
scale of the planned Hyper-Kamiokande to observe the oscillations in
the electron antineutrino luminosity at a source distance of only
$1\,\mathrm{kpc}$. IceCube, on the other hand, would in principle have
the sensitivity to detect the oscillations at distances of
$\sim$$1-2\,\mathrm{kpc}$, but lacks the necessary timing resolution.

Each model simulation carried out in this study was run once with and
once without postbounce neutrino leakage. Comparing the two cases, we
find that deleptonization and neutrino cooling/heating in the region
behind the stalled shock have a strong effect on the radius of the
stalled shock, on the convective dynamics in the region behind the
shock, and on the associated GW signal.  The GW signal from convective
dynamics dominates the GW emission in nonrotating and slowly rotating
cores, but its detailed shape is impossible to predict exactly, due to
its inherently stochastic nature and its dependence on the unknown
magnitude and location of seed perturbations and the detailed
thermodynamics of the collapsed core. In moderately rapidly and
rapidly rotating models (with $T/|W|$ at bounce $\gtrsim$$1.5\%$), the
GW emission is dominated by the dynamics of the inner core in which
neutrinos are trapped and leak out only on a diffusion timescale. The
overall GW signal of these models shows little sensitivity to
postbounce neutrino leakage, which lends credence to the results of
Dimmelmeier~\emph{et al.}~\cite{dimmelmeier:08}, who neglected
postbounce leakage.

Investigating the dependence of rotating collapse dynamics and GW
signal on the progenitor star zero-age-main-sequence mass, we
performed simulations with a $12$-$M_\odot$ and a $40$-$M_\odot$
presupernova model drawn from \cite{woosley:07}. For each progenitor
model, we carried out six simulations, varying the precollapse
rotation from zero to very rapid and setting up rotation in a way to
ensure that both progenitors were endowed with nearly the same angular
momentum as a function of enclosed mass.

The results of this progenitor comparison show that the GW signal of
rapidly rotating core collapse and the early postbounce evolution is
essentially independent of progenitor structure and thermodynamics and
is determined primarily by the mass and angular momentum of the inner
core at bounce. This finding is in contrast to previous work in which
progenitor dependence was studied, but in which rotation was
parameterized either by the precollapse central angular velocity
$\Omega_{c,\text{initial}}$ \cite{dimmelmeier:08} or by choosing an
initial value of $T/|W|$. Both $\Omega_{c,\text{initial}}$ and initial
$T/|W|$ are not directly observable in nature and both lead to inner
core angular momentum that depends sensitively on how far the inner
regions of the presupernova models have already collapsed when
rotation is imposed at the time of mapping from the stellar evolution
code into the collapse code. This introduces an essentially arbitrary
factor in the rotational setup and makes it impossible to infer the
true precollapse rotational configuration from GW (and neutrino)
observations.  In previous work, this systematic problem with
$\Omega_{c,\text{initial}}$ and initial $T/|W|$ as parameters for
rotation was not fully realized, which has lead to misinterpretations
of simulation results.

\vskip.25cm Two aspects of rotating core collapse evolution not
addressed in this study are the dependence of the dynamics and
resulting GW signal on the nuclear EOS and on the degree of
precollapse differential rotation. The dependence on the former has
been studied to some extent by \cite{dimmelmeier:08,scheidegger:10b},
who found quantitative, but not qualitative, differences in the
emitted GW signals that may be recovered by GW 
observations of a nearby event \cite{roever:09}. Since the frequency
of the fundamental quadrupole pulsation mode depends on PNS structure,
one would expect at least a slight variation with EOS of the
early-postbounce GW and neutrino signal frequencies in rapidly
spinning models. This is an interesting possibility that should be
considered in future work.  Strong differential rotation, though
currently not deemed likely in precollapse iron cores, could
potentially lead to significant qualitative differences that so far
have been explored only with simpler approaches
\cite{dimmelmeier:02,kotake:03,ott:04,ott:09}.  Furthermore, for
application of simulated GW signals in advanced GW detector data
analysis (e.g., \cite{roever:09,logue:12}), extensive fine-grained
parameter studies with probably hundreds of simulated models are
necessary.  The twelve new waveforms resulting from this study, which
we make available at \cite{ottcatalog}, are too few and too coarsely
sample the parameter space to provide useful guides for GW data
analysis.

The work presented in this paper marks a sizable, though
incremental, step towards realistic 3D GR models of rotating core
collapse and postbounce supernova evolution and reliable predictions
of GW and neutrino signals.  It complements the recent groundbreaking
study of Kuroda~\emph{et al.}~\cite{kuroda:12}, who performed 3D GR
simulations at modest resolution to $\sim$$100\,\mathrm{ms}$ after
bounce without symmetry constraints and with a single nonrotating
progenitor, using similar microphysical detail and neutrino
approximations. 

Since our focus here has been on the collapse and early postbounce
evolution during which 3D and magnetohydrodynamic (MHD) effects are
likely small \cite{ott:07prl,scheidegger:10b,burrows:07b}, the most
severe limitation of our work is the reliance on parameterized
deleptonization during collapse and on the energy-average neutrino
leakage scheme for deleptonization, cooling, and heating after bounce.
While we expect the qualitative features and quantitative trends to be
robust and hold when a more accurate neutrino treatment is employed,
the technical limitations of our simulations must be taken into
account when considering the results of this study. In particular, the newly
identified correlation between GW signal and neutrino luminosities in
rapidly rotating cores must receive further scrutiny, for example, by
the 2D conformally-flat GR approach of M\"uller~\emph{et
  al.}~\cite{mueller:12a} that implements neutrino-transport in a
ray-by-ray two-moment approach with a variable Eddington factor
determined by a full solution of the Boltzmann equation.  This
approach would allow one to also extract the low-frequency GW signal due
to anisotropic neutrino emission, which is likely to be of sizable
amplitude due to the large rotationally-induced flux anisotropy in
rotating models (e.g., \cite{ott:08,ott:09}), but could not be
considered here due to the limitations of our leakage scheme.

Our future work in 3D GR will be directed towards extending our
simulations to full 3D and including MHD to study 3D rotational
instabilities and magnetic-field-driven dynamics. We will also explore
improved treatments of neutrino transport that are computationally
efficient and scale to full 3D simulations, e.g., the Monte Carlo
approach to neutrino transport laid out by Abdikamalov~\emph{et
  al.}~\cite{abdikamalov:12}.


\acknowledgments

We are happy to acknowledge helpful exchanges with Y.~Chen, C.~Cutler,
L.~Dessart, I.~Hawke, W.~Kastaun, F.~L\"offler, C. Meakin, P.~M\"osta,
N.~Stergioulas, P. Ajith, and D.~Tsang.  This work is supported
by the National Science Foundation under grant numbers AST-0855535,
OCI-0905046, and OCI-0941653, and
by the Sherman Fairchild Foundation. We wish to thank Chris Mach for
support of the group servers at TAPIR on which much of the code
development and testing was carried out. Results presented in this
article were obtained through computations on the Caltech compute
cluster ``Zwicky'' (NSF MRI award No.\ PHY-0960291), on the NSF XSEDE
network under grant TG-PHY100033, on machines of the Louisiana Optical
Network Initiative under grant loni\_numrel07, and at the National
Energy Research Scientific Computing Center (NERSC), which is
supported by the Office of Science of the US Department of Energy
under contract DE-AC03-76SF00098.

\end{document}